%% file: ttHf_paper.tex
\documentclass[cernpreprint,texlive=2011,txfonts,UKenglish]{atlasdoc} 
\pdfoutput=1
\usepackage[subfigure=true,biblatex=false]{latex/atlaspackage} 
\usepackage{latex/atlasphysics} 
\usepackage{latex/atlasjournal} 
\usepackage{latex/atlasmisc} 
\usepackage{latex/atlasparticle} 
\usepackage{latex/atlasunit} 
\usepackage{latex/atlasxref} 
\usepackage{multirow}
\usepackage{rotating}
\usepackage{relsize}
\usepackage{siunitx}
\usepackage{dcolumn}
\usepackage{a4wide}
\newcolumntype{\pm}{D{*}{\,\pm\,}{-1,-1}}

\usepackage[numbers,sort&compress]{natbib}

\graphicspath{{logos/}{plots/}}

\newcommand\ttbt{\ensuremath{ttb}}
\newcommand\ttct{\ensuremath{ttc}}
\newcommand\ttlt{\ensuremath{ttl}}
\newcommand\ttbbt{\ensuremath{ttbb}}
\newcommand\ttbXt{\ensuremath{ttbX}}
\newcommand\ttcXt{\ensuremath{ttcX}}
\newcommand\ttlXt{\ensuremath{ttlX}}

\newcommand{\ttbb}{\ensuremath{t\bar{t}b\bar{b}}}

\newcommand{\sttbb}{\ensuremath{\sigma_{t\bar{t}b\bar{b}}}}

\newcommand\ttv{\ensuremath{t\bar{t}V}}

\newcommand\ttz{\ensuremath{t\bar{t}Z}}
\newcommand\tth{\ensuremath{t\bar{t}H}}

\newcommand{\beq}{\begin{equation}}
\newcommand{\eeq}{\end{equation}}

\newcommand{\AtlasCoordFootnote}{%
ATLAS uses a right-handed coordinate system with its origin at the nominal interaction point (IP)
in the centre of the detector and the $z$-axis along the beam pipe.
The $x$-axis points from the IP to the centre of the LHC ring,
and the $y$-axis points upwards.
Cylindrical coordinates $(r,\phi)$ are used in the transverse plane, 
$\phi$ being the azimuthal angle around the $z$-axis.
The pseudorapidity is defined in terms of the polar angle $\theta$ as $\eta = -\ln \tan(\theta/2)$.
Angular distance is measured in units of $\Delta R \equiv \sqrt{(\Delta\eta)^{2} + (\Delta\phi)^{2}}$.}

\newcommand{\ttbLjets}{950 $\pm$ 70~(stat.) $^{+240}_{-190}$~(syst.)~fb}
\newcommand{\ttbDil}{50 $\pm$ 10~(stat.) $^{+15}_{-10}$~(syst.)~fb}
\newcommand{\ttbbCc}{19.3 $\pm$ 3.5~(stat.) $\pm$ 5.7~(syst.)~fb}
\newcommand{\ttbbFit}{13.5 $\pm$ 3.3~(stat.)~$\pm$ 3.6~(syst.)~fb}
\newcommand{\ttbbFitRatio}{1.30 $\pm$ 0.33~(stat.) $\pm$ 0.28~(syst.)\%}

\AtlasTitle{Measurements of fiducial cross-sections for \ttbar\ production with one or two additional $b$-jets in $pp$ collisions at $\sqrt{s}$=8 \TeV~using the ATLAS detector}

\author{The ATLAS Collaboration}

\AtlasAbstract{
Fiducial cross-sections for $\ttbar$ production with one or two additional $b$-jets are reported, using an integrated luminosity of 20.3 fb$^{-1}$ of proton--proton collisions at a centre-of-mass energy of 8 \TeV~at the Large Hadron Collider, collected with the ATLAS detector.
The cross-section times branching ratio for $\ttbar$ events with at least one additional $b$-jet is measured to be \ttbLjets\ in the lepton-plus-jets channel and \ttbDil\ in the  $e \mu$ channel. 
The cross-section times branching ratio for events with at least two additional $b$-jets is measured to be \ttbbCc\ in the dilepton channel ($e \mu$,\,$\mu\mu$, and \,$ee$) using a method based on tight selection criteria, and \ttbbFit\ using a looser selection that allows the background normalisation to be extracted from data.
The latter method also measures a value of \ttbbFitRatio\ for the ratio of \ttbar\ production with two additional $b$-jets to \ttbar\ production with any two additional jets.
All measurements are in good agreement with recent theory predictions.
}

\PreprintIdNumber{CERN-PH-EP-2015-200}
\AtlasJournal{EPJC}

\begin{document}

\maketitle

\tableofcontents
\clearpage

\input{intro}

\input{fiducial_definition}

\input{detector}

\input{samples}

\input{object_reconstruction}

\input{sys_uncertainties}

\input{mv1c_template}

\input{analysis_method_ttb_ljets}

\input{analysis_method_ttbb_cc}
\input{analysis_method_ttbb_fit}

\input{results}

\section{Conclusions}\label{sec:Conclusions}
Measurements in the fiducial phase space of the detector of the cross-sections for the production of \ttbar\ events with one or two additional $b$-jets are performed in proton--proton collisions at a centre-of-mass energy of 8 \TeV~at the LHC. 
The results are based on a dataset corresponding to an integrated luminosity of 20.3 fb$^{-1}$, collected with the ATLAS detector. 
The cross-section times branching ratio for top pair events with at least one additional $b$-jet is measured to be \ttbLjets\ in the lepton-plus-jets channel and \ttbDil\ in the $e \mu$ channel. 
The cross-section times branching ratio with at least two additional $b$-jets is measured to be \ttbbCc\ in the dilepton channel ($e \mu$,\,$\mu\mu$, and\,$ee$) using a method based on tight selection criteria, and \ttbbFit\ using a looser selection which allows extraction of the background normalisation from data.
A measurement of the ratio of \ttbar\ production with two additional $b$-jets to \ttbar\ production with any two additional jets is also performed; this ratio is found to be \ttbbFitRatio.
The measurements are found to agree within their uncertainties with NLO+PS calculations of the $pp\rightarrow\ttbb$ process, as well as with merged LO+PS calculations of $pp\rightarrow t\bar{t}+\leq 3$ jets, favouring the predictions obtained with soft renormalisation/factorisation scales. The measurements are shown to be sensitive to the description of $g\rightarrow b\bar{b}$ splitting in the parton shower, with the most extreme {\sc Pythia}~8 model being disfavoured by the measurements.

\section*{Acknowledgements}
\input{Acknowledgements}


\bibliographystyle{atlasBibStyleWithTitle}
\bibliography{ttHf_paper}

\newpage

\input{atlas_authlist}

\end{document}

%% file: intro.tex
\section{Introduction}

The measurement of top quark pair (\ttbar) production in association with one or more jets containing \mbox{$b$-hadrons} (henceforth referred to as $b$-jets) is important in providing a detailed understanding of quantum chromodynamics (QCD). 
The most accurate theoretical predictions for these processes are fixed-order calculations at next-to-leading order (NLO) accuracy~\cite{NLO1,NLO2,NLO3} in perturbative QCD (pQCD), which have been matched to a parton shower \cite{NLOPS1,NLOPS2,NLOPS3}. 
These calculations have significant uncertainties from missing higher-order terms~\cite{Alwall:2014hca,Garzelli:2014aba}, making direct experimental measurements of this process desirable.
The measurement of such cross-sections in fiducial phase-spaces, defined to correspond as closely as possible to the acceptance of the ATLAS detector, can be compared to theoretical predictions using the same fiducial requirements. This minimises theoretical extrapolations to phase-space regions that are not experimentally measurable.

Moreover, following the discovery of the Higgs boson~\cite{ATLASHiggs,CMSHiggs}, the Standard Model prediction for the top quark Yukawa coupling can be tested via a measurement of the \tth\ associated production cross-section.
Due to the large Higgs branching ratio to $b$-quarks, the $\tth \to t\bar{t}b\bar{b}$ channel is promising, but suffers from a large and poorly constrained background of events with top pairs and additional $b$-jets from QCD processes~\cite{Khachatryan:2014qaa,ttHMEMCMS, atlasttH}. 
 
Measurements of \ttbar\ production with additional heavy-flavour jets have been performed by ATLAS at $\sqrt{s}= 7$~\TeV~\cite{ATLASttbb7TeV} and CMS at $\sqrt{s}$ = 8~\TeV~\cite{CMSttbb8TeV,CMSttbb8TeV2}.
The ATLAS measurement reported a ratio of heavy flavour to all jets produced in association with a \ttbar\ pair where heavy flavour includes both bottom jets as well as charm jets. 
The CMS measurement is a fiducial measurement of events with two leptons and four or more jets, of which at least two are identified as containing a $b$-hadron.

This paper presents measurements of fiducial cross-sections for \ttbar\ production in association with one or two additional $b$-jets. 
Because the top quark decays almost exclusively to a $b$-quark and a $W$ boson, these processes have three or four $b$-jets in the final state.
The particle-level objects are required to be within the detector acceptance of $|\eta|< 2.5$, where $\eta$ is the pseudorapidity.\footnote{\AtlasCoordFootnote} The jets are required to have transverse momenta above 20~\GeV~and the electrons and muons to have transverse momenta above 25~\GeV.
The lepton-plus-jets and dilepton ($e \mu$) channels\footnote{Unless otherwise specified, ``leptons'' refers exclusively to electrons and muons. The top quark pair production channels are labelled according to the decay of the two $W$ bosons. The lepton-plus-jets channel refers to events where one $W$ boson from a top quark decays to hadrons, the other to an electron or muon (either directly or via a $\tau$ lepton). Dilepton events are those in which both $W$ bosons decay to an electron or muon.}  are used to perform two measurements of the cross-section for the production of \ttbar\ events with at least one additional $b$-jet.
In both cases, the signal cross-section is extracted from a fit to a multivariate discriminant used to identify $b$-tagged jets~\cite{MV1}. 
The lepton-plus-jets channel has a higher acceptance times branching ratio, but suffers from a significant background of events in which the $W$ boson decays to a $c$- and a light quark. 

Two analysis techniques are used in the dilepton channel ($ee$, $\mu \mu$ and $e \mu$) to measure a cross-section for the production of \ttbar\ events with two additional $b$-jets. 
The first, referred to as the cut-based analysis, applies very tight selection criteria including a requirement of four $b$-tagged jets. This analysis results in a high signal-to-background ratio and relies on the Monte Carlo (MC) estimates of the background, including the \ttbar\ background with additional jets containing $c$-quarks ($c$-jets) or only light quarks and gluons (light jets). 
The second applies a looser selection and extracts the signal cross-section from a fit to the distribution of a multivariate $b$-jet identification discriminant. 
This second method, referred to as the fit-based analysis, confirms the validity of the background predictions used in the cut-based approach, and offers a measurement of the ratio of cross-sections for events with two additional $b$-jets and all events with two additional jets.

The fiducial measurements are made considering both electroweak  (e.g. from $Z$ boson decays) and QCD production of the additional $b$-quarks as signal. 
In order to compare to NLO pQCD theory predictions, the measurements are also presented after subtracting the electroweak processes, \ttv\ ($V$ corresponding to a $W$ or $Z$ boson) and \tth .

The paper is organised as follows.
First, the definitions of the fiducial regions are given in Section~\ref{sec:FiducialDefinition}.
The ATLAS detector is briefly described in Section~\ref{sec:atlas}, 
followed in Section~\ref{sec:DataMCSamples} by a description of the data and simulated samples used.
Section~\ref{sec:ObjectReconstruction} describes the reconstruction of physics objects in the detector and presents the event selection used.
The sources of systematic uncertainties affecting the measurements are described in Section~\ref{sec:SystematicUncertainties}. 
Section~\ref{sec:analysis_methods} describes the analysis techniques used to extract the cross-sections and their uncertainties.
The final cross-sections are presented in Section~\ref{sec:Results} and compared to recent theoretical predictions. 
Finally, Section~\ref{sec:Conclusions} gives brief conclusions.

%% file: fiducial_definition.tex
\section{Measurement definition}\label{sec:FiducialDefinition}

This section details the particle-level fiducial phase-space definitions.
Particle-level object definitions that are common to all measurements are described in Section~\ref{sec:particle_objects}. The particle-level event selection is then discussed in Section~\ref{sec:particle_event_def}, describing first the fiducial selection used to define the cross-section, and then, where relevant, the selection used to define the templates that are fit to the data.

\subsection{Particle-level object definitions}\label{sec:particle_objects}
The particle-level definition of objects is based on particles with a proper lifetime $\tau_{\mathrm{particle}} > 3\times 10^{-11}$~s. 
The definitions used here follow very closely previous ATLAS \ttbar\ fiducial definitions~\cite{ttjets}.
Fiducial requirements are placed only on jets and charged leptons.

\noindent \textbf{Electrons and muons:} 
Prompt electrons and muons, i.e. those that are not hadron decay products, are considered for the fiducial lepton definition.
Electrons and muons are dressed by adding to the lepton the four-vector momenta of photons within a cone of size $\Delta R=0.1$ around it. 
Leptons are required to have $\pt > 25$ \GeV~and $|\eta| < 2.5$.

\noindent \textbf{Jets:} Jets are obtained by clustering all stable particles, except the leptons, dressed with their associated photons, and neutrinos that are not hadron decay products, using the anti-$k_t$ algorithm~\cite{ref:Cacciari2008,ref:Cacciari2006,ref:fastjet} with a radius parameter $R=0.4$.
Particles from the underlying event are included in this definition, whereas particles from additional inelastic proton--proton collisions (pile-up) are not included. 
The products of hadronically decaying $\tau$ leptons are thus included within jets. Photons that were used in the definition of the dressed leptons are excluded from the jet clustering.
Particle jets are required to have $\pt> 20$ \GeV~and $|\eta| < 2.5$. 
The \pt\ threshold for particle-level jets is optimised to reduce the uncertainty of the measurement; it is chosen to be lower than the 25 \GeV~threshold used for reconstructed jets (see Section~\ref{sec:ObjectReconstruction}), as jets with a true \pt\ just below the reconstruction threshold may satisfy the event selection requirement due to the jet energy resolution.
This effect is enhanced by the steeply falling \pt\ spectra for the additional jets.
A similar choice is not necessary for electrons and muons due to their better energy resolution.

\noindent \textbf{Jet flavour identification:} 
A jet is defined as a $b$-jet by its association with one or more $b$-hadrons with $\pt >$ 5~\GeV.
To perform the matching between $b$-hadrons and jets, the magnitudes of the four-momenta of $b$-hadrons are first scaled to a negligible value (in order to not alter normal jet reconstruction), and then the modified $b$-hadron four-momenta are included in the list of stable particle four-momenta upon which the jet clustering algorithm is run, a procedure known as {\it ghost-matching}~\cite{Cacciari:2008gn}.
If a jet contains a $b$-hadron after this re-clustering, it is identified as a $b$-jet; 
similarly, if a jet contains no $b$-hadron but is ghost-matched to a $c$-hadron with $\pt > 5$~\GeV, it is identified as a $c$-jet. 
All other jets are considered light-flavour jets.

\noindent \textbf{Overlap between objects:} In order to ensure isolation of all objects considered, events are rejected if any of the jets satisfying the fiducial requirements lie within $\Delta R=0.4$ of a dressed, prompt lepton.

\subsection{Fiducial event selection}\label{sec:particle_event_def}
The fiducial object definitions given above are used to classify events as signal or background. This is described in Section~\ref{sec:signal_fiducial}. Section~\ref{sec:template_fiducial} defines the templates used in the fit-based measurements. 

\subsubsection{Signal event selection}\label{sec:signal_fiducial}
The signal definitions are related to the fiducial definition of either a lepton-plus-jets or a dilepton \ttbar\ decay topology with at least one or at least two extra jets. 
The classification is based on the number of leptons and the number and flavour of the jets passing the fiducial object selection.
Cross-section measurements are reported in the following three fiducial phase-spaces:
\begin{itemize}
\item {\bf \ttbt\ lepton-plus-jets} refers to \ttbar\ events with exactly one lepton and at least five jets, of which at least three are $b$-jets;
\item {\bf \ttbt\ $e \mu$} refers to \ttbar\ events with one electron, one muon, and at least three $b$-jets;
\item {\bf \ttbbt\ dilepton } refers to \ttbar\ events with two leptons and least four $b$-jets.
\end{itemize}

For the $ttbb$ fiducial region, additional requirements are placed on the invariant mass of the lepton pair.
For all flavours of lepton pairs, the invariant mass of the two leptons ($m_{\ell\ell}$) must be above 15~\GeV. In events with same-flavour leptons, $m_{\ell\ell}$ must also satisfy $|m_{\ell\ell} - m_{Z} | > 10$~\GeV, where $m_{Z}$ is the mass of the $Z$ boson.
Table~\ref{tab:fid_sel} summarises the fiducial definition of all three phase-spaces.

\begin{table}[h!]
\centering
\begin{tabular}{l ccc}
\hline\hline
Fiducial & $ttb$  & $ttb$  & $ttbb$   \\
Requirement & lepton-plus-jets & $e \mu$ & dilepton  \\ \hline
$N_{\mathrm{leptons}}$ ($\pt> 25$ \GeV, $|\eta| < 2.5$)& $1$ & $2$ & $2$  \\
Lepton flavours & $e$ and $\mu$ & $e \mu$ only & $ee$, $\mu\mu$ and $e\mu$ \\
$m_{\ell\ell} > 15$~\GeV & - & - & yes \\
$|m_{ee / \mu\mu}-91 \mathrm{\GeV}|>10$~\GeV & - & - & yes \\
$N_{\mathrm{jets}}$ ($\pt> 20$ \GeV, $|\eta| < 2.5$) & $\geq 5$ & $\geq 3$ & $\geq 4$ \\
$N_{b-\mathrm{jets}}$ & $\geq 3$  & $\geq 3$ & $\geq 4$ \\
$\Delta R_{\ell,j}>0.4$ & yes & yes & yes \\
\hline \hline
\end{tabular}
\caption{Summary of the three sets of fiducial selection criteria employed for the $ttb$ and $ttbb$ cross-section measurements. 
The jet--lepton isolation ($\Delta R_{\ell,j}$) requires $\Delta R > 0.4$ between any of the jets and the leptons.}
\label{tab:fid_sel}
\end{table}

\subsubsection{Template definitions}\label{sec:template_fiducial}
The measurements based on fits determine the signal and background contributions using templates of the $b$-tagging discriminant for the various categories of events.
Because $b$-jets, $c$-jets and light jets give different distributions for the discriminant, the non-signal \ttbar\ events are split according to the flavour of the additional jet(s) in the event. 

In particular, the $ttb$ analyses define the signal template (\ttbt) using the same requirements on the jets as used for the cross-section definition, and similar templates are defined for $c$-jets (\ttct) and light jets (\ttlt).
With two additional jets, the $ttbb$ fit-based measurement has a larger number of possible flavour combinations. 
The templates of different combinations are merged if they have similar shapes and if they are produced through similar processes.
This results in four templates: \ttbbt, \ttbXt, \ttcXt\ and \ttlXt.

In addition, because the lepton kinematics do not significantly affect the distributions of the $b$-jet discriminant, 
the dilepton fit measurements do not include the lepton requirements in the template definitions.
For these analyses, a correction for the fiducial acceptance of the leptons thus needs to be applied ($f_{\mathrm{fid}}$).
The $ttb$ lepton-plus-jets analysis uses the same lepton requirements in defining the templates as are used for the signal definition.

Table~\ref{tab:signal_definition} shows the complete set of criteria used in the fiducial definitions of the various templates.
For the lepton-plus-jets analysis, contributions from $W \rightarrow cq$ ($q = s, d$) decays where the $c$-hadron is matched to one of the fiducial jets are included in the \ttct\ template; this contribution is found to dominate over that from \ttbar\ with additional heavy flavour.

The $ttbb$ cut-based measurement does not make use of templates for fitting. Events are considered as signal if they meet the definition of $ttbb$ in Section~\ref{sec:signal_fiducial}; all other \ttbar\ events are considered background.

\begin{table}[h!]
\begin{center}
\begin{tabular}{l|l}
\hline\hline
Shorthand notation  & Particle-level event requirements\\
 for the templates & \\\hline
\multicolumn{2}{c}{$ttb$ lepton-plus-jets}\\ \hline
 \ttbt\  &  $n_{\mathrm{leptons}} =1$, $n_{\mathrm{jets}} \geq 5$ and  $n_{b\mathrm{-jets}} \geq 3$ \\
 \ttct\   		&  $n_{\mathrm{leptons}} =1$, $n_{\mathrm{jets}} \geq 5$ and  $n_{b\mathrm{-jets}}=2$ and $n_{c\mathrm{-jets}}\geq 1$ \\
 \ttlt\               & other events \\
\hline\hline
\multicolumn{2}{c}{$ttb$ $e \mu$}\\ \hline
 \ttbt\    		&  $n_{\mathrm{jets}} \geq 3$ and  $n_{b\mathrm{-jets}}\geq 3$ \\
  \ttct\    		&  $n_{\mathrm{jets}} \geq 3$ and  $n_{b\mathrm{-jets}}\leq2$ and $n_{c\mathrm{-jets}}\geq 1$ \\
  \ttlt\             	& other events \\
\hline\hline
\multicolumn{2}{c}{$ttbb$ dilepton fit-based}\\ \hline
 \ttbbt\   		&  $n_{\mathrm{jets}} \geq 4$ and  $n_{b\mathrm{-jets}}\geq 4$ \\
 \ttbXt\   		& $n_{b\mathrm{-jets}}=3$  \\
 \ttcXt\  		&  $n_{b\mathrm{-jets}}= 2$ and $n_{c\mathrm{-jets}} \geq 1 $  \\
 \ttlXt\               & other events \\ \hline\hline
\end{tabular}
\caption{Particle-level definitions used to classify selected \ttbar\ events into templates for the likelihood fits. The categories depend on the number of jets and number of $b$- and $c$-jets within the fiducial region.}
\label{tab:signal_definition}
\end{center}
\end{table}

%% file: detector.tex
\section{ATLAS detector}
\label{sec:atlas}
The ATLAS detector~\cite{atlas-detector} at the LHC covers nearly the entire solid angle around the collision point.
It consists of an inner tracking detector surrounded by a thin superconducting solenoid, electromagnetic and hadronic calorimeters,
and a muon spectrometer incorporating three large superconducting toroid magnets.
The inner-detector system (ID) is immersed in a \SI{2}{\tesla} axial magnetic field 
and provides charged-particle tracking in the range $|\eta| < 2.5$.

A high-granularity silicon pixel detector covers the vertex region and typically provides three measurements per track, 
the first hit being normally in the innermost layer.
This pixel detector is important for the reconstruction of displaced vertices used to identify jets containing heavy-flavour hadrons.
It is followed by a silicon microstrip tracker, which has four layers in the barrel region.
These silicon detectors are complemented by a transition radiation tracker,
which enables radially extended track reconstruction up to $|\eta| = 2.0$. 
The transition radiation tracker also provides electron identification information 
based on the fraction of hits (typically 30 in total) above a higher energy-deposit threshold corresponding to transition radiation.
The ID reconstructs vertices with spatial resolution better than 0.1 mm in the direction longitudinal to the beam for vertices with more than ten tracks.

The calorimeter system covers the pseudorapidity range $|\eta| < 4.9$.
Within the region $|\eta|< 3.2$, electromagnetic calorimetry is provided by barrel and 
endcap high-granularity lead/liquid-argon (LAr) electromagnetic calorimeters,
with an additional thin LAr presampler covering $|\eta| < 1.8$,
to correct for energy loss in material upstream of the calorimeters.
Hadronic calorimetry is provided by a steel/scintillating-tile calorimeter,
segmented into three barrel structures within $|\eta| < 1.7$, and two copper/LAr hadronic endcap calorimeters.
The solid angle coverage is completed with forward copper/LAr and tungsten/LAr calorimeter modules
optimised for electromagnetic and hadronic measurements respectively.

The muon spectrometer (MS) comprises separate trigger and
high-precision tracking chambers measuring the deflection of muons in a magnetic field generated by superconducting air-core toroids.
The precision chamber system covers the region $|\eta| < 2.7$ with three layers of monitored drift tubes,
complemented by cathode strip chambers in the forward region, where the background is highest.
The muon trigger system covers the range $|\eta| < 2.4$ with resistive-plate chambers in the barrel, and thin-gap chambers in the endcap regions.

A three-level trigger system is used to select interesting events~\cite{atlas-trigger-2010}.
The Level-1 trigger is implemented in hardware and uses a subset of detector information
to reduce the event rate to a design value of at most \SI{75}{\kHz}.
This is followed by two software-based trigger levels which together reduce the event rate to about 400~Hz.

%% file: samples.tex
\section{Data samples and MC simulations}
\label{sec:DataMCSamples}

\subsection{Data samples}
The results are based on proton--proton collision data collected with the ATLAS experiment at the LHC at a centre-of-mass energy of $\sqrt{s} = 8$~\TeV~in 2012. 
Only events collected under stable beam conditions with all relevant detector subsystems operational are used.  
Events are selected using single-lepton triggers with $\pt$ thresholds of 24 or 60~\GeV\ for electrons and 24 or 36~\gev\ for muons.  
The triggers with the lower $\pt$ threshold include isolation requirements on the candidate lepton
in order to reduce the trigger rate to an acceptable level.
The total integrated luminosity available for the analyses is $20.3~\ifb$. 

\subsection{Signal and background modelling}
\label{sec:SimulatedSamples}

The sample composition for all analyses is dominated by \ttbar events.
Contributions from other processes arise from $W$+jets, $Z$+jets, single top ($t$-channel, $Wt$ and $s$-channel), dibosons ($WW,WZ,ZZ$) and events with one or more non-prompt or fake leptons from decays of hadrons. 
In these measurements, \ttv  (where $V$ corresponds to a $W$ or $Z$ boson) and \tth\ events
that pass the fiducial selection are considered as part of the signal. 
Results with those processes removed are also provided to allow direct comparison to theory predictions at NLO in pQCD matched to parton showers (see Section~\ref{sec:theory_pred}).
All backgrounds are modelled using MC simulation except for the non-prompt or fake lepton background, which is obtained from data for the \ttbt\ lepton-plus-jets and \ttbt\ $e\mu$ analyses, as described below.

\noindent{\bf {\ttbar:}} 
The nominal sample used to model \ttbar\ events was generated using the {\sc PowhegBox} (version 1, r2330) NLO generator~\cite{powheg,powbox1,powbox2}, with the NLO {\sc CT10} parton distribution function (PDF)~\cite{ct10} assuming a top quark mass of $172.5$~\GeV. 
It was interfaced to {\sc Pythia} 6.427~\cite{PythiaManual} with the {\sc CTEQ6L1}~\cite{cteq6l1} PDF and the Perugia2011C~\cite{perugia} settings for the tunable parameters (hereafter referred to as tune). 
The {\sc hdamp} parameter of {\sc PowhegBox}, which controls the \pt\ of the first additional emission beyond the Born configuration, was set to $m_{\mathrm{top}}=172.5$~\GeV.
The main effect of this is to regulate the high-\pt\ emission against which the \ttbar\ system recoils.
In Figures~\ref{fig:controlplots_ljets} and ~\ref{fig:controlplots_dil}, tables of event yields, and comparison to predictions, the \ttbar\ sample is normalised to the theoretical calculation of $253^{+13}_{-15}$~pb performed at next-to-next-to leading order (NNLO) in QCD that includes resummation of next-to-next-to-leading logarithmic (NNLL) soft gluon terms with {\sc Top++}2.0~\cite{ref:xs1,ref:xs2,ref:xs3,ref:xs4,ref:xs5,ref:xs6}. 
The quoted uncertainty includes the scale uncertainty and the uncertainties from PDF and $\alpha_{\rm S}$ choices.

\noindent{\bf{\ttv:}} 
The samples of $t\bar{t}V$ with up to one additional parton were generated with the {\sc MadGraph v5} generator (v1.3.33)~\cite{madgraph} with the {\sc CTEQ6L1} PDF set. {\sc Pythia} 6.426 with the AUET2B tune~\cite{auet2b} was used for showering. The top quark production and decay was performed in {\sc MadGraph} and $t\bar{t} + Z/\gamma^{*}$ interference was included. The \ttv\ samples are normalised to the NLO cross-section predictions~\cite{ttbarVxs1,ttbarVxs2}.

\noindent{\bf{\tth:}}
The \tth\ process was simulated using NLO matrix elements for $pp\rightarrow \tth$ provided by the {\sc HELAC-Oneloop} package \cite{Helac}, interfaced to {\sc Pythia} 8.175 \cite{pythia8} through {\sc PowhegBox} \cite{Alioli:2010xd}, also known as the {\sc Powhel} approach \cite{Powhel-ttH}. The matrix-element calculation was performed using the {\sc CT10} PDF set and the parton shower used the {\sc AU2CT10} tune~\cite{au2}. The sample is normalised to the NLO cross-section prediction and uses the SM values for the Higgs boson branching ratios~\cite{Heinemeyer:2013tqa}.

\noindent{\bf{$W/Z$+jets:}}
 Samples of $W$+jets and $Z/\gamma^{*}$+jets were generated using the {\sc Alpgen v2.14}~\cite{alpgen} leading-order (LO) generator and the {\sc CTEQ6L1} PDF set~\cite{cteq6}. Parton shower and fragmentation were modelled with {\sc Pythia} 6.426~\cite{PythiaManual}. To avoid double-counting of partonic configurations generated by both the matrix-element calculation and the parton-shower evolution, a parton--jet matching scheme (``MLM matching")~\cite{mlm} was employed. 
The $W/Z$+jets samples were generated with up to five additional partons, separately for production in association with $b$-quarks, $c$-quarks and light quarks.
The overlap between events with heavy-flavour quarks obtained from the matrix element and the parton showers was removed using a scheme based on angular separation between the heavy quarks.
The $W/Z$+jets backgrounds are normalised to the inclusive NNLO theoretical cross-section~\cite{vjetsxs}.
In the dilepton channel, a data-driven method is used to validate the $Z$+jets normalisation. A region enriched in $Z$+jets events is defined by inverting the requirement $|m_{ee / \mu\mu}-91 \mathrm{\GeV}|>10$~\GeV. The data are found to agree with the prediction in all lepton channels.

\noindent{\bf{Dibosons:}} 
Samples of $WW/WZ/ZZ$+jets were generated using {\sc Alpgen v2.14}~\cite{alpgen}. Parton shower and fragmentation were modelled with {\sc Herwig 6.520}~\cite{herwig}.
{\sc Sherpa 1.4.3}~\cite{Gleisberg:2008ta, Hoeche:2009rj, Gleisberg:2008fv, Schumann:2007mg} samples including massive $b$- and $c$-quarks with up to three additional partons were used to cover the $WZ$ channel with the $Z$ decaying to hadrons, which was not taken into account in the {\sc Alpgen} samples.
All diboson samples are normalised to their NLO theoretical cross-sections~\cite{dibosonxs,dibosonxs2} as calculated with MCFM~\cite{MCFM}; the NLO PDF set {\sc MSTW2008} was used for all decay channels.

\noindent{\bf{Single top:}}
Background samples of single top quarks corresponding to the $t$-channel, $s$-channel and $Wt$ production mechanisms were generated with {\sc PowhegBox} (version 1, r2330)~\cite{powheg,powbox1,powbox2} using the {\sc CT10} PDF set~\cite{ct10}. All samples were interfaced to {\sc Pythia} 6.426~\cite{PythiaManual} with the {\sc CTEQ6L1} set of parton distribution functions and the Perugia2011C tune. 
In the dilepton channels, only the $Wt$ process is considered.
Overlaps between the \ttbar\ and $Wt$ final states were removed according to the inclusive Diagram Removal scheme~\cite{SingletopWtDRScheme}.
 The single-top-quark samples are normalised to the approximate NNLO theoretical cross-sections~\cite{stopxs,stopxs_2,stopxs_3} using the {\sc MSTW2008} NNLO PDF set. 

All event generators using {\sc Herwig 6.520}~\cite{herwig} were also interfaced to {\sc Jimmy} v4.31~\cite{jimmy} to simulate the underlying event. 
The samples that used {\sc Herwig} or {\sc Pythia} for showering and hadronisation were interfaced to {\sc Photos}~\cite{PhotosPaper} for modelling of the QED final-state radiation and {\sc Tauola}~\cite{TauolaPaper} for modelling the decays of $\tau$ leptons. The $t\bar{t}H$ sample was interfaced to {\sc Photos++}.
All samples were simulated taking into account the effects of multiple $pp$ interactions based on the pile-up conditions in the 2012 data. The pile-up interactions are modelled by overlaying simulated hits from events with exactly one inelastic (signal) collision per bunch crossing with hits from minimum-bias events that are produced with {\sc Pythia} 8.160 using the A2M tune \cite{au2} and the {\sc MSTW2008} LO PDF \cite{Martin:2009iq}. Finally the samples were processed through a simulation~\cite{atlas_sim} of the detector geometry and response using {\sc Geant4}~\cite{geant}.
All simulated samples were processed through the same reconstruction software as the data. Simulated events are corrected so that the object identification efficiencies, energy scales and energy resolutions match those determined in data control samples.
The alternative \ttbar\ samples described in Section~\ref{sec:syst_ttbarmodel}, used for evaluating systematic uncertainties, were instead processed with the {\sc ATLFAST-II}~\cite{atlas_sim} simulation. 
This employs a parameterisation of the response of the electromagnetic and hadronic calorimeters, and {\sc Geant4} for the other detector components.
The nominal \ttbar\ sample is processed with both {\sc Geant4} and {\sc ATLFAST-II}; the latter is used when calculating the generator uncertainties.

Table~\ref{tab:NBparam} provides a summary of basic settings of the 
MC samples used in the analysis. 
The alternative \ttbar\ samples used to evaluate the \ttbar\ generator uncertainties are described in Section~~\ref{sec:syst_ttbarmodel}.

\begin{table}[h!]
\centering     
\begin{small}
\begin{tabular}{l l l l l }
\hline\hline       
Sample & Generator & PDF & Shower & Normalisation \\
\hline
\ttbar\ &  {\sc PowhegBox} (version 1, r2330) & CT10 & {\sc Pythia} 6.427 & NNLO+NNLL\\
$W$ + jets &  {\sc Alpgen} v2.14 & CTEQ6L1 &  {\sc Pythia} 6.426 & NNLO \\
$Z$ + jets &  {\sc Alpgen} v2.14 & CTEQ6L1 &  {\sc Pythia} 6.426 & NNLO \\
Single top $t$-channel &  {\sc PowhegBox} (version 1, r2330)  & CT10 &  {\sc Pythia} 6.426 & approx. NNLO\\
Single top $s$-channel &  {\sc PowhegBox} (version 1, r2330)  & CT10 &  {\sc Pythia} 6.426 & approx. NNLO\\
Single top $Wt$ channel &  {\sc PowhegBox} (version 1, r2330)  & CT10 &  {\sc Pythia} 6.426 & approx. NNLO\\
$WZ$ (excluding $Z\rightarrow q\bar{q}$) &  {\sc Alpgen} v2.14 & CTEQ6L1 &   {\sc Herwig} 6.520 &  NLO \\
$WZ$ ($Z\rightarrow q\bar{q}$) &  {\sc Sherpa} 1.4.3 & CT10 &  {\sc Sherpa} 1.4.3  &  NLO \\
$WW$, $ZZ$ &  {\sc Alpgen} v2.14 & CTEQ6L1 &   {\sc Herwig} 6.520 &  NLO \\
\ttv &  {\sc MadGraph} v5 (v1.3.33) & CTEQ6L1 &  {\sc Pythia}  6.426 &  NLO\\
\tth & {\sc  Powhel} & CT10 &  {\sc Pythia} 8.175 &  NLO\\
 \hline
\hline
\end{tabular}
\caption{Summary of the Monte Carlo event generators used in the analyses. 
Generators used only for evaluating systematic uncertainties are not included.}
\label{tab:NBparam}
\end{small}
\end{table}

\subsection{Backgrounds with fake or non-prompt leptons} 
\label{sec:QCD}

Events with fewer prompt leptons than required may satisfy the selection criteria if one or more jets are mis-identified as isolated leptons, or if the jets include leptonic decays of hadrons which then satisfy lepton identification and isolation requirements. Such cases are referred to as fake leptons.

In the lepton-plus-jets channel, this background is estimated from data using the so-called {\it matrix method}~\cite{matrix_method_CONF}.
A sample enhanced in fake leptons is selected by removing the lepton isolation requirements and, for electrons, loosening the identification criteria (these requirements are detailed in Section~\ref{sec:ObjectSelection}).
Next, the efficiency for these ``loose'' leptons to satisfy the tight criteria is measured in data, separately for prompt and for fake leptons. 
For prompt leptons it is taken from a sample of $Z$ boson decays, while for fake leptons it is estimated from events with low missing transverse momentum or high lepton impact parameter.
With this information the number of fake leptons satisfying the tight criteria can be calculated. 

In the $ttb$ $e \mu$ analysis, this background is estimated from data using events where the two leptons have electrical charges with the same sign. 
Processes which contain two prompt leptons with the same sign, such as $t\bar{t}W$, and cases of lepton charge mis-identification, are subtracted from the same-sign data using MC simulation.
In the $ttbb$ measurements, the background is less important, as the higher jet multiplicity requirement means fewer additional jets available to be mis-identified as leptons.
In this case the background is estimated from the simulation samples described above.

\subsection{Predictions for \ttbar\ with additional heavy flavour}\label{sec:theory_pred}
The measured fiducial cross-sections are compared to a set of theory predictions obtained with the generators shown in Table~\ref{tab:TheoryGenerators}. 
In each case the fiducial phase-space cuts are applied using {\sc Rivet} 2.2.1 \cite{Buckley:2010ar}.

Two generators are used which employ NLO \ttbb\ matrix elements with the top quarks being produced on-shell.
A {\sc MadGraph5\_aMC@NLO} sample was generated in the massive 4-flavour scheme (4FS), using two different functional forms for the renormalisation and factorisation scales: $\mu=m_{\mathrm{top}}^{1/2}\left(p_{\rm T}(b)p_{\rm T}(\bar{b})\right)^{1/4}$ (the BDDP~\cite{NLO1} form), and $\mu=\frac{1}{4}H_{\rm T}=\frac{1}{4} \sum_{i} \sqrt{m_i^2+p_{\mathrm{T},i}^2}$, where the sum runs over all final-state particles. 
A {\sc Powhel} sample was generated as described in Ref.~\cite{NLOPS1}, with the top quark mass set to 173.2 \GeV. 
The renormalisation and factorisation scales were set to  $\mu=\frac{1}{2}H_{\rm T}$, with the sum in this case running over all final-state particles in the underlying Born configuration.
In contrast to {\sc MadGraph5\_aMC@NLO}, this sample employed the 5-flavour scheme (5FS), which unlike the 4FS treats $b$-quarks as being massless and contains a resummation of logarithmically enhanced terms from collinear $g\rightarrow b\bar{b}$ splittings \cite{Maltoni:2012pa}.
In order to regularise the divergence associated with gluon splitting into a pair of massless $b\bar{b}$ quarks, the transverse momentum of each $b$-quark, and the invariant mass of the $b\bar{b}$ pair, were all required to be greater than 2 \GeV. 
This implies that the 5FS calculation does not cover the entire phase-space measured by the $ttb$ analyses. However, the missing events, in which a second $b$-quark is produced with \pt\ below 2 \GeV, or two $b$-quarks have invariant mass below 2 \GeV, are expected to contribute only a small amount to the fiducial cross-section. The prediction for the $ttbb$ fiducial cross-section is unaffected by the generator cuts.
Both the {\sc MadGraph5\_aMC@NLO} and {\sc Powhel} samples used {\sc Pythia} 8.205 \cite{Sjostrand:2014zea} with the Monash tune \cite{Skands:2014pea} for the parton shower.

The cross-sections are also compared to predictions in which the additional $b$-quarks are not present in the matrix-element calculation and are only created in the parton shower.
The {\sc PowhegBox} sample is the same one used for the nominal \ttbar\ prediction, described in Section~\ref{sec:SimulatedSamples}.
A merged sample containing a \ttbar\ final state with up to three additional partons ($b$, $c$, or light) was generated with {\sc MadGraph5} interfaced to {\sc Pythia} 6.427 with the Perugia2011C~\cite{perugia} tune.
Finally, in order to assess the effect of the different descriptions of the $g\rightarrow b\bar{b}$ splitting in the parton shower, a sample consisting of LO \ttbar\ matrix elements was generated with {\sc Pythia} 8.205 \cite{Sjostrand:2014zea} using the ATTBAR tune \cite{ATL-PHYS-PUB-2015-007}. The inclusive cross-section of the sample was normalised to the NNLO+NNLL result \cite{ref:xs1,ref:xs2,ref:xs3,ref:xs4,ref:xs5,ref:xs6}.
{\sc Pythia}~8 offers several options for modelling $g\rightarrow b\bar{b}$ splittings in the final-state parton showers, which may be accessed by varying the {\sc Timeshower:weightGluonToQuark} (wgtq) parameter~\cite{TimeShowerWebpage}. 
Differences between the models arise by neglecting (wgtq5) or retaining (wgtq3, wgtq6) the mass-dependent terms in the $g\rightarrow b\bar{b}$ splitting kernels. 
Differences also arise with respect to the treatment of the high-$m_{b\bar{b}}$ region, with specific models giving an enhanced or suppressed $g\rightarrow b\bar{b}$ rate. The model corresponding to wgtq3 was chosen to maximise this rate.
Finally, some of the models (wgtq5, wgtq6) offer the possibility to choose $\mathrm{sgtq\cdot} m_{b\bar{b}}$ instead of the transverse momentum as the argument of $\alpha_{\rm S}$ in the $g\rightarrow b\bar{b}$ vertices. Here sgtq refers to the {\sc TimeShower:scaleGluonToQuark} parameter, and is allowed to vary in the range $0.25\leq \mathrm{sgtq}\leq 1$, with larger values giving a smaller $g\rightarrow b\bar{b}$ rate and vice versa. 
For the model wgtq5, sgtq was set to 1, a combination that minimises the $g\rightarrow b\bar{b}$ rate, while for wgtq6, sgtq was set to 0.25.

\begin{table}[h!]
\centering    
\begin{tabular}{l c c c c c }
\hline       
Sample & Generator & Shower & PDF & $b$ mass [\GeV] & Tune \\
\hline
\ttbb &   {\sc MadGraph5\_aMC@NLO}  &  {\sc Pythia} 8.205 & CT10f4 & 4.8 & Monash \\
\ttbb &   {\sc Powhel}  &  {\sc Pythia} 8.205 & CT10nlo & 0 & Monash \\
\ttbar+$\leq$3 partons &   {\sc MadGraph5}  &  {\sc Pythia} 6.427 & CT10& 4.8 & Perugia2011C  \\
\ttbar &   {\sc Pythia} 8.205  &  {\sc Pythia} 8.205 & CTEQL1 & 4.8 & ATTBAR \\
\ttbar & {\sc PowhegBox} & {\sc Pythia} 6.427 &  CT10 & 0 &  Perugia2011C \\
 \hline\hline
\end{tabular}
\caption{Details of the theoretical cross-section calculations. For {\sc MadGraph5\_aMC@NLO}, two different functional forms are used for the renormalisation and factorisation scales. Additionally, the leading-order {\sc Pythia} calculations were done with three different options for the $g\rightarrow b\bar{b}$ splitting, as described in the text. The {\sc PowhegBox} sample is the one used for the nominal \ttbar\ prediction in the analyses. }
\label{tab:TheoryGenerators}
\end{table}

%% file: object_reconstruction.tex
\section{Object and event selection}
\label{sec:ObjectReconstruction}

\subsection{Object reconstruction}
\label{sec:ObjectSelection}
A description of the main reconstruction and identification criteria applied for electrons, muons, jets and $b$-jets 
is given below.

\noindent \textbf{Electrons:} Electron candidates~\cite{ElectronPerformance} are reconstructed from energy clusters in the electromagnetic calorimeter that are matched to reconstructed tracks in the inner detector.  The electrons are required to have $\et >25\gev$ and $|\eta_{\rm cluster}| < 2.47$.  Candidates in the electromagnetic calorimeter barrel/endcap transition region $1.37 < |\eta_{\rm cluster}| < 1.52$ are excluded. 
The longitudinal impact parameter of the track with respect to the primary vertex, $|z_{0}|$, is required to be less than 2 mm.
Electrons must satisfy tight quality requirements based on the shape of the energy deposit and the match to the track to distinguish them from hadrons.
Additionally, isolation requirements are imposed based on nearby tracks or calorimeter energy deposits. These requirements depend on the electron kinematics and are derived to give an efficiency that is constant with respect to the electron $\et$ and $\eta$.
The cell-based isolation uses the sum of all calorimeter cell energies within a cone of $\Delta R = 0.2$ around the electron direction while
the track-based isolation sums all track momenta within a cone of $\Delta R=0.3$; in both cases the track momentum itself is excluded from the calculation.
A set of isolation selection criteria with an efficiency of 90\% for prompt electrons in $Z \rightarrow ee$ events is used in the $ttb$ analyses.
Due to the reduced fake lepton background in the $ttbb$ analyses, a looser 98\% efficient set of selection criteria is used.

\noindent \textbf{Muons:} Muon candidates are reconstructed by matching tracks formed in the muon spectrometer and inner detector.
The final candidates are refit using the complete track information from both detector systems, and are required to have $\pt >25\gev$, $|\eta|<2.5$, and $|z_{0}|<2$\,mm.
Muons must be isolated from nearby tracks, using a cone-based algorithm with cone size $\Delta R_{\mathrm{iso}} = 10\gev / p_{\rm T}^{\mu}$. All tracks with momenta above 1~\GeV, excluding the muon's track, are considered in the sum. The ratio of the summed track transverse momenta to the muon $p_{\rm T}$ is required to be smaller than 5\%, corresponding to a 97\% selection efficiency for prompt muons from $Z\rightarrow \mu\mu$ decays.
If a muon and an electron are formed from the same track, the event is rejected.

\noindent \textbf{Jets:} Jets are reconstructed with the anti-$k_t$ algorithm~\cite{ref:Cacciari2008,ref:Cacciari2006,ref:fastjet} with a radius parameter $R=0.4$, using calibrated topological clusters~\cite{atlas-detector} built from energy deposits in the calorimeters. 
Prior to jet finding, a local cluster calibration scheme is applied to correct the topological cluster energies for the non-compensating response of the calorimeter, dead material, and out-of-cluster leakage~\cite{ATLASJetEnergyMeasurement}. 
The corrections are obtained from simulations of charged and neutral particles.  After energy calibration, jets are required to have $\pt > 25$~\GeV~and $|\eta| < 2.5$. 
To avoid selecting jets from secondary interactions, a jet vertex fraction (JVF) cut is applied~\cite{jvf}. 
The variable is defined as the ratio of two sums of the $\pt$ of tracks associated with a given jet and that satisfy $\pt>1 \gev$.  In the numerator, the sum is restricted to tracks compatible with the primary vertex, while in the denominator the sum includes all such tracks.
A requirement that its value be above 0.5 is applied to jets with $\pt <50\gev$, $|\eta|<2.4$, and at least one associated track.

%
During jet reconstruction, no distinction is made between identified electrons and other energy deposits. Therefore, if any of the jets lie within $\Delta R$ = 0.2 of a selected electron, the single closest jet is discarded in order to avoid double-counting electrons as jets. After this, electrons or muons within $\Delta R$ = 0.4 of a remaining jet are removed.

\noindent \textbf{$b$-tagged jets:} Jets are identified as likely to originate from the fragmentation of a $b$-quark ($b$-tagged) using multivariate techniques that combine information from the impact parameters of associated tracks and topological properties of secondary and tertiary decay vertices reconstructed within the jet~\cite{MV1}.
The multivariate algorithms are trained either using only light-flavour jets as background (the ``MV1'' algorithm), or additionally including charm jets in the background to improve the charm jet rejection (the ``MV1c'' algorithm).
The efficiency of identification in simulation is corrected to that measured in data, separately for each flavour of jet~\cite{MV1,btagging2}.
For the analyses using a binned fit of the $b$-tagging discriminant, the probability for a simulated jet to lie in a particular bin is corrected using data.

\subsection{Event selection}
\label{sec:EventSelection}

To ensure that events originate from proton collisions, events are required to have at least one reconstructed vertex with at least five associated tracks.

Events are required to have exactly one or exactly two selected leptons in the lepton-plus-jets and dilepton measurements, respectively. At least one of the leptons must be matched to the trigger object which triggered the event.
For the $ttb$ $e \mu$ measurement, only events with one electron and one muon are considered. 
To increase the number of events in the $ttbb$ measurements, all three lepton flavour combinations ($ee$, $\mu\mu$ and $e\mu$) are considered.
Additional lepton requirements are applied in the $ttbb$ analyses to remove the backgrounds from $Z/\gamma^{*}$, $\Upsilon$ and $J/\psi$ decays. 
The invariant mass of the two leptons must satisfy $m_{\ell\ell} > 15$~\GeV~and, for events with same-flavour leptons ($ee$ or $\mu\mu$), must also satisfy $|m_{\ell\ell}-91 \mathrm{\GeV}|>10$~\GeV.

The lepton-plus-jets $ttb$ analysis requires at least five jets, at least two of which must be $b$-tagged. For this analysis, $c$-jet rejection is important so the MV1c $b$-tagging algorithm is used, at a working point with 80\% efficiency for $b$-jets from top quark decays. This working point is optimised to give the lowest total expected uncertainty on the measurement.
The $ttb$ $e \mu$ and $ttbb$ fit-based dilepton analyses require at least three jets, two of which have to be $b$-tagged.
The same $b$-tagging algorithm and working point as in the lepton-plus-jets analysis is used to improve the separation between $b$- and $c$-jets.
The $ttbb$ cut-based analysis requires exactly four $b$-tagged jets; for this analysis the MV1 algorithm is used at a working point with 70\% efficiency for $b$-jets from top decays. For this analysis, the tighter working point is chosen to reduce the background as much as possible, while the MV1 algorithm is chosen since the impact of the $c$-jet background on the analysis is less important.
Table~\ref{tab:event_selection} summarises the selection criteria applied to the analyses.

\begin{table}[h]
\begin{center}
\begin{tabular}{l cccc}
\hline\hline
Requirement & $ttb$  & $ttb$  & $ttbb$  & $ttbb$ \\
 & Lepton-plus-jets & $e \mu$ &  Cut-based &  Fit-based \\ \hline
$N_{\mathrm{leptons}}$ & $1$ & $2$ & $2$ & $2$ \\
Electron isolation efficiency & 90\% & 90\% & 98\% & 98\% \\
$m_{\ell\ell} > 15$~\GeV & - & - & yes & yes\\
$|m_{ee / \mu\mu}-91 \mathrm{\GeV}|>10$~\GeV & - & - & yes & yes\\
$N_{\mathrm{jets}}$ & $\geq 5$ & $\geq 3$ & $\geq 4$ & $\geq 4$ \\
$N_{b-\mathrm{jets}}$ & $\geq 2$  & $\geq 2$ & $4$ & $\geq 2$ \\
$b$-tagging algorithm & MV1c @ 80\%  & MV1c @ 80\%  & MV1 @ 70\%  & MV1c @ 80\% \\ 
\hline \hline
\end{tabular}
\caption{Summary of the main event selection criteria applied in the various channels. Other requirements which are common to all channels, including muon isolation, are described in the text.}\label{tab:event_selection}
\end{center}
\end{table}

\begin{figure}[t!]
\centering
\includegraphics[width=0.49\textwidth]{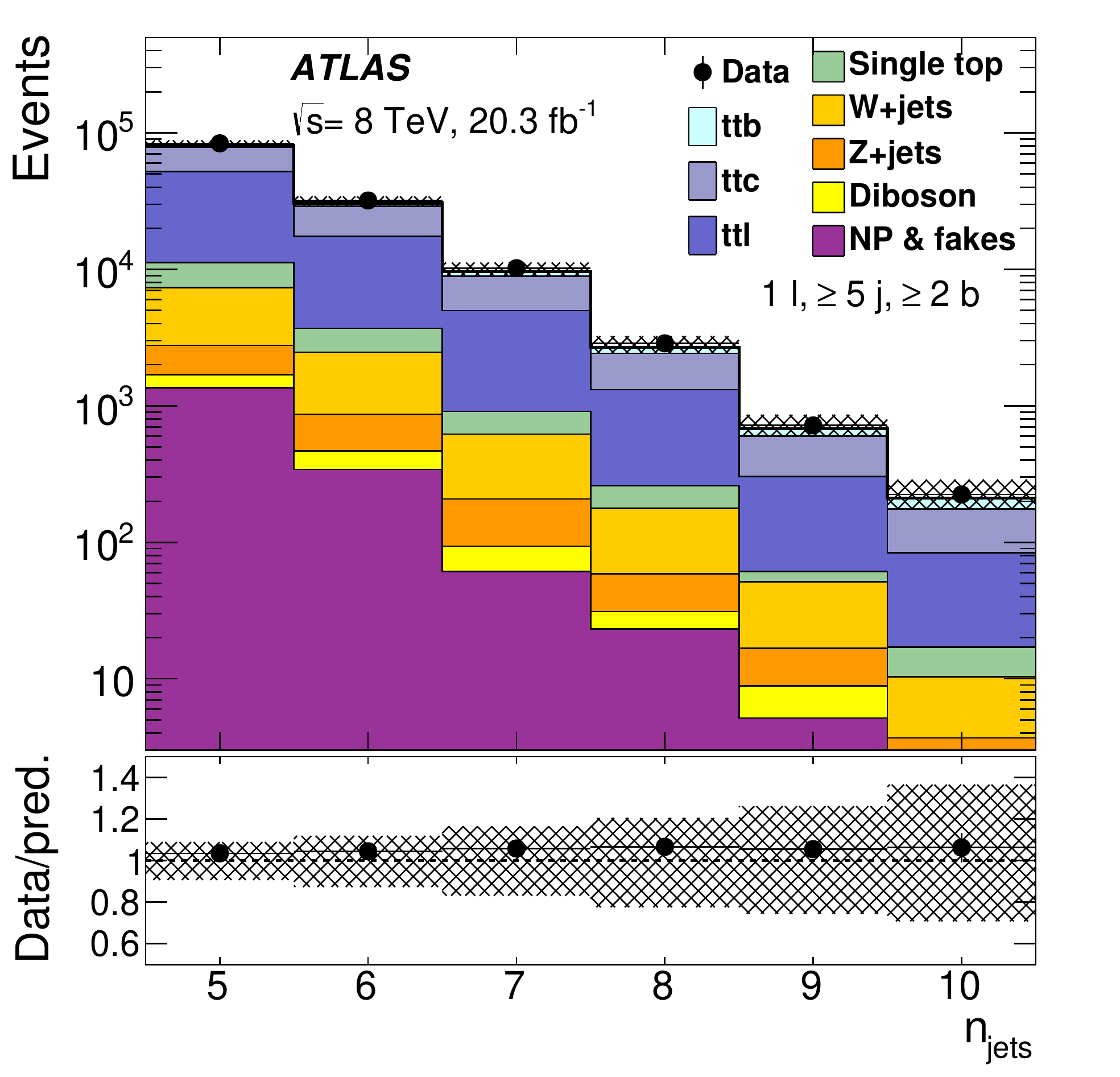} 
\includegraphics[width=0.49\textwidth]{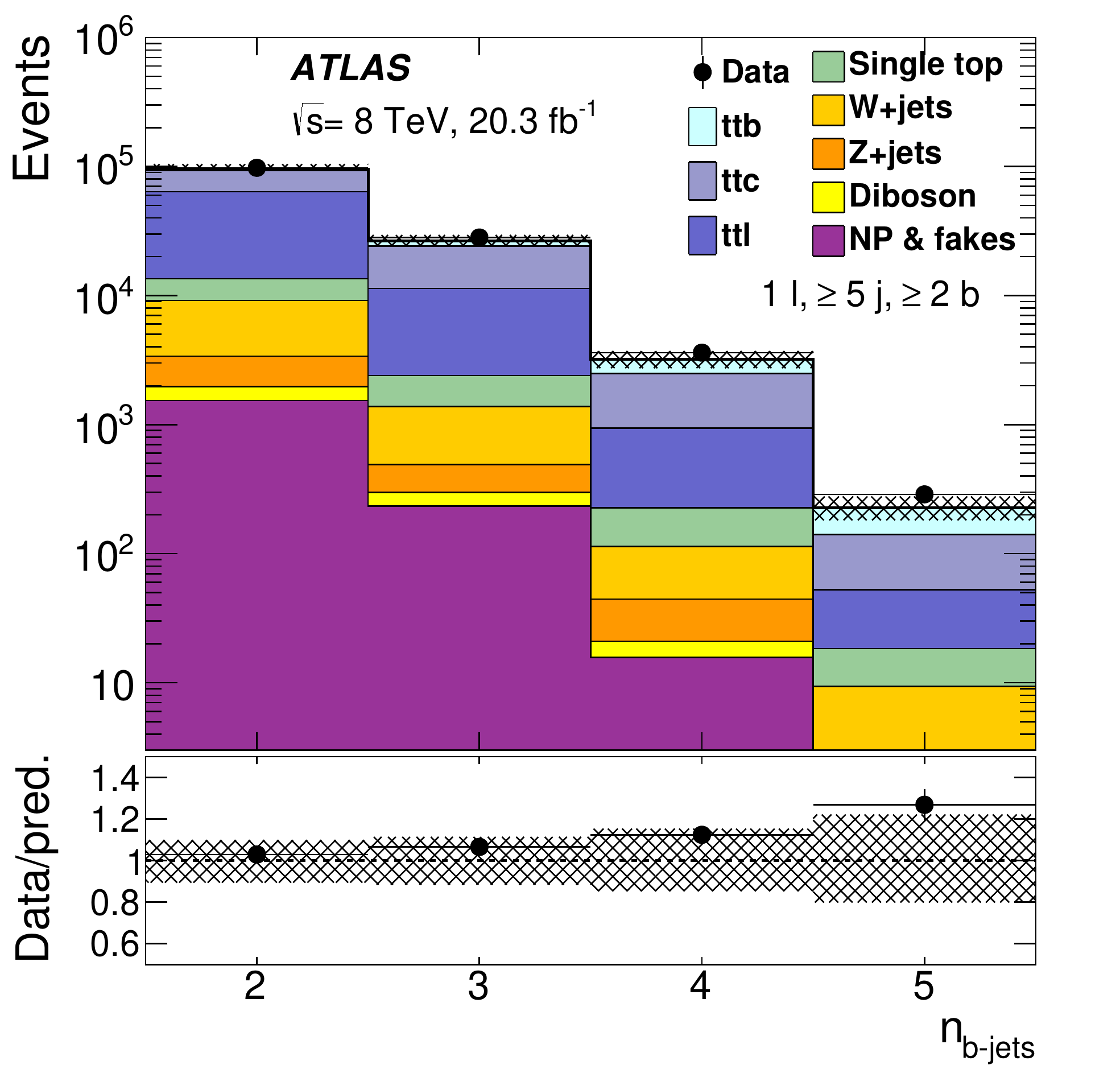} 
\includegraphics[width=0.49\textwidth]{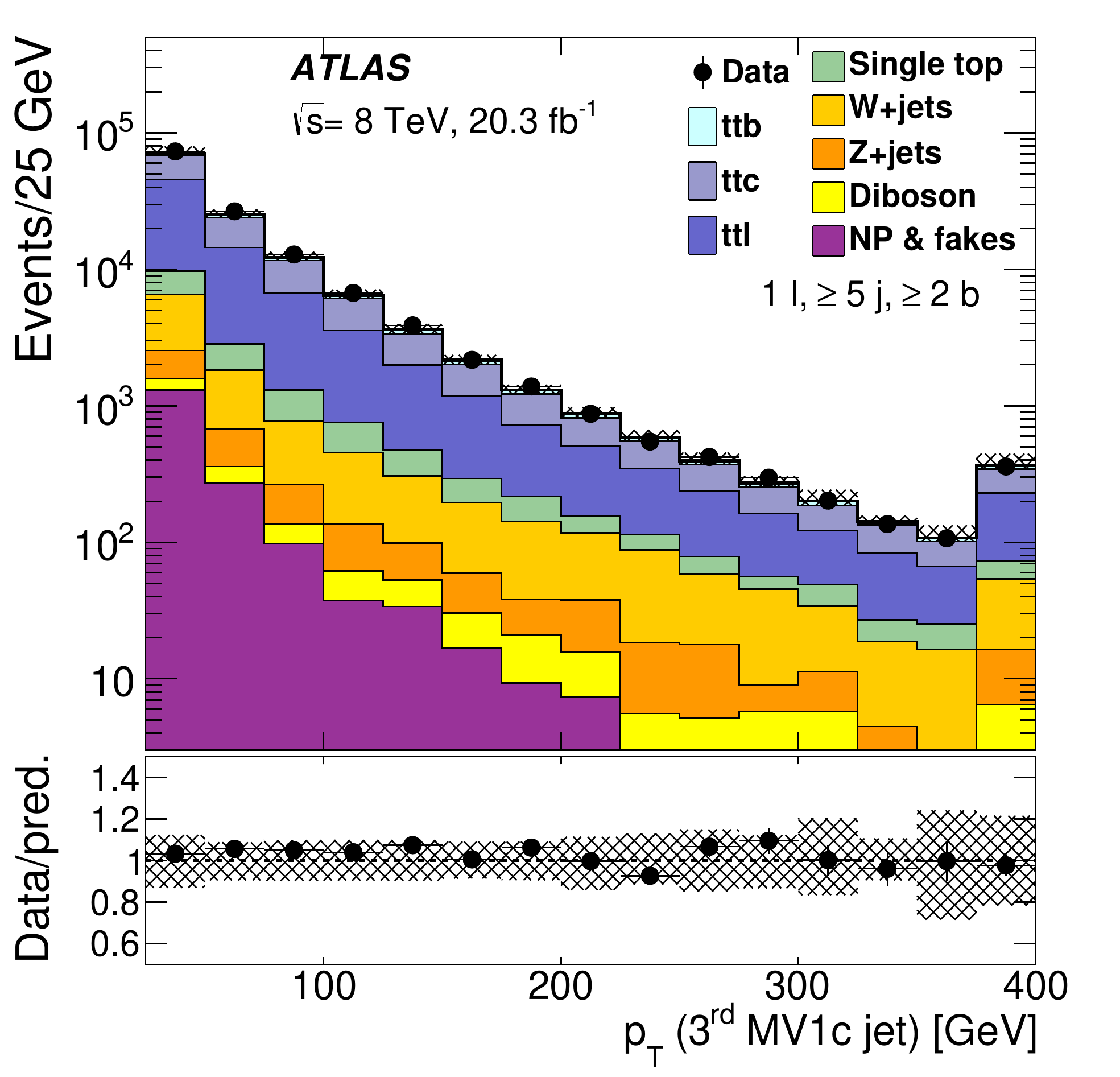} 
\caption{Jet multiplicity, $b$-tagged jet multiplicity, and transverse momentum \pt\ of the jet with the third highest MV1c value in the lepton-plus-jets channel. Events are required to have at least five jets, at least two $b$-tagged jets and one lepton. 
The data are shown as black points with their statistical uncertainty. The stacked distributions are the nominal predictions from Monte Carlo simulation; the hashed area shows the total uncertainty on the prediction. 
The bottom sub-plot shows the ratio of the data to the prediction. The non-prompt and fake lepton backgrounds are referred to as `NP \& fakes'.
The last bin of the distribution includes the overflow.}
\label{fig:controlplots_ljets}
\end{figure}

\begin{figure}[t!]
\centering
\includegraphics[width=0.49\textwidth]{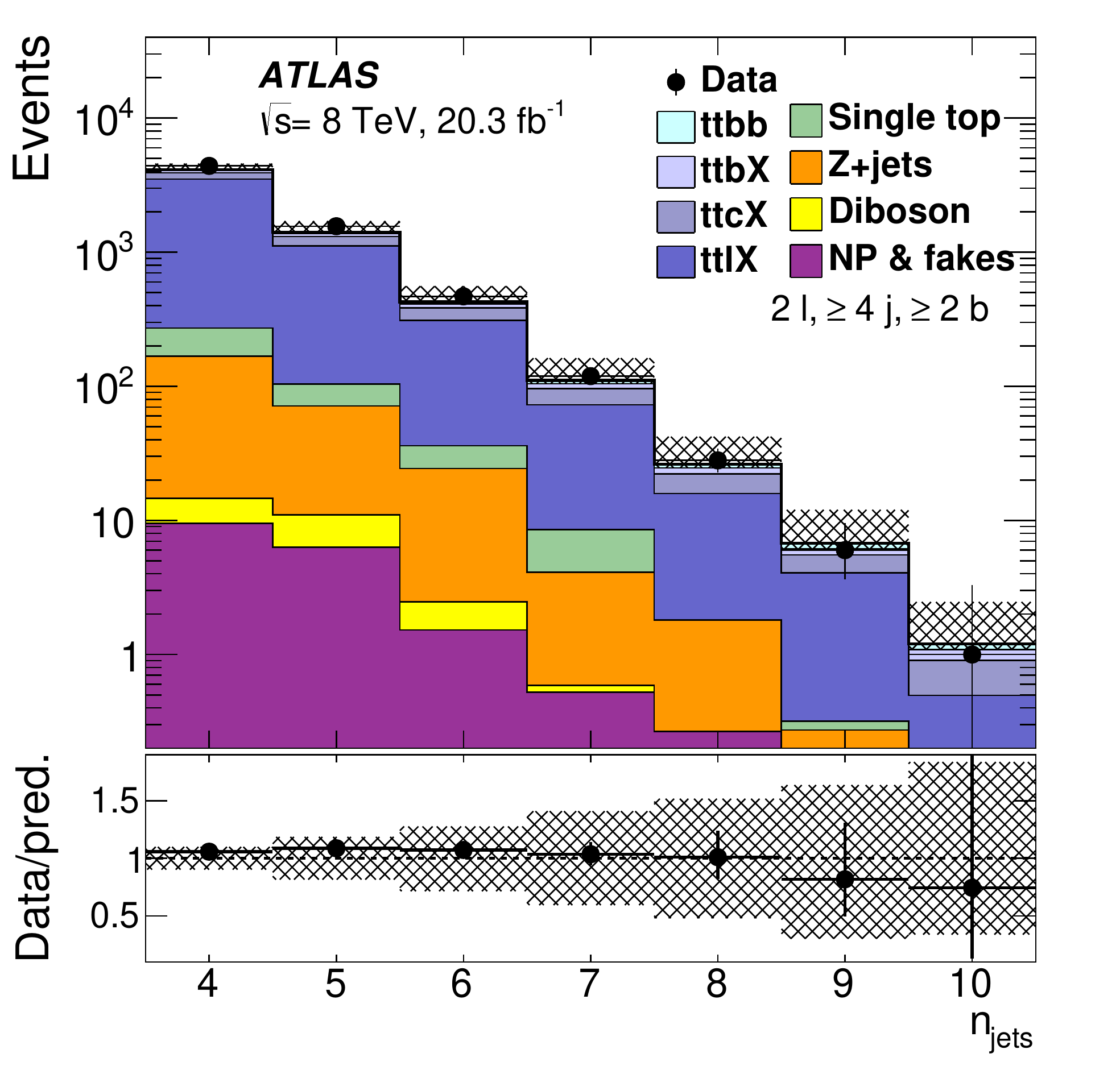} 
\includegraphics[width=0.49\textwidth]{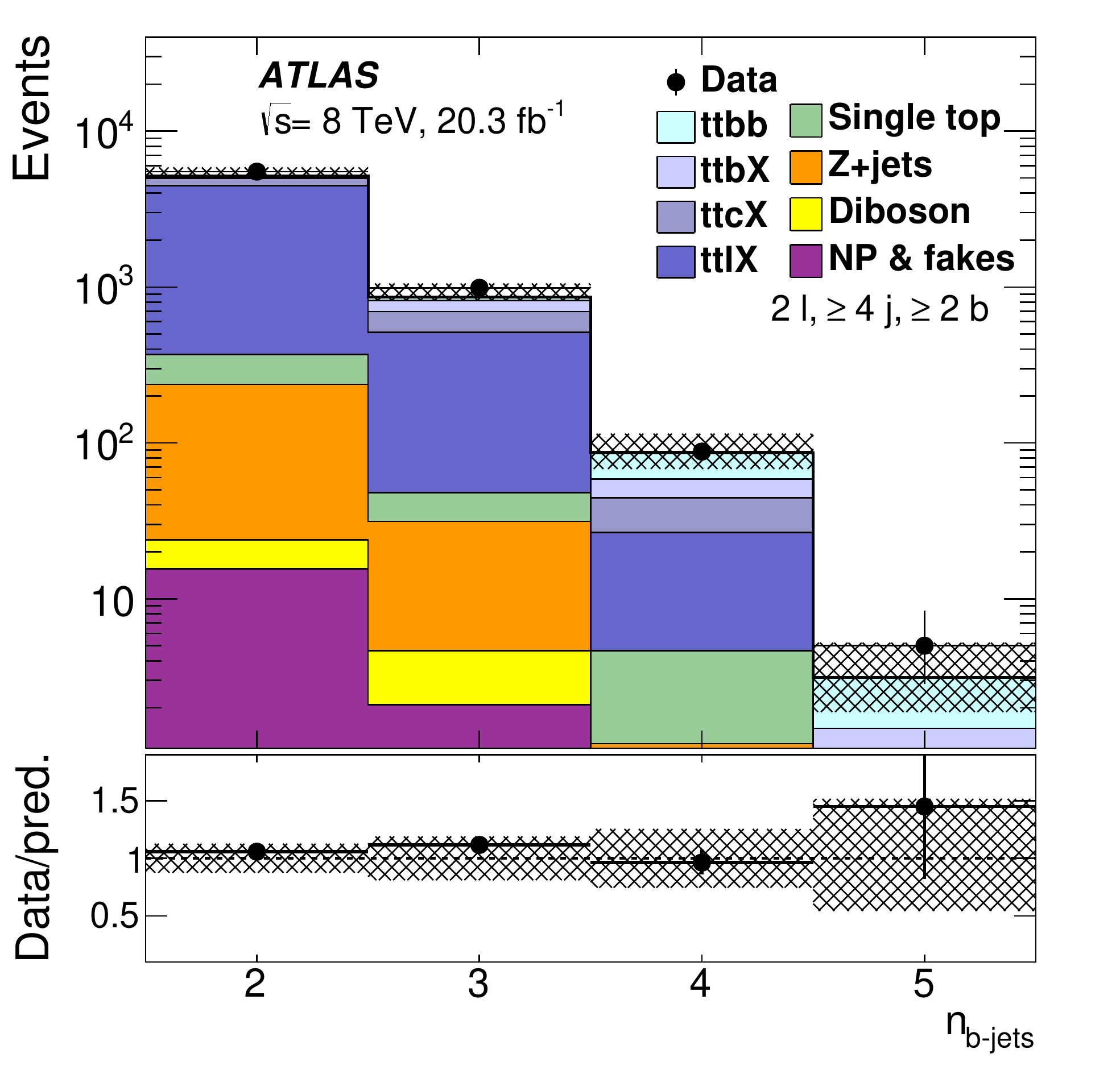} 
\includegraphics[width=0.49\textwidth]{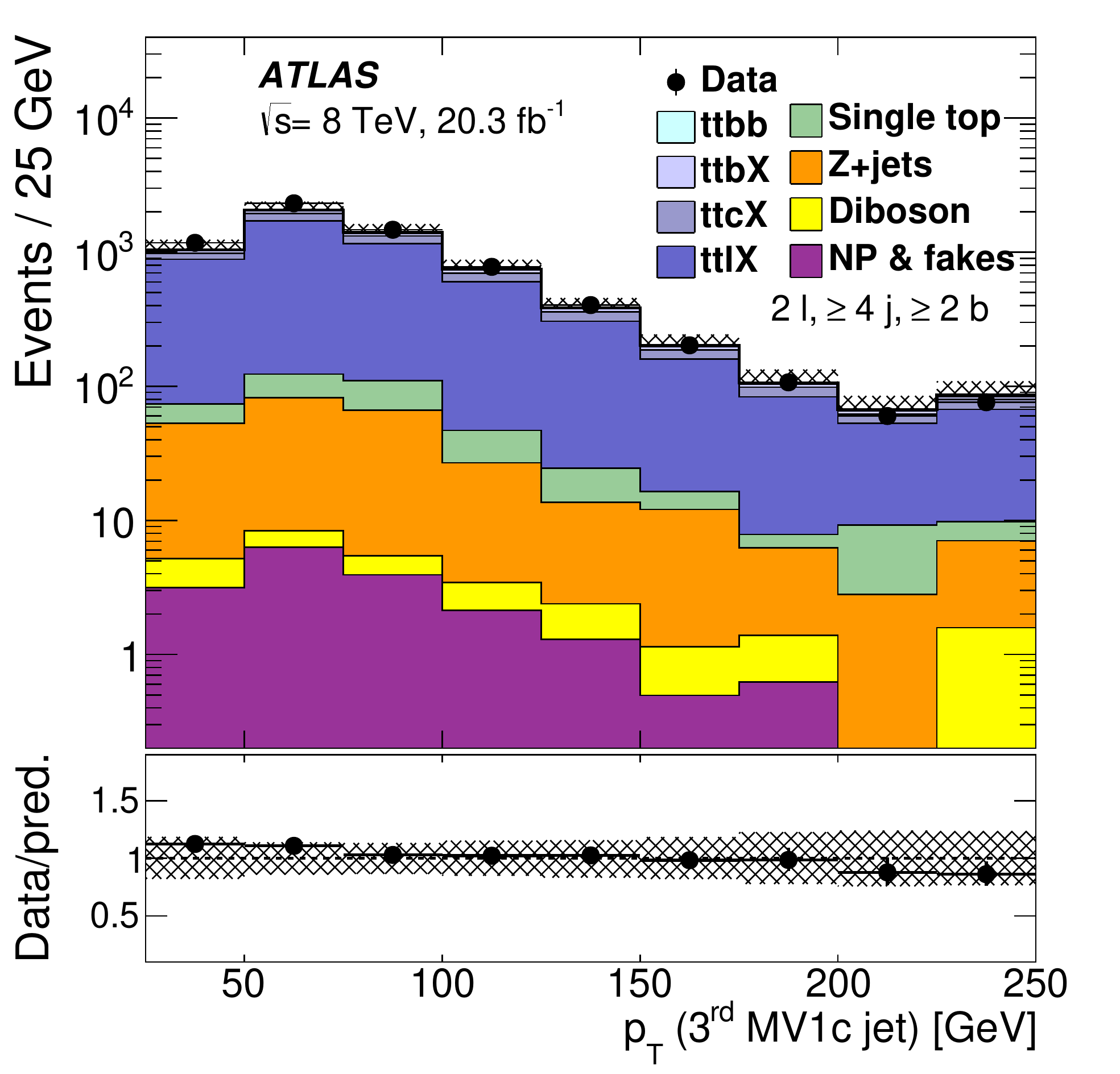} 
\includegraphics[width=0.49\textwidth]{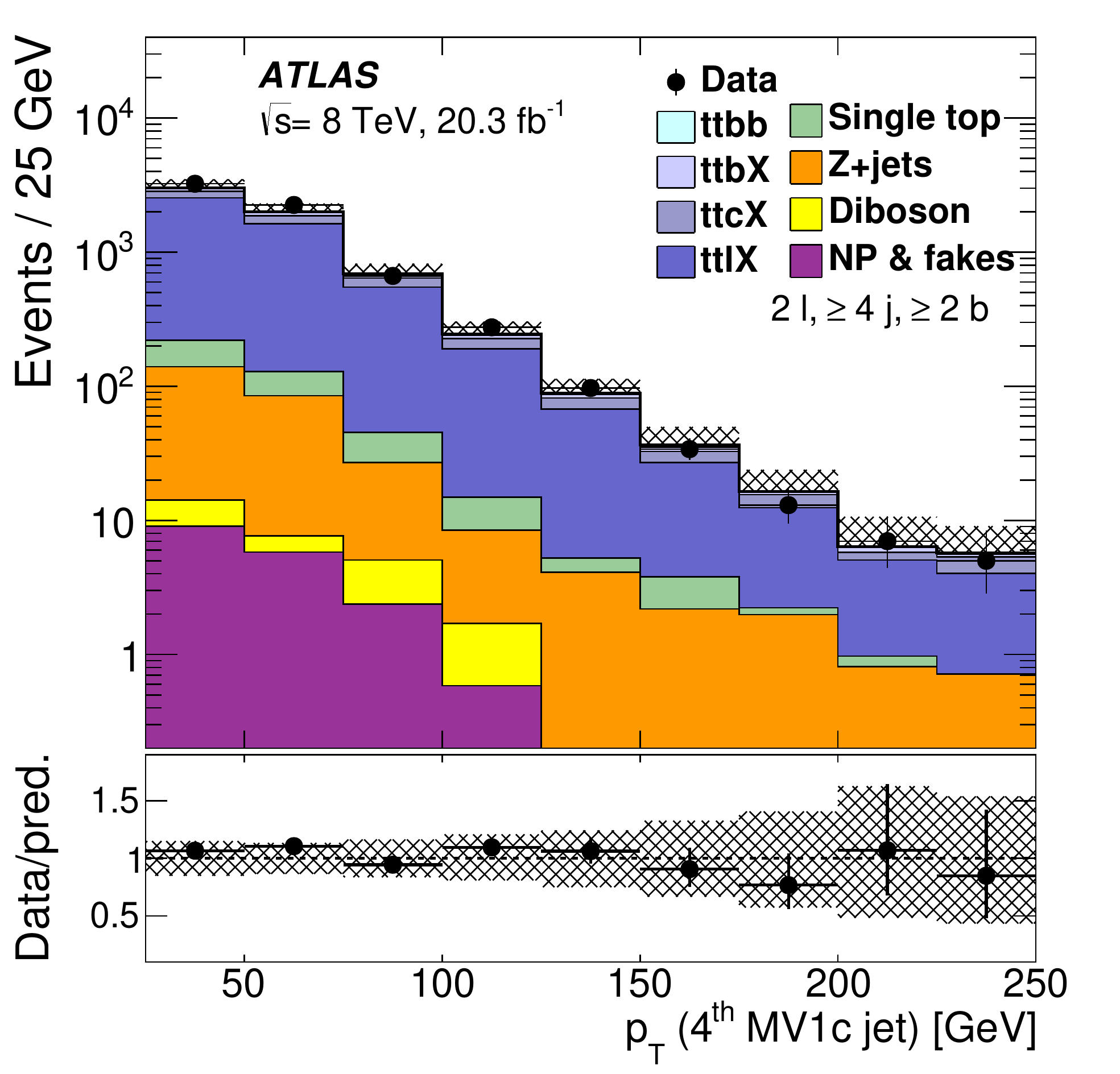} 
\caption{Jet multiplicity, $b$-tagged jet multiplicity, and transverse momentum \pt\ of the jets with the third and fourth highest MV1c values, in the dilepton channel using the $ttbb$ fit-based selection; events are required to have at least four jets, two $b$-tagged jets and two leptons ($ee$, $e\mu$ or $\mu\mu$).
The data are shown in black points with their statistical uncertainty. The stacked distributions are the nominal predictions from Monte Carlo simulation; the hashed area shows the total uncertainty on the prediction. 
The bottom sub-plot shows the ratio of the data to the prediction.
The non-prompt and fake lepton backgrounds are referred to as `NP \& fakes'.
The last bin of the distribution includes the overflow.}
\label{fig:controlplots_dil}
\end{figure}

After these selection criteria are applied, the number of observed and expected events are shown in Table~\ref{tab:yields_ttb} for the $ttb$ analyses and Table~\ref{tab:yields_ttbb} for the $ttbb$ analyses.
For all but the $ttbb$ cut-based analysis, the samples are dominated by \ttbar\ events with an additional light or charm jet.
In all cases the data agree with the expectation within the systematic uncertainties described in Section~\ref{sec:SystematicUncertainties}.
The kinematics in all channels are also found to be well-modelled. 
As an example, Figure~\ref{fig:controlplots_ljets} shows the jet multiplicity, $b$-tagged jet multiplicity, and \pt\ distribution of the jet with the third highest MV1c weight in the lepton-plus-jets selection. 
Figure~\ref{fig:controlplots_dil} shows the $b$-tagged jet multiplicity along with the \pt\ distribution of the jets with the third and fourth highest MV1c values in the dilepton selection. The jet \pt\ distributions in Figures~\ref{fig:controlplots_ljets} and~\ref{fig:controlplots_dil} correspond to the jets that are used in the fit to the distributions of the $b$-tagging discriminant MV1c (see Section~\ref{sec:mv1c_template}).

\begin{table}[h!]
\begin{center}
\begin{tabular}{l \pm \pm }
\hline\hline
 Component &\multicolumn{1}{c}{Lepton-plus-jets} & \multicolumn{1}{c}{$ttb$ $e \mu$} \\ \hline
 \ttbar\ 				& 108600*7500  	&  6620*710 \\
 \hspace{0.5cm}      \ttbt\ 		& 5230*330 		& 286*27 \\
 \hspace{1cm}     \ttv\ signal	        & 67*67     	&  3.6*3.6 \\
 \hspace{1cm}      \tth\ signal	        & 140*140    	& 10*10 \\
 \hspace{0.5cm}     \ttct\ 		& 43300*3000 		& 629*57 \\
 \hspace{0.5cm}    \ttlt\ 		& 60100*6800 		& 5700*630\\
$W$+jets 				&  6700*3500 		& \multicolumn{1}{c}{-}  \\
Single top				&  5490*760 		& 216*58 \\
$Z$+jets  				&  1640*860 		& 20*11   \\
Diboson   				&   510*140	                &  8.8*3.3 \\
Fake and non-prompt leptons   		&   1790*890		& 50*25 \\\hline
Total prediction 			& 124800*8400   	& 6910*720 \\\hline  
Data     				&  \multicolumn{1}{c}{129743} 		& \multicolumn{1}{c}{7198}   \\  
\hline\hline
\end{tabular}
\caption{The number of observed and expected events in the $ttb$ lepton-plus-jets and $e \mu$ analysis signal regions. Indented sub-categories indicate that they are subsets of the preceding category. 
The uncertainty represents the total uncertainty (pre-fit) on the Monte Carlo samples, or on data events in the case of the fake and non-prompt leptons. 
In the $ttb$ $e \mu$ channel, only the $Z\rightarrow \tau \tau$ contribution is included in $Z$+jets; the rest is accounted for in the fake lepton component, as is $W$+jets. 
The breakdown of the \ttbar\ sample into the fiducial sub-samples is given, using the template definitions. 
For illustration, the contributions to $ttb$ from \ttv\ and \tth\ are also shown.}
\label{tab:yields_ttb}
\end{center}
\end{table}

\begin{table}[h!]
\begin{center}
\begin{tabular}{l \pm \pm }
\hline\hline
Component &  \multicolumn{1}{c}{Cut-based} &  \multicolumn{1}{c}{Fit-based} \\ \hline
\ttbar\  					& 23.8*7.2 	        &  5750*850 \\
\hspace{0.5cm}     \ttbbt\ 	        	& 17.1*4.8     	&  110*35 \\
 \hspace{1cm}     \ttv\ signal	                & 0.59*0.59   	&  2.7*2.7 \\ 
 \hspace{1cm}     \tth\ signal	                & 1.6*1.6   	&  7.7*7.7 \\
 \hspace{0.5cm}     \ttbXt\	 	        & 4.1*2.7 		&  280*93 \\
 \hspace{0.5cm}     \ttcXt\ 		        & 2.4*1.0	        &  730*350 \\
 \hspace{0.5cm}     \ttlXt\ 		        & 0.30*0.39	        &  4630*670 \\ 
Single top					&   0.41*0.51 	&  150*57 \\ 
$Z$+jets  					&   0.82*0.96 	&  240*46 \\
Diboson   					&    \multicolumn{1}{c}{<0.1}                &  10.9*3.9 \\
Fake and non-prompt leptons                     &    \multicolumn{1}{c}{<0.1}                &  18.1*9.1 \\\hline
Total prediction 				&   25.1*7.2	& 6180*890 \\\hline
Data     					&   \multicolumn{1}{c}{37} &   \multicolumn{1}{c}{6579} \\
\hline\hline
\end{tabular}
\caption{The number of observed and expected events in the two $ttbb$ analysis signal regions. Indented sub-categories indicate that they are subsets of the preceding category. The uncertainty represents the total uncertainty (pre-fit) on the Monte Carlo samples, or on data events in the case of the fake and non-prompt leptons. 
The breakdown of the \ttbar\ sample into the fiducial sub-samples is given, using the template definitions. 
For illustration, the contributions to $ttbb$ from \ttv\ and \tth\ are also shown.}
\label{tab:yields_ttbb}
\end{center}
\end{table}

%% file: sys_uncertainties.tex
\section{Systematic uncertainties}
\label{sec:SystematicUncertainties}
				   
Several sources of systematic uncertainty are considered that can affect the normalisation of signal and background and/or the shape of their corresponding final discriminant distributions, where relevant. 
Individual sources of systematic uncertainty are considered as correlated between physics processes and uncorrelated with all other sources. 
The following sections describe each of the systematic uncertainties considered in these analyses. 
The uncertainties quoted are illustrative only and the effect of that uncertainty depends on the channel and analysis method used.
All analyses use relative normalisation uncertainties. 
Section~\ref{sec:analysis_methods} details the method by which the uncertainties are included in each analysis and discusses their impact on the measurements.

\subsection{Luminosity uncertainty} Using beam-separation scans performed in November 2012,  a luminosity uncertainty of 2.8\% for $\sqrt{s}=8\tev$ analyses was derived applying the methodology of Ref.~\cite{lumi}. This uncertainty directly affects the cross-section calculation, as well as all background processes determined from MC simulation.

\subsection{Physics objects}
\label{sec:syst_objects}
In this section, uncertainties relevant to the reconstruction of leptons, jets, and $b$-tagging are described. 

\noindent{\bf{Lepton reconstruction, identification and trigger:}}
\label{sec:syst_lepID}
The reconstruction and identification efficiency of electrons and muons, their isolation, as well as the efficiency of the triggers used to record the events, differ slightly between data and simulation.  Correction factors are derived using tag-and-probe techniques on $Z\to \ell^+\ell^-$ ($\ell=e,\mu$) data and simulated samples to correct the simulation for these discrepancies~\cite{ATLAS-CONF-2014-032, PERF-2014-05}.
These have $\sim$1\% uncertainty on all simulated samples.

\noindent{\bf{Lepton momentum scale and resolution}}
The accuracy of the lepton momentum scale and resolution in simulation is checked using reconstructed distributions of the $Z\to \ell^+\ell^-$ and $J/\psi \to \ell^+\ell^-$ masses~\cite{PERF-2014-05, PERF-2013-05}. In the case of electrons, $E/p$ studies using $W\to e\nu$ events are also used. Small discrepancies between data and simulation are observed and corrected for.
In the case of muons, momentum scale and resolution corrections are only applied to the simulation, while for electrons these corrections are applied to data and simulation. Uncertainties on both the momentum scale and resolutions in the muon spectrometer and the tracking systems are considered, and varied separately. 
These uncertainties have an effect of less than 0.5\% on most samples, but up to 1\% on a few of the smaller backgrounds.

\noindent{\bf{Jet reconstruction efficiency:}}
\label{sec:syst_jre}
The jet reconstruction efficiency is found to be about 0.2\% lower in the simulation than in data for jets with \pt\ below $30\gev$, and consistent with data for higher jet \pt. To evaluate the systematic uncertainty due to this small inefficiency, 0.2\% of the jets with $\pt$ below $30\gev$ are removed randomly and all jet-related kinematic variables
are recomputed. The event selection is repeated using the modified selected jet list.  
These uncertainties have less than a 0.5\% effect on the acceptance of all samples.

\noindent{\bf{Jet vertex fraction efficiency:}}
\label{sec:syst_jvf}
The efficiency for each jet to satisfy the jet vertex fraction requirement is measured in $Z(\to \ell^+\ell^-)$+1-jet events in data and simulation, selecting separately events enriched in hard-scatter jets and events enriched in jets from other proton interactions in the same bunch crossing (pile-up).  
The corresponding uncertainty is evaluated in the analysis by changing the nominal JVF cut value.  
This uncertainty has less than a 1\% effect on the signal sample, and up to 5\% effect on the other samples~\cite{jvf,jvf2}.

\noindent{\bf{Jet energy scale:}}
\label{sec:syst_jes}
The jet energy scale (JES) and its uncertainty have been derived by combining information from test-beam data, LHC collision data and simulation~\cite{ATLASJetEnergyMeasurement, PERF-2012-01}.  The jet energy scale uncertainty is split into 22 uncorrelated sources, each of which can have different jet $\pt$ and $\eta$ dependencies.
The largest of these components is the uncertainty specifically related to $b$-jets, which yields an uncertainty of $1.2-2.5$\% on the fiducial cross-section measurements.

\noindent{\bf{Jet energy resolution:}}
\label{sec:syst_jer}
The jet energy resolution (JER) has been measured  separately for data and simulation using two {\em in situ} techniques~\cite{PERF-2011-04}. 
The expected fractional $\pt$ resolution for a given jet is measured as a function of its $\pt$ and pseudorapidity. A systematic uncertainty is defined as the difference in quadrature between the JER for data and simulation and is applied as an additional smearing to the simulation. This uncertainty is then symmetrised.
This uncertainty has a 2--4\% effect on the acceptance of most samples.

\noindent{\bf{Flavour tagging uncertainty:}}
\label{sec:syst_btag}
The efficiencies for $b$, $c$ and light jets to satisfy the $b$-tagging criteria have been evaluated in data, and corresponding correction factors have been derived for jets in simulation~\cite{MV1,btagging2}.
These scale factors and their uncertainties are applied to each jet depending on its flavour and $\pt$. In the case of light-flavour jets, the corrections also depend on jet $\eta$. 
The scale factors for $\tau$ jets are set to those for $c$ jets and an additional extrapolation uncertainty is considered. 
For the fit-based analyses, the effect on the shape of the MV1c templates is considered.
A covariance matrix is formed describing how each source of uncertainty in the scale factor measurement affects each $\pt$ bin.
This matrix is diagonalised, leading to a set of statistically independent eigenvectors for each jet.
The result is 24 uncorrelated uncertainties affecting the $b$-jet efficiency, 16 uncorrelated sources each for the $c$-jets and $\tau$-jets, and 48 uncorrelated sources affecting the light jets. 
The effect of these uncertainties depends on the analysis and the sample in question. 
The $b$-tagging uncertainties are typically largest for the $ttbb$ channels, having an effect of up to 10\%. 
The uncertainty on the measurement from varying the $c$-jet and light jet mis-tagging rates is usually less than 1\%, but may be larger for individual backgrounds.
The uncertainties associated with $\tau$ jets are less than 0.5\% for all samples.

\subsection{Uncertainties on \ttbar modelling}
\label{sec:syst_ttbarmodel}
A number of systematic uncertainties affecting the modelling of \ttbar\ production are considered. 
In particular, systematic uncertainties due to the choice of parton shower and hadronisation models, the choice of generator, the choice of scale, the parton distribution function (PDF), and the inclusion of \ttv\ and \tth\ events are considered.  
These systematic uncertainties are treated as fully correlated between the various components of \ttbar\ (e.g. between $\ttbXt$, $\ttcXt$ and $\ttlXt$).
The effect of assuming these uncertainties to be uncorrelated among the \ttbar\ components
is found to yield slightly smaller uncertainties on the measured cross-sections. 
As many of these uncertainties originate from similar physics processes, they are taken to be correlated. 

\noindent{\bf{Parton shower:}}
An uncertainty due to the choice of parton shower and hadronisation model is derived by comparing events produced by {\sc Powheg} interfaced with {\sc Pythia 6.427} to {\sc Powheg} interfaced with {\sc Herwig 6.520}.
The {\sc PowhegBox} parameter {\sc hdamp} was set to infinity for this comparison for both samples. The difference between the samples is symmetrised to give the total uncertainty.

\noindent{\bf{Generator:}}
An uncertainty due to the choice of generator is derived by comparing a \ttbar\ sample generated with {\sc MadGraph} interfaced to {\sc Pythia 6} to a sample generated by {\sc PowhegBox}+{\sc Pythia}~6. 
The {\sc MadGraph} sample considered was produced with up to three additional partons. It used the CT10 PDF and was showered with {\sc Pythia 6.427}. 
The difference between the samples is symmetrised to give the total uncertainty.

\noindent{\bf{Initial- and final-state radiation:}}
An uncertainty on the amount of additional radiation is determined using samples generated with {\sc MadGraph} interfaced to {\sc Pythia 6} but where the renormalisation and factorisation scales are doubled or halved in the matrix element and parton shower simultaneously, which covers the variations allowed by the ATLAS measurement of \ttbar\ production with a veto on additional central jet activity~\cite{gap_fraction}.
The uncertainty is taken as half of the difference between the samples with higher and lower scales, relative to the central {\sc MadGraph} prediction. 

\noindent{\bf{Parton distribution function:}}
The PDF and $\alpha_{\mathrm{S}}$ uncertainties are calculated using the PDF4LHC recommendations~\cite{ref:pdf4lhc} considering the full envelope of the variations of the {\sc MSTW2008} 68\% CL NLO~\cite{mstw1,mstw2}, {\sc CT10 NLO}~\cite{ct10,ct102} and {\sc NNPDF2.3 5f FFN}~\cite{nnpdf} PDF sets. 
Due to limitations in the information available in the {\sc Powheg} event record, this systematic uncertainty is evaluated on a \ttbar\ MC sample generated with {\sc MC@NLO}~\cite{mcatnlo_1,mcatnlo_2,mcatnlo_3} using {\sc Herwig 6.520} for the parton shower, {\sc AUET2} for the underlying-event tune and {\sc CT10} as the nominal PDF.

\noindent{\bf{Variation of \ttv\ and \tth\ contributions:}}
The signal in these analyses includes contributions from \ttv\ and \tth\ in addition to QCD \ttbb\ production.
The relative proportion of these processes affects the fraction of $ttbb$ events within the $ttb$ templates, and the fractions of $ttcc$ within the $ttc$ and $ttcX$ templates. 
It additionally affects the calculation of the fiducial efficiency, due to the different kinematics of the $b$-jets.
In order to avoid making assumptions on the processes being measured, the effect of doubling or removing \ttv\ and \tth\ is considered as an uncertainty.

Table~\ref{tab:ttbar_sys} summarises the MC samples used to evaluate the systematic uncertainties on the \ttbar\ modelling.

\begin{table}[h!]
\centering     
\begin{small}
\begin{tabular}{l l l l }
\hline       
Uncertainty & Generator & PDF & Shower \\
\hline 
Nominal &  {\sc PowhegBox}  & CT10 & {\sc Pythia 6.427} \\ 
PDF variations & {\sc MC@NLO} & CT10, & {\sc Herwig 6.520} \\
 & & MSTW2008 and  & \\
 &  & NNPDF2.3 &  \\
Parton shower &  {\sc PowhegBox}  & CT10 & {\sc Herwig 6.520}  \\
Generator & {\sc MadGraph} & CT10 & {\sc Pythia 6.427}  \\ 
Additional radiation  $(\times 2,\,\times 1/2)$&{\sc MadGraph} & CT10 & {\sc Pythia 6.427}  \\ 
 \hline
\hline
\end{tabular}
\caption{Summary of the Monte Carlo event generator parameters for the \ttbar\ samples used to evaluate the modelling uncertainties. For all {\sc PowhegBox} samples version 1, r2330 is used. For {\sc MSTW2008} the 68\% CL at NLO is used.}
\label{tab:ttbar_sys}
\end{small}
\end{table}

\subsection{Uncertainties on the non \ttbar\ backgrounds}
\label{sec:syst_norm}

An uncertainty of 6.8\% is assumed for the theoretical cross-section of single top production~\cite{stopxs,stopxs_2}.
For the $Wt$ channel, the diagram-removal scheme is applied in the default sample, in which all doubly-resonant NLO diagrams that overlap with the \ttbar\ definition are removed~\cite{mcatnlo_3}.
The difference between this and an alternative scheme, inclusive diagram subtraction, where the cross-section contribution from Feynman diagrams containing two quarks is subtracted, is considered as a systematic uncertainty.

Normalisation uncertainties for $W$+jets and $Z$+jets backgrounds are set conservatively to 50\%. 
The uncertainty on the diboson background rate is taken to be 25\%. 
In the lepton-plus-jets and $ttb$ $e \mu$ analyses, a conservative uncertainty of 50\% is used on the number of fake and non-prompt lepton events. 
Because the data samples are dominated by \ttbar\ events, the effect of all of these uncertainties on the final result is small.

%% file: mv1c_template.tex
\section{Analysis methods}\label{sec:analysis_methods}
The common components of the cross-section extraction for all analyses are presented in Section~\ref{sec:xs_extraction}.
Three of the four measurements presented make use of the distribution of the multivariate discriminant used for $b$-jet identification. These distributions are presented in Section~\ref{sec:mv1c_template}.
The profile likelihood fits applied in the measurements of the cross-section for $ttb$ production in the lepton-plus-jets and $e \mu$ channels are presented in Section~\ref{sec:analysis_method_ttb_ljets}.
The extraction of the cross-section for $ttbb$ in the cut-based approach is presented in Section~\ref{sec:analysis_method_ttbb_cc}. 
This is followed in Section~\ref{sec:analysis_method_ttbb_fit} by the description of the measurement of the same process using a template fit.

\subsection{Cross-section extraction}\label{sec:xs_extraction}
The cross-sections for fiducial $ttb$ and $ttbb$ production ($\sigma^{\mathrm{fid}}$) are obtained from the best estimate of the number of signal events ($N_{\mathrm{sig}}$), the fiducial efficiency ($\epsilon_{\mathrm{fid}}$), and, where relevant, the correction for the absence of leptons in the fiducial region used in the templates ($f_{\mathrm{fid}}$).
The method to determine $N_{\mathrm{sig}}$ is analysis specific and described in detail in each respective analysis section below.
The fiducial efficiency is the probability for an event in the fiducial region of the templates to meet all reconstruction and selection criteria.
The correction factor $f_{\mathrm{fid}}$ is defined as the fraction of selected events satisfying the template definition that also meet the fiducial signal definition. 
It is only needed for the $ttb$ $e \mu$ and $ttbb$ dilepton fit analyses, which do not include the lepton requirements in the template definitions as documented in Table~\ref{tab:signal_definition}; the $ttb$ lepton-plus-jets analysis uses the same fiducial criteria for defining the signal and building the templates, while the $ttbb$ cut-based does not make use of templates.
The cross-section is given by

\beq\label{eq:xsdef}
\sigma^{\mathrm{fid}}=
\frac{N_{\mathrm{sig}} \cdot f_{\mathrm{fid}}}{\mathcal{L}\cdot\epsilon_{\mathrm{fid}}},
\eeq

where $\mathcal{L}$ is the integrated luminosity.

The values for $\epsilon_{\mathrm{fid}}$ and $f_{\mathrm{fid}}$ are given in Table~\ref{tab:fid_eff}. 
While the cut-based $ttbb$ analysis has the highest signal-to-background ratio, due to the high requirement on the number of $b$-tagged jets (at least four instead of at least two), the fiducial acceptance is much smaller than in the other channels.

\begin{table}[h!]
\begin{center}
\begin{tabular}{l | c c c  c}
\hline\hline
 Parameters & $ttb$ & $ttb$ & $ttbb$ & $ttbb$  \\
  & lepton-plus-jets & $e \mu$ & cut-based & fit-based \\
\hline
$\epsilon_{\mathrm{fid}}$  	& 0.360$\pm$0.002 & 0.358 $\pm$ 0.006 & 0.0681$\pm$0.0036  & $0.399\pm 0.008$  \\
$f_{\mathrm{fid}}$ 			& 1 & 0.969 $\pm$ 0.003 & - & $0.900\pm 0.007$  \\
\hline\hline
\end{tabular}
\caption{The fiducial efficiency ($\epsilon_{\mathrm{fid}}$) and leptonic fiducial acceptance ($f_{\mathrm{fid}}$) for all analyses.
The uncertainties quoted include only the uncertainty due to the limited number of MC events.}\label{tab:fid_eff}
\end{center}
\end{table}

\subsection{Multivariate discriminant for $b$-jet identification}\label{sec:mv1c_template}
The event selection for the three template fit analyses requires the presence of two or more $b$-tagged jets. Relatively loose working points are chosen with $b$-tagging efficiencies of $\sim$80\%, using the MV1c multivariate algorithm, because this allows for high efficiency and good signal-to-background separation.

The distribution of the MV1c discriminant for jets with the third highest, or third and fourth highest, MV1c weights is found to have significant shape differences between the \ttbar\ components.
The $b$-tagging probability distribution for these jets has, on average, high values for $ttb$ and $ttbb$ events, intermediate values for events with additional $c$-jets, and low values for \ttbar\ events with only additional light jets.

The MV1c distribution is calibrated to data in five exclusive bins.
These bin edges correspond to the equivalent cuts on the $b$-jet identification with efficiencies of approximately 80\%, 70\%, 60\%, and 50\% for $b$-jets from top quark decays.

\begin{figure}[ht!]
\centering
\begin{tabular}{cc}
\includegraphics[width=0.46\textwidth]{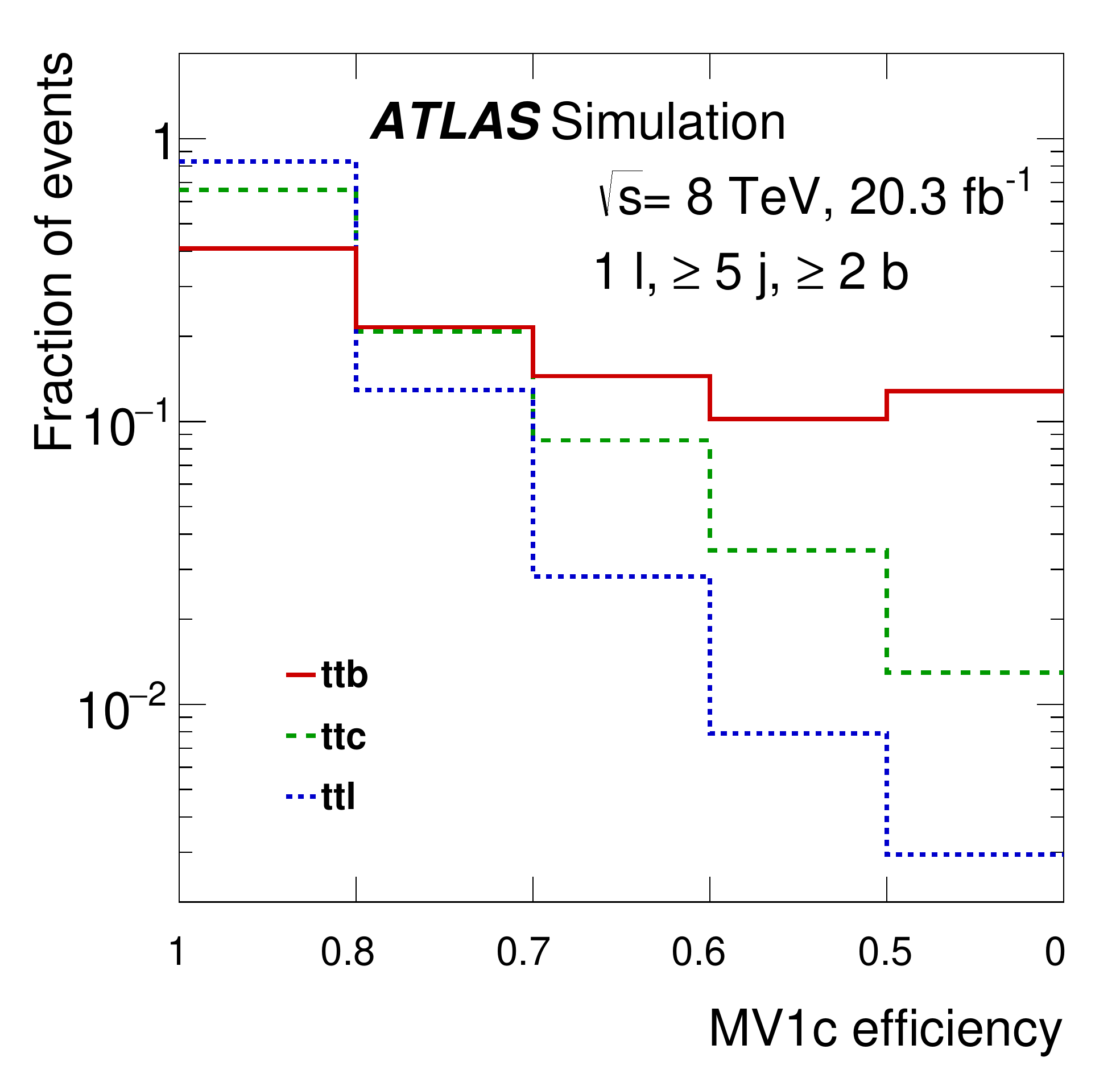} 
\includegraphics[width=0.46\textwidth]{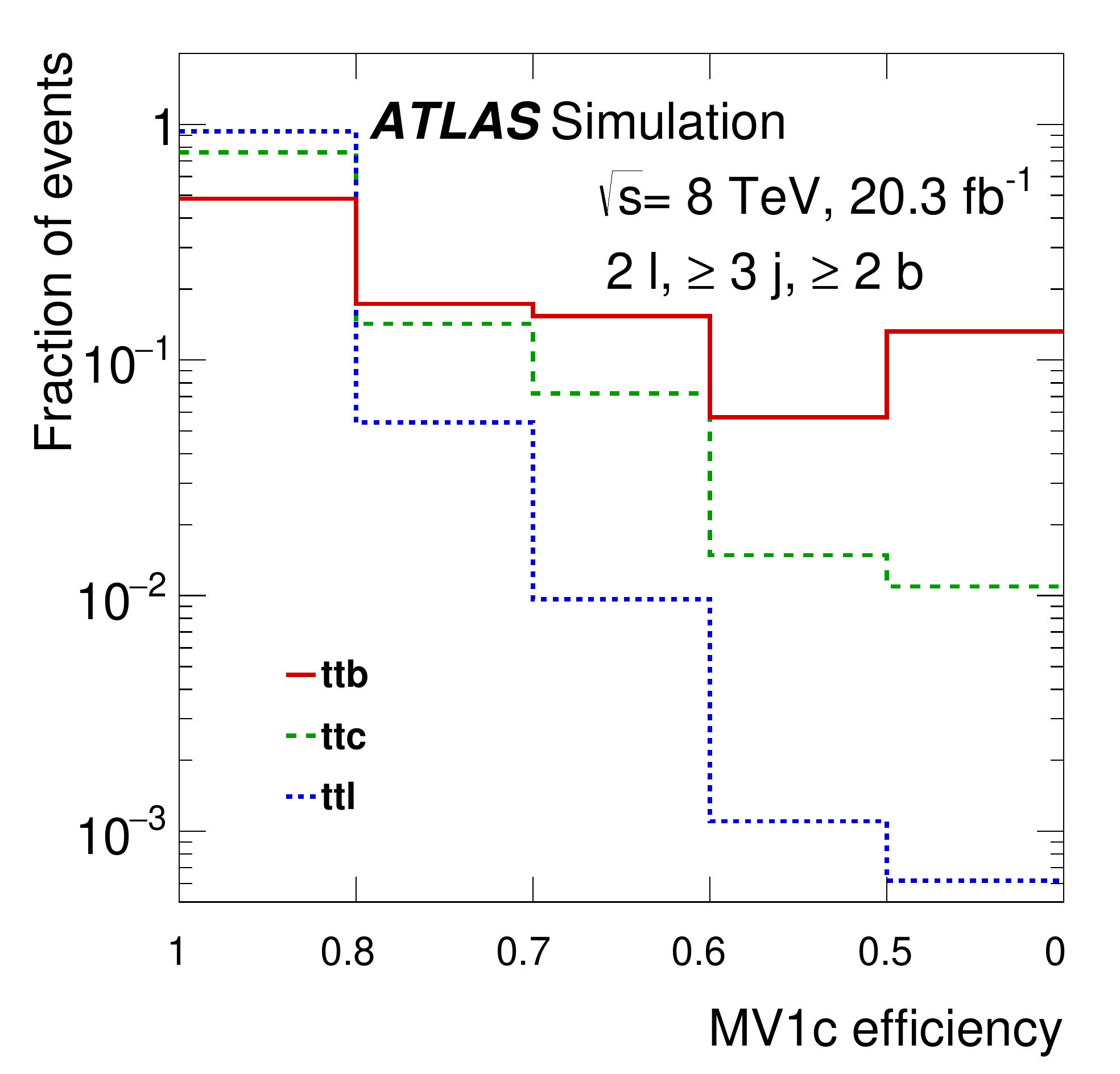} 
\end{tabular}
\caption{Distribution of the MV1c discriminant for the jet with the third highest MV1c weight in the
lepton-plus-jets (left) and $ttb$ $e \mu$ (right) channels. 
The \ttbt\ signal distribution is compared to the distributions for backgrounds with an additional charm jet (\ttct)
and backgrounds with only additional light jets (\ttlt).
The bin edges correspond to the $b$-tagging efficiency of the MV1c weight.
The plots are normalised such that the sum over the bins is equal to unity.
The statistical uncertainty of these distributions is negligible.} 
\label{fig:MV1cPDF}
\end{figure}

The discriminant used in the $ttb$ analyses consists of the distribution of the MV1c of the jet with the third highest MV1c weight, in the five calibrated bins. The templates used for the lepton-plus-jets and $ttb$ $e \mu$ analyses are shown in the left and right plots of Figure~\ref{fig:MV1cPDF}, respectively. 

\begin{figure}[h!]
\centering
\begin{tabular}{cc}
\includegraphics[width=0.46\textwidth]{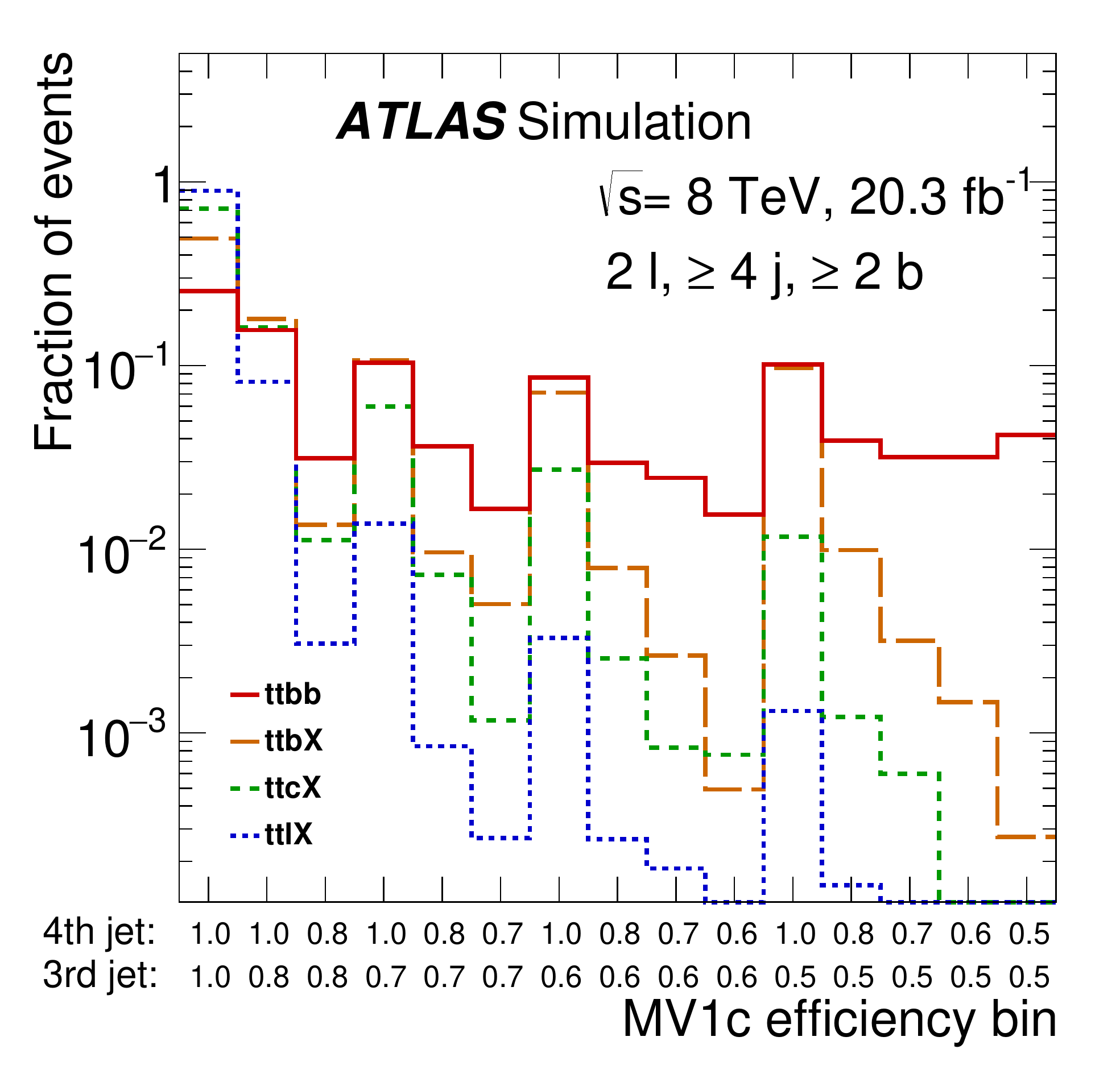} 
\end{tabular}
\caption{Distributions of the third and fourth highest MV1c weight among jets for \ttbbt\ signal, \ttbXt, \ttcXt\ and \ttlXt\ background. 
The bins are labelled with the upper edge of the efficiency point of the third highest and fourth highest MV1c scores in the event. The order of the bins does not affect the cross-section measurement, for this figure the bins have been ordered by decreasing MV1c efficiency point of the fourth and third MV1c score.
The plots are normalised such that the sum over the bins is equal to unity. 
The statistical uncertainty of these distributions is negligible.}
\label{fig:MV1c34PDF}
\end{figure}

For the dilepton $ttbb$ fit analysis, the MV1c distributions for the jets with third and fourth highest MV1c weights are used. Since these are ordered, the weight of the fourth jet is by construction smaller than that of the third, resulting in 15 possible bins of the discriminant. The distribution of the templates used in the fit is shown in Figure~\ref{fig:MV1c34PDF}.

%% file: analysis_method_ttb_ljets.tex
\subsection{Profile likelihood fit to extract the $ttb$ cross-sections}\label{sec:analysis_method_ttb_ljets}

In the lepton-plus-jets and $ttb$ $e \mu$ channels, the numbers of events in the \ttbt, \ttct\ and \ttlt\ categories are obtained by fitting to data the templates of the third highest MV1c weight. 
The fit is performed combining the events from both $e$+jets and $\mu$+jets into a single set of templates for the lepton-plus-jets analysis. 

A binned likelihood function is constructed as the product of Poisson probability terms over all bins considered in the analysis. 
This likelihood depends on the signal-strength parameters, which are independent multiplicative factors of the MC predictions for \ttbt, \ttct\ and \ttlt\ production cross-sections, henceforth referred to as $\mu_{\ttbt}$, $\mu_{\ttct}$ and $\mu_{\ttlt}$.
The nominal prediction ($\mu=1$) for each analysis is obtained from the {\sc PowhegBox} \ttbar\ sample. 
No constraints are applied to the values of these parameters. 
Nuisance parameters (denoted $\theta$) are used to encode the effect of the various sources of systematic uncertainty on the signal and background expectations; these are implemented in the likelihood function with multiplicative Gaussian or log-normal priors. 
The likelihood is then maximised with respect to the full set of $\mu$ and $\theta$ parameters. 
The values of these parameters after maximisation are referred to as $\hat\mu$ and $\hat\theta$.
The cross-section from Eq. (\ref{eq:xsdef}) can be re-written as:

\beq\label{eq:xsdef_profile_fit}
\nonumber
\sigma^{\mathrm{fid}}=
\frac{N_{\mathrm{sig}}(\hat\mu, \hat \theta) \cdot f_{\mathrm{fid}}}{\mathcal{L}\cdot\epsilon_{\mathrm{fid}}( \hat \theta)}\,.
\eeq

The effects from the systematic uncertainties on both the shape and normalisation of the templates are considered, as well as the effect on the fiducial efficiency.
In the $ttb$ $e \mu$ analysis, the uncertainty on $ f_{\mathrm{fid}}$ is also taken into account.
The impact of each systematic uncertainty on these different quantities are considered as correlated.

Due to the large number of nuisance parameters considered, the likelihood fit only includes uncertainties with at least a 0.5\% effect on the event yield, or shape uncertainties that cause a relative variation of more than 0.5\% between two bins. This simplification changes the final result or uncertainty by less than 1\% and significantly reduces the execution time.

The shape variations for the PDF uncertainties on \ttbar\ in the lepton-plus-jets analysis are found to be negligible, therefore only the largest variation in acceptance is considered.
In the $ttb$ $e \mu$ analysis, the PDF uncertainty is evaluated outside of the profile likelihood fit. For each eigenvector of each PDF set, new nominal templates are obtained for each of the components and a statistics-only fit to the Asimov dataset~\cite{asimov} obtained using the central value of the {\sc MC@NLO} prediction is done. The relative difference between the fitted cross-section and the one obtained from the nominal {\sc MC@NLO} is considered as the PDF uncertainty of that eigenvector. The envelope of all eigenvectors is then considered as the PDF uncertainty and added in quadrature to the total uncertainty obtained from the full profile likelihood fit.

\begin{figure}[t!]
\centering
\includegraphics[width=0.46\textwidth]{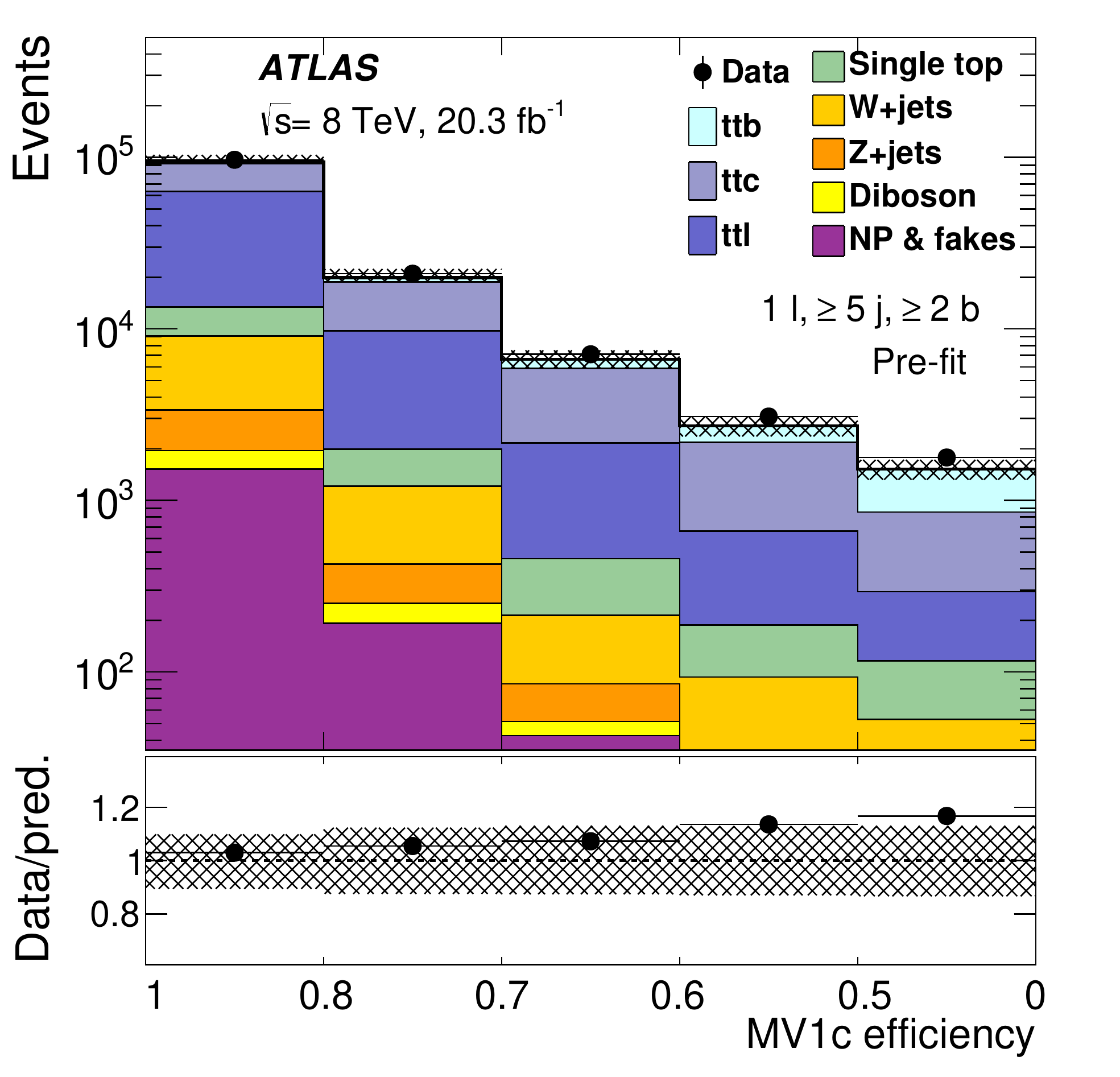} 
\includegraphics[width=0.46\textwidth]{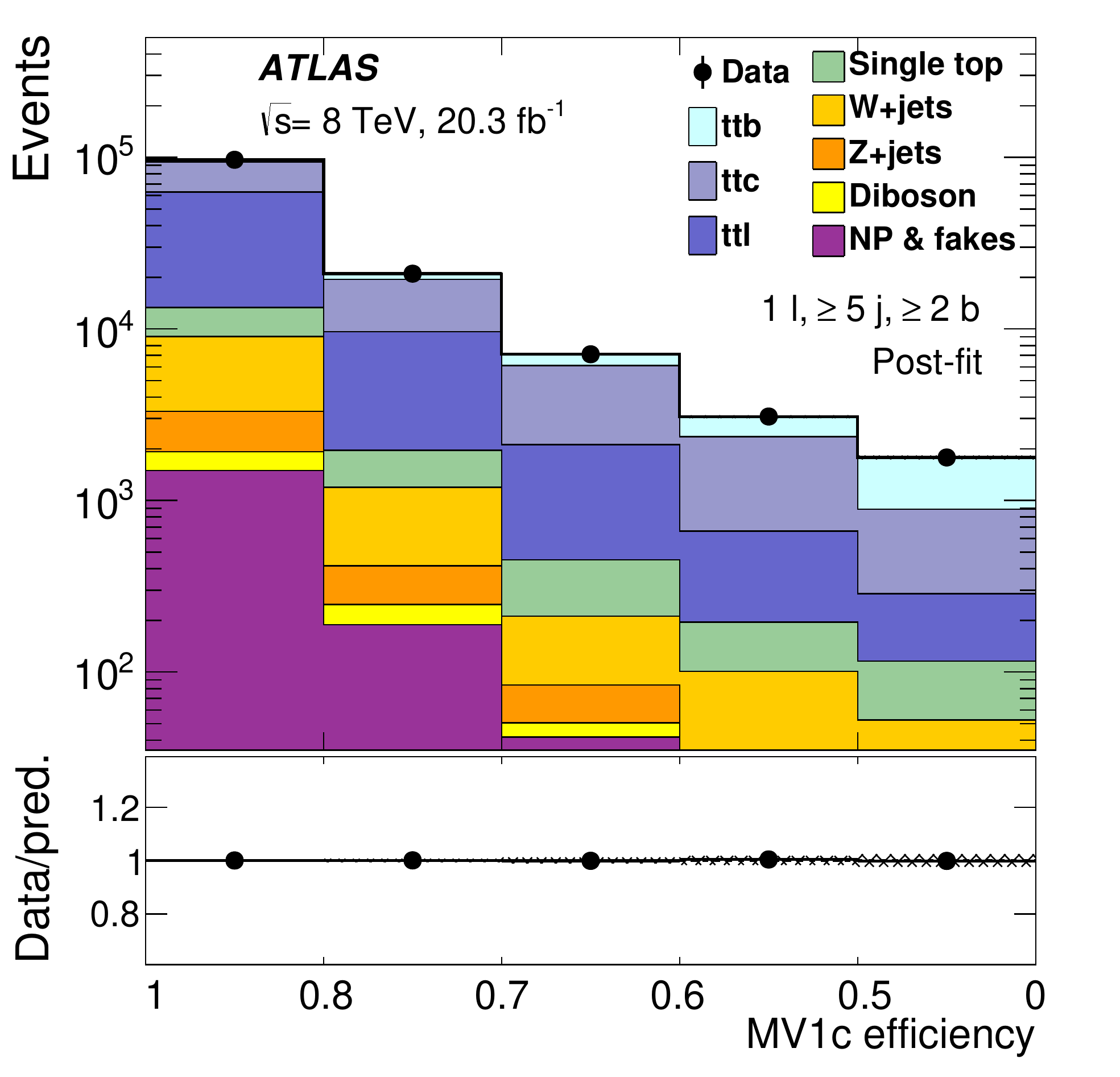} 
\includegraphics[width=0.46\linewidth]{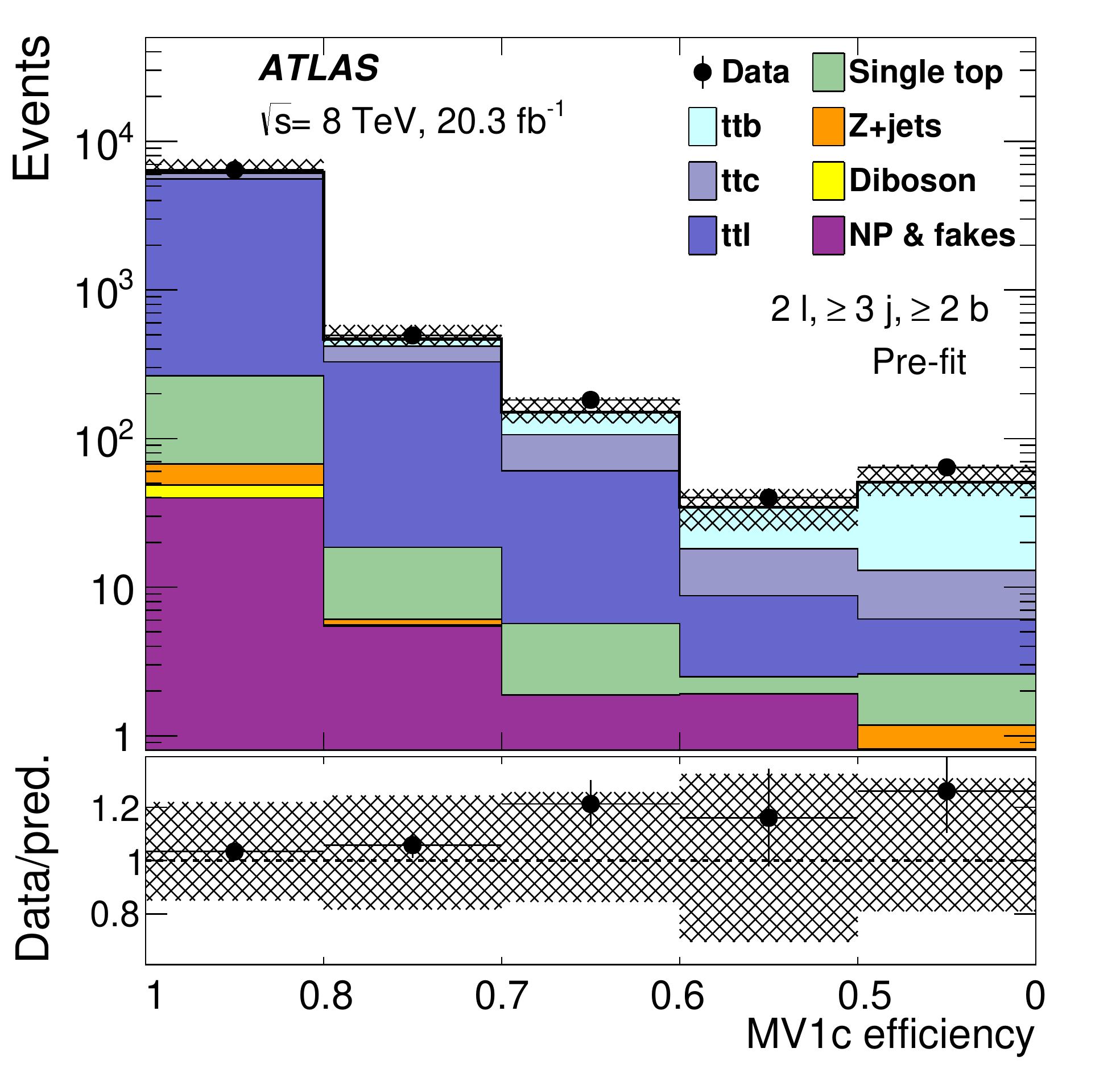}
\includegraphics[width=0.46\linewidth]{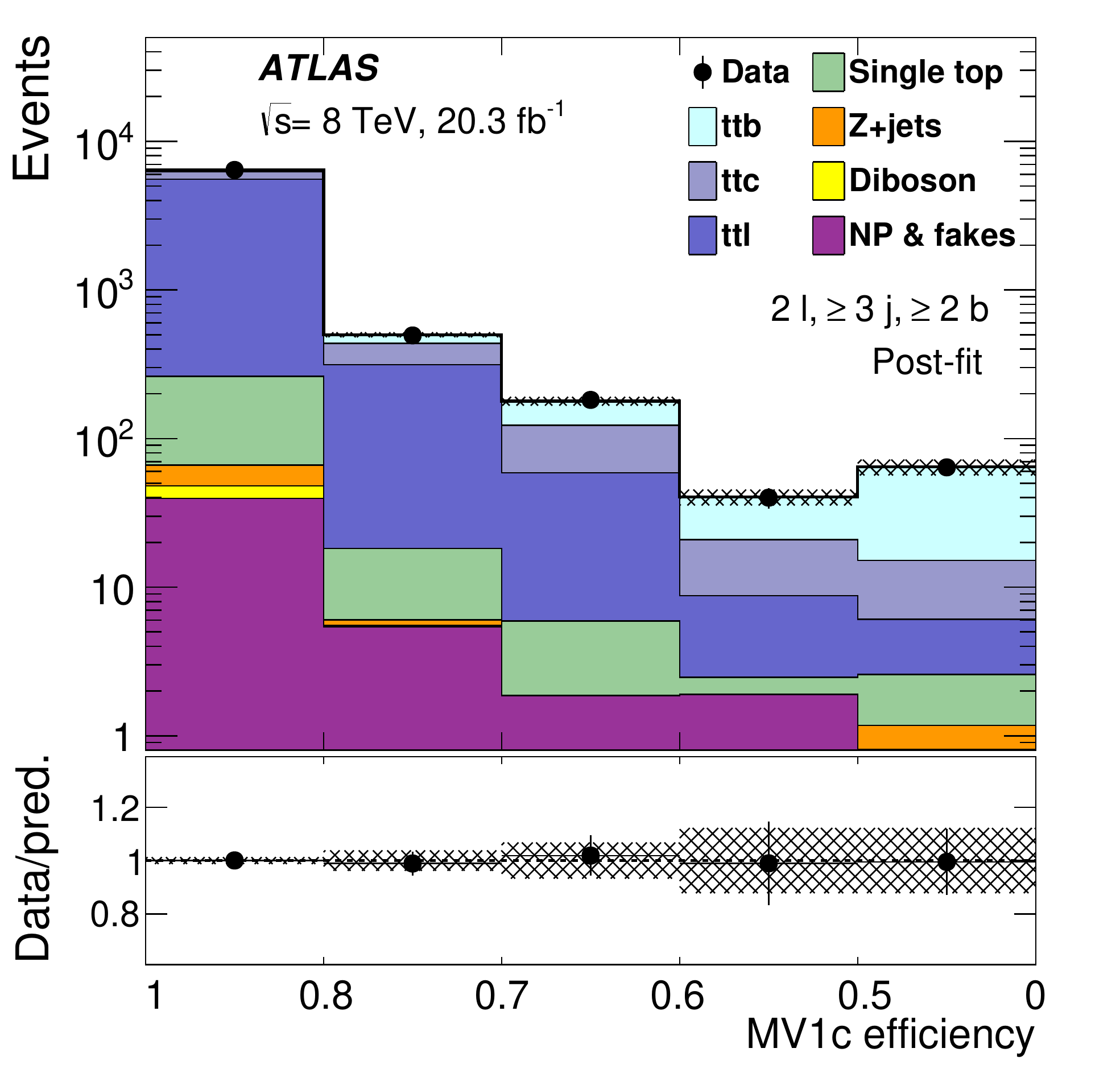}
\caption{The MV1c distribution of jets with the third highest MV1c weight in the lepton-plus-jets analysis (top) and $ttb$ $e \mu$ analysis (bottom) for all signal and background components. 
The data are compared to the nominal predictions (Pre-fit) (left), and to the output of the fit (Post-fit) (right).
The points include the statistical uncertainty on the data.
The hashed area shows the uncertainty on the total prediction. 
The non-prompt and fake lepton backgrounds are referred to as `NP \& fakes'.}
\label{fig:mv1c_ljets}
\end{figure}

Figure~\ref{fig:mv1c_ljets} shows the MV1c distribution used to fit the \ttbt\ signal strength in the lepton-plus-jets analysis (top) and $ttb$ $e \mu$ analysis (bottom). The left figure shows the predictions from simulation and the uncertainty band from the sum in quadrature of the impact of each source of uncertainty. The right plot shows the fitted results and the final uncertainty on the total prediction, which is largely driven by the size of the available MC samples.
Table~\ref{tab:FitResults_ljets} shows the fitted values of the parameters of interest. The Asimov dataset is used to provide expected results. The total uncertainty on the measurement is found to be similar to the expected one in both analyses and the fitted \ttbt\ signal strength in both analyses is higher than one, but still compatible with unity within uncertainties.
The impact of the \ttct\ and \ttlt\ backgrounds on the measurement may be assessed by considering the correlation of $\mu_{\ttbt}$ with $\mu_{\ttct}$ or $\mu_{\ttlt}$ within the likelihood function.
In the $ttb$ $e \mu$ analysis, the correlation is $-0.5$ between $\mu_{\ttbt}$ and $\mu_{\ttct}$, and $+0.5$ between $\mu_{\ttbt}$ and $\mu_{\ttlt}$; in the lepton-plus-jets analysis, the correlation is $+0.1$ in both cases.

\begin{figure}[h!]
\centering
\includegraphics[width=0.49\textwidth]{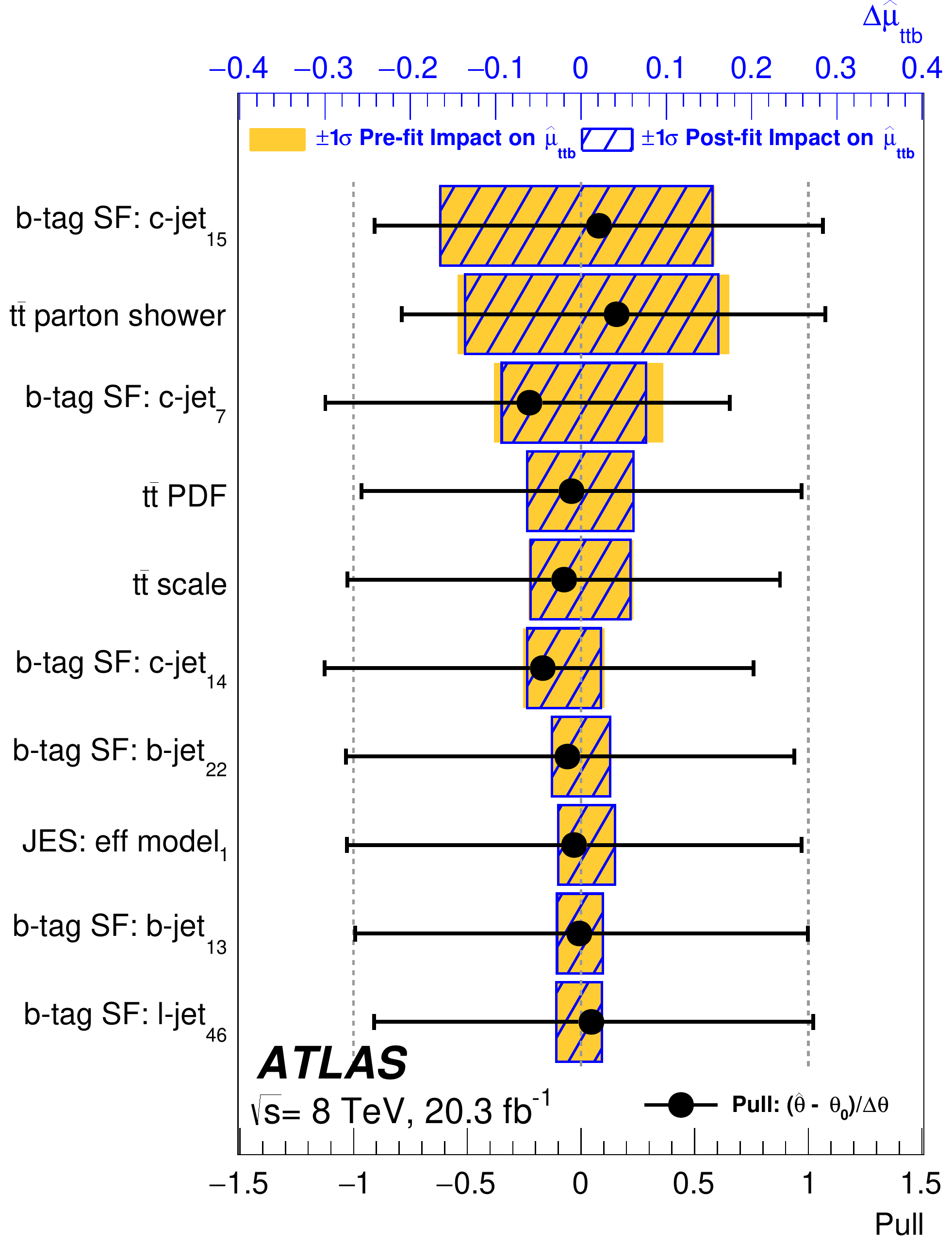} 
\includegraphics[width=0.49\linewidth]{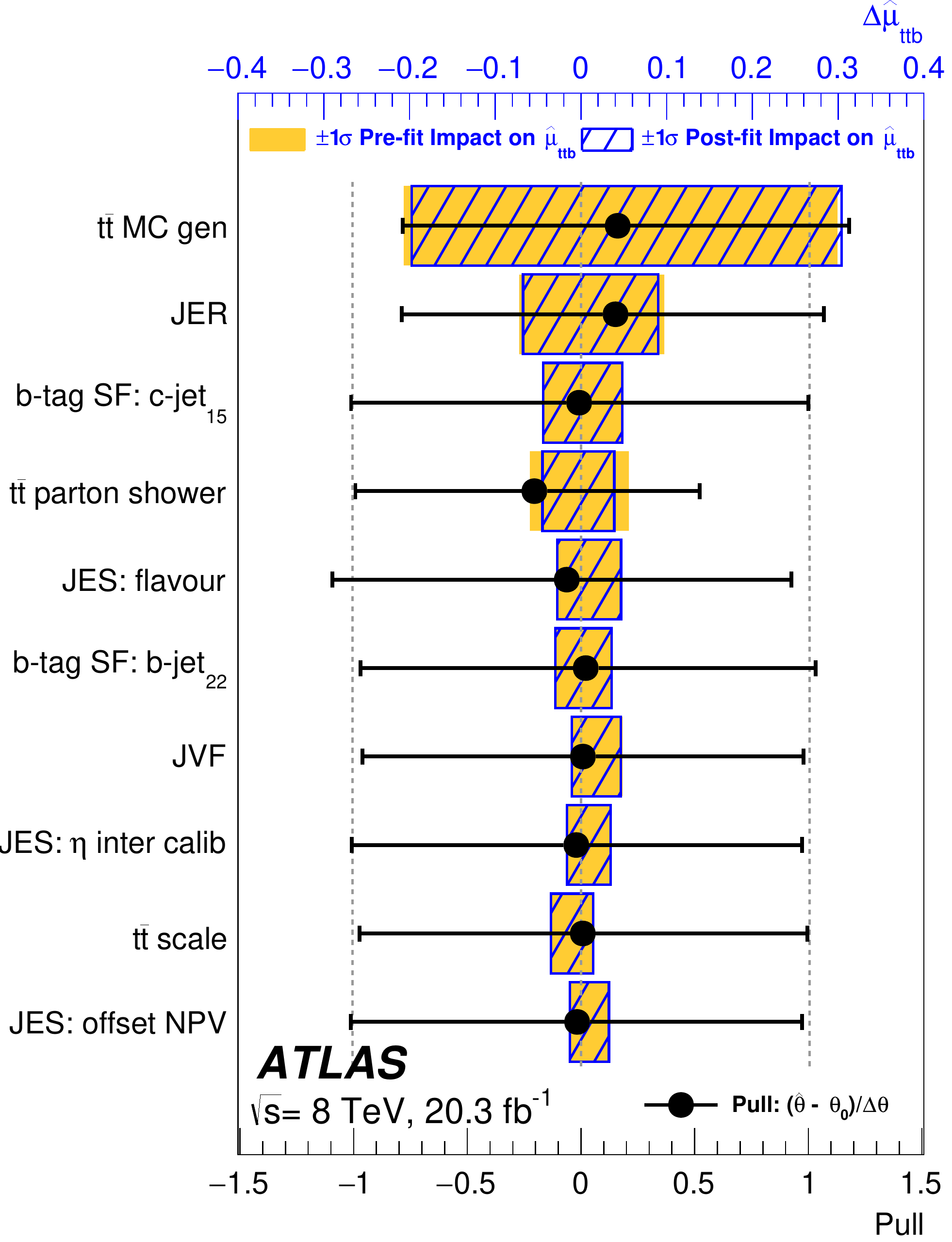}
\caption{Effect of the uncertainty on the fitted value of $\mu_{\ttbt}$ ($\Delta \hat \mu_{\ttbt}$) and pull of the dominant nuisance parameters in the lepton-plus-jets (left) and $ttb$ $e \mu$ analyses (right). 
The shaded and hashed areas refer to the top axis:
the shaded bands show the initial impact of that source of uncertainty on the precision of $\mu_{\ttbt}$; 
the hatched areas show the impact on the measurement of that source of uncertainty, after the profile likelihood fit at the $\pm 1 \sigma$ level. 
The points and associated error bars show the fitted value of the nuisance parameters and their errors and refer to the bottom axis; a mean of zero and a width of 1 would imply no constraint due to the profile likelihood fit.
Dashed lines are shown at 0 and $\pm$ 1 for reference.
Only the ten highest ranked uncertainties on $\mu_{\ttbt}$ are shown.
The index on the $b$-tagging uncertainties refers to the fixed, but arbitrary position in the list of eigenvectors associated with each jet flavour.
}
\label{fig:ranking_ljets}
\end{figure}

The effect of the dominant uncertainties on the fitted signal strength is illustrated in Figure~\ref{fig:ranking_ljets}. The post-fit effect on $\mu_{\ttbt}$ is calculated by fixing the corresponding nuisance parameter at $\hat{\theta}\pm \sigma_{\theta}$, where $\hat{\theta}$ is the fitted value of the nuisance parameter and $\sigma_{\theta}$ is its post-fit uncertainty, and performing the fit again. 
The difference between the default and the modified $\hat{\mu}_{\ttbt}$, $\Delta\hat{\mu}_{\ttbt}$, represents the effect on $\mu_{\ttbt}$ of this particular uncertainty. The dominant uncertainties on both of these measurements are from \ttbar\ modelling and $b$-tagging uncertainties affecting the $c$-jets.
In the lepton-plus-jets analysis, due to the large fraction of \ttbar\ events where the $W$-boson decays to a $c$-quark and a light quark, the effect of the $b$-tagging uncertainties on the $c$-jets is large. 
Other significant contributions come from the effect of $b$-tagging on $b$-jets and light jets, and the jet energy scale and resolution.
The generator comparison shows a large effect on both the template shapes and normalisations; it is the dominant uncertainty for the $ttb$ $e \mu$ analysis, while for the lepton-plus-jets analysis it is smaller due to a cancellation in these effects.

Table~\ref{tab:sys_all} shows a summary of the uncertainties grouped into categories. The effect of each uncertainty is obtained as above and all sources of uncertainty within a category are added in quadrature to obtain the category uncertainty. The total uncertainty in the table is the uncertainty obtained from the full fit, and is therefore not identical to the sum in quadrature of each component, due to the correlations induced between the uncertainties by the fit. Nonetheless, these correlations are small enough that the difference is less than 3\% in both analyses.
In order to obtain separate estimates for the statistical and systematic components of the total uncertainty in both profile likelihood fit analyses, the statistical component of the uncertainty is evaluated by fixing all nuisance parameters to their fitted values and re-evaluating the uncertainty on the fit.

\begin{table}[h!]
\begin{center}
\small
\renewcommand{\arraystretch}{1.2}
\begin{tabular}{l|c|c | c | c }
	\hline\hline
	Fit parameter &  \multicolumn{2}{c|}{Lepton-plus-jets} & \multicolumn{2}{c}{$ttb$ $e \mu$}  \\ \hline
	&  Asimov & data &  Asimov & data \\ \hline
	$\mu_{\ttbt}$ 	 & $1.00\,^{+0.27}_{-0.24}  $ 	&  $ 1.32\,^{+0.35}_{-0.27}$ 	& 1.00 $^{+0.40}_{-0.30}$ &  1.30 $^{+0.47}_{-0.35}$ \\
	$\mu_{\ttct}$  	 & $1.00\,^{+0.23}_{-0.21} $  	&  $ 1.08\,^{+0.31}_{-0.16} $ 	& 1.00 $^{+0.64}_{-0.72}$&  1.40 $^{+0.70}_{-0.78}$ \\
	$\mu_{\ttlt}$  	 & $1.00\,^{+0.19}_{-0.17} $  	& $  1.00\,^{+0.18}_{-0.18}  $	& 1.00 $^{+0.13}_{-0.11}$ &  1.00 $^{+0.14}_{-0.11}$ \\
	\hline\hline
\end{tabular}
\caption{Fitted values for the parameters of interest for the signal strength for \ttbt, \ttct\ and \ttlt\ in the lepton-plus-jets and $ttb$ $e \mu$ analyses. Both the results from the Asimov dataset and the values obtained from the fits to data are shown. The uncertainties quoted are from the total statistical and systematic uncertainties.}
\label{tab:FitResults_ljets}
\end{center}
\end{table}

\begin{table}[htb]
\begin{center}
\begin{small}
\begin{tabular}{l|c | c | c | c | c }
\hline\hline
& $\sigma_{ttb}^{\mathrm{fid}}$		&  $\sigma_{ttb}^{\mathrm{fid}}$& $\sigma_{ttbb}^{\mathrm{fid}}$ & $\sigma_{ttbb}^{\mathrm{fid}}$ & $R_{ttbb}$\\
  & Lepton-plus-jets			& $ttb$ $e \mu$ & Cut-based &  Fit-based & Fit-based \\
  Source  		  		& uncertainty  	&  uncertainty & uncertainty  	&  uncertainty& uncertainty \\
&  (\%) 	&   (\%) &  (\%) 	&   (\%) &  (\%)\\
\hline
 Total detector 					& +17.5 $-$14.4	    & +11.6 $-$8.0  & \multicolumn{1}{l|}{\hskip 1em $\pm$14.5}      & +11.9 $-$13.1  & +10.9 $-$12.5 \\
 \hspace{1cm}     Jet (combined) 	                & +3.9 $-$2.7	    & +10.1 $-$6.1  & \multicolumn{1}{l|}{\hskip 1em $\pm$5.5}       & +6.0 $-$8.5    & +8.7 $-$10.7  \\
 \hspace{1cm}     Lepton 			        & $\pm$0.7	    & +1.0 $-$0.5   & \multicolumn{1}{l|}{\hskip 1em $\pm$2.0} 	     & +2.4 $-$2.7    & +0.8 $-$1.6   \\ 
  \hspace{1cm}    $b$$-$tagging effect on $b$$-$jets    & +4.4 $-$4.0	    & +3.6 $-$3.1   & \multicolumn{1}{l|}{\hskip 1em $\pm$12.9}	     & +9.4 $-$9.0    & +6.0 $-$5.8   \\ 
 \hspace{1cm}     $b$$-$tagging effect on $c$$-$jets    & +16.2 $-$13.4	    & +4.0 $-$3.6   & \multicolumn{1}{l|}{\hskip 1em $\pm$1.7} 	     & $\pm$ 1.4    & +1.2 $-$1.3   \\ 
 \hspace{1cm}     $b$$-$tagging effect on light jets    & +3.1 $-$2.0	    & +1.9 $-$2.0   & \multicolumn{1}{l|}{\hskip 1em $\pm$4.3}	     & +3.3 $-$2.9    & +2.2 $-$1.9   \\ \hline
 Total \ttbar\ modelling 			        & +13.1 $-$13.7	    & +23.8 $-$16.1 & \multicolumn{1}{l|}{\hskip 1em $\pm$23.8}	     & \multicolumn{1}{l|}{\hskip 1em $\pm$21.7}    & \multicolumn{1}{l}{\hskip 1em $\pm$16.1}   \\  
 \hspace{1cm}     Generator  		                & +1.1 $-$1.4	    & +23.3 $-$15.1 & \multicolumn{1}{l|}{\hskip 1em $\pm$16.9}      & \multicolumn{1}{l|}{\hskip 1em $\pm$17.4}    & \multicolumn{1}{l}{\hskip 1em $\pm$12.4}   \\ 
 \hspace{1cm}     Scale choice 		                & $\pm4.3$	    & +1.1 $-$2.7   & \multicolumn{1}{l|}{\hskip 1em $\pm$14.2}	     & \multicolumn{1}{l|}{\hskip 1em $\pm$9.5}     & \multicolumn{1}{l}{\hskip 1em $\pm$6.0}    \\ 
 \hspace{1cm}     Shower/hadronisation                  & +11.4 $-$12.1	    & +3.0 $-$3.4   & \multicolumn{1}{l|}{\hskip 1em $\pm$8.2} 	     & \multicolumn{1}{l|}{\hskip 1em $\pm$8.7}     & \multicolumn{1}{l}{\hskip 1em $\pm$7.1}    \\ 
 \hspace{1cm}     PDF 			                & +4.7 $-$4.5	    & $\pm$3.3      & \multicolumn{1}{l|}{\hskip 1em $\pm$3.3} 	     & \multicolumn{1}{l|}{\hskip 1em $\pm$0.8}     & \multicolumn{1}{l}{\hskip 1em $\pm$4.1}    \\ \hline 
Removing/doubling \ttv\ and \tth\                       & $\pm0.4$	    & +1.1 $-$0.9   & \multicolumn{1}{l|}{\hskip 1em $\pm$1.5} 	     & +3.1 $-$2.7    & +3.0 $-$ 2.6  \\
 Other backgrounds 				        & $\pm0.8$	    & +0.9 $-$0.8   & \multicolumn{1}{l|}{\hskip 1em $\pm$1.6}	     & +3.5 $-$3.3    & $\pm 2.5$   \\
 MC sample size 					&  $<1$		    & $<1$	    & \multicolumn{1}{l|}{\hskip 1em $\pm$9.6} 	     & $\pm 7.4$    & $\pm$7.4    \\ 
 Luminosity 					        & $\pm$2.8	    & $\pm$2.8      & \multicolumn{1}{l|}{\hskip 1em $\pm$3.2} 	     & $\pm 2.9$    & $\pm$0.1    \\ \hline
 Total systematic uncertainty		                & +25.5 $-$19.2	    & +30.5 $-$19.9 & \multicolumn{1}{l|}{\hskip 1em $\pm$29.5}	     & +26.4 $-$26.9  & +21.1 $-$21.9 \\ \hline
 Statistical uncertainty 			        & $\pm$7.1	    & +19.2 $-$17.9 & \multicolumn{1}{l|}{\hskip 1em $\pm$18.4}      & $\pm$24.6    & $\pm$25.2   \\ \hline 
 Total uncertainty 				        & +26.5 $-$20.5	    & +36.0 $-$26.8 & \multicolumn{1}{l|}{\hskip 1em $\pm$35.2}	     & +36.1 $-$36.4  & +32.9 $-$33.4 \\
\hline\hline
\end{tabular}
\caption{Effect of the various sources of uncertainty on the \ttbt\ and \ttbbt\ cross-section measurements in the lepton-plus-jets and dilepton channels. The uncertainties on the $R_{ttbb}$ ratio measurement in the dilepton fit analysis are also shown. Asymmetric uncertainties are shown when relevant. For the fit-based measurements, the individual and total uncertainties are evaluated from the fit to the data.}\label{tab:sys_all}
\end{small}
\end{center}
\end{table}

\clearpage

%% file: analysis_method_ttbb_cc.tex
\subsection{$ttbb$ cross-section from cut-based analysis}\label{sec:analysis_method_ttbb_cc}

This $ttbb$ measurement uses an event counting method in the dilepton channel to extract the cross-section.
Events with at least four identified $b$-jets are considered. 

The estimate of the number of signal events is obtained from the total number of observed events passing the final selection ($N_{\mathrm{data}}$) and the estimate of the number of background events.
A distinction is made between background processes which contain two top and two bottom quarks, but do not pass the fiducial selection (referred to as non-fiducial background), and backgrounds from all other processes (referred to as non-$ttbb$).
In order to avoid making any assumptions about the cross-section for $ttbb$ processes, the prediction for the non-fiducial background is not taken directly from simulation; instead, simulation is used to determine the fractions of $ttbb$ events that are signal and non-fiducial background.
In particular, the fraction of particle-level $ttbb$ events that pass the fiducial selection, $f_{\mathrm{sig}}$,  is defined as 

\beq\label{eq:fsig}
\nonumber
f_{\mathrm{sig}} = \frac{N_{\mathrm{sig}}}{N_{\mathrm{sig}} + N^{\mathrm{non-fiducial}}_{ttbb}}\,.
\eeq

The cross-section from Eq. (\ref{eq:xsdef}) can then be re-written as 

\beq\label{eq:xsdef1}
\nonumber
\sttbb^{\mathrm{fid}}=
\frac{(N_{\mathrm{data}} - N_{\mathrm{non}-ttbb}) \cdot f_{\mathrm{sig}}}
{\mathcal{L}\cdot\epsilon_{\mathrm{fid}}}\,.
\eeq

In order to classify background events as non-fiducial or non-$ttbb$, an attempt is made to match the four reconstructed $b$-jets to particle-level jets.\footnote{The matching is carried out by considering the closest particle-level jet lying $\Delta R\leq 0.4$ from the reconstructed jet.}
If two or more of the reconstructed $b$-tagged jets match light-flavour or charm particle-level jets, then the event is classified as non-$ttbb$, otherwise it is considered as $ttbb$ non-fiducial.

The prediction for the non-$ttbb$ backgrounds is taken from simulation. The prediction has been validated by repeating the calculation with different definitions of the signal region, 
based on the $b$-jets with the fourth-highest value in MV1.
These alternative signal regions vary in the fraction of non-$ttbb$ backgrounds from less than 1\% to more than 50\%. Nonetheless, the measured cross-sections among the regions agree within their statistical uncertainties, giving confidence that the Monte Carlo simulation provides a sufficient description of these backgrounds.

For the calculations of $\epsilon_{\mathrm{fid}}$ and $f_{\mathrm{sig}}$, both electroweak (\ttz\ and \tth) and QCD production are considered, weighted according to their theoretical cross-sections.
The values of the parameters $N_{\mathrm{data}}$, $N_{\mathrm{non}-ttbb}$, $\epsilon_{\mathrm{fid}}$, and $f_{\mathrm{sig}}$ are shown in Table~\ref{tab:DL_xs}, together with their uncertainties.

\newcommand\myCommand{(stat.) $^{+1.5}_{-1.7}$ (syst.)}

\begin{table}[htb]
\begin{center}
\begin{tabular}{l| r@{ }l@{ }l@{ }l }
\hline\hline
 Parameter &  \multicolumn{4}{c}{Value}  \\
\hline
$N_{\mathrm{data}}$ &  \multicolumn{4}{c}{37}  \\
$N_{\mathrm{non}-ttbb}$ & 3.9 & $\pm$ 1.0 & (stat.) $^{+ 1.5}_{- 1.7}$ & (syst.) \\
$f_{\mathrm{sig}}$ & 0.806 & $\pm$ 0.060 & (stat.) $\pm$ 0.061 & (syst.) \\
$\epsilon_{\mathrm{fid}}$ (\%) & 6.8 & $\pm$ 0.4 & (stat.) $^{+ 1.5}_{- 0.9}$ & (syst.)   \\
\hline\hline
\end{tabular}
\caption{The number of observed data events $N_{\mathrm{data}}$, the predicted non-$ttbb$ background $N_{\mathrm{non-ttbb}}$, the signal fraction $f_{\mathrm{sig}}$, and the fiducial efficiency $\epsilon_{\mathrm{fid}}$ in the $ttbb$ cut-based measurement. The numbers include \ttv\ and \tth\ as signal.}\label{tab:DL_xs}
\end{center}
\end{table}

Each source of systematic uncertainty is propagated to the cross-section measurement in a coherent way by varying simultaneously the effect on the background prediction, on $f_{\mathrm{sig}}$ and on $\epsilon_{\mathrm{fid}}$, where applicable.
A symmetrisation of the uncertainties is carried out; for uncertainties for which the positive and negative variations differ (in absolute value) by less than 0.5\%, the larger of the two is used for both variations.
The middle column of Table~\ref{tab:sys_all} shows the effect of the dominant sources of uncertainty on this cross-section measurement.

%% file: analysis_method_ttbb_fit.tex
\subsection{Maximum-likelihood fit to extract the $ttbb$ cross-section}\label{sec:analysis_method_ttbb_fit}
The looser event selection used in this analyses allows a template fit to be performed in the 15 populated bins of the MV1c distribution for the jets with the third and fourth highest MV1c values.
A maximum-likelihood fit to the nominal templates of \ttbbt, \ttbXt, \ttcXt, \ttlXt\ and non-\ttbar\ background is carried out to extract the number of signal events in each category.
Systematic uncertainties are not included in the likelihood. The cross-section is then extracted directly from Eq. (\ref{eq:xsdef}).

This analysis also allows an extraction not only of the \ttbbt\ signal but also of the \ttbXt, \ttcXt, \ttlXt\ contributions and of the ratio of $ttbb$ to the total $ttjj$ yield: 

\beq
\nonumber
R_{ttbb} = \frac{\sigma_{ttbb}}{\sigma_{ttjj}}\,,
\eeq

where $ttjj$ refers to \ttbar\ production with two additional jets. The cross-section for $ttjj$ is obtained by correcting the \ttbbt, \ttbXt, \ttcXt\ and \ttlXt\ cross-sections, which are calculated for events with three or four particle-level jets, to the fraction with four jets only.  For \ttbbt\ the fiducial efficiency and fraction as documented in Table~\ref{tab:fid_eff} are used; for \ttbXt , \ttcXt\ and \ttlXt\ the fiducial efficiencies and fractions are shown in Table~\ref{tab:fid_eff_auxiliary}.

\begin{table}[h!]
\begin{center}
\begin{tabular}{l | c c c}
\hline\hline
 Parameter  & \ttbXt & \ttcXt & \ttlXt \\
\hline
$\epsilon_{\mathrm{fid}}$  	& 0.197 $\pm$ 0.003 & 0.177 $\pm$ 0.002 & 0.0355 $\pm$ 0.0001    \\
$f_{\mathrm{fid}}$ 	        & 0.898 $\pm$ 0.005 & 0.899 $\pm$ 0.003 & 0.902 $\pm$ 0.001  \\
\hline\hline
\end{tabular}
\caption{The fiducial efficiency ($\epsilon_{\mathrm{fid}}$) and leptonic fiducial acceptance ($f_{\mathrm{fid}}$) for the \ttbXt, \ttcXt\ and \ttlXt\ categories as used in the $ttbb$ fit-based analysis.
The uncertainties quoted include only the uncertainty due to the limited number of MC events.}\label{tab:fid_eff_auxiliary}
\end{center}
\end{table}

\begin{figure}[h!]
\centering
\includegraphics[width=0.46\linewidth]{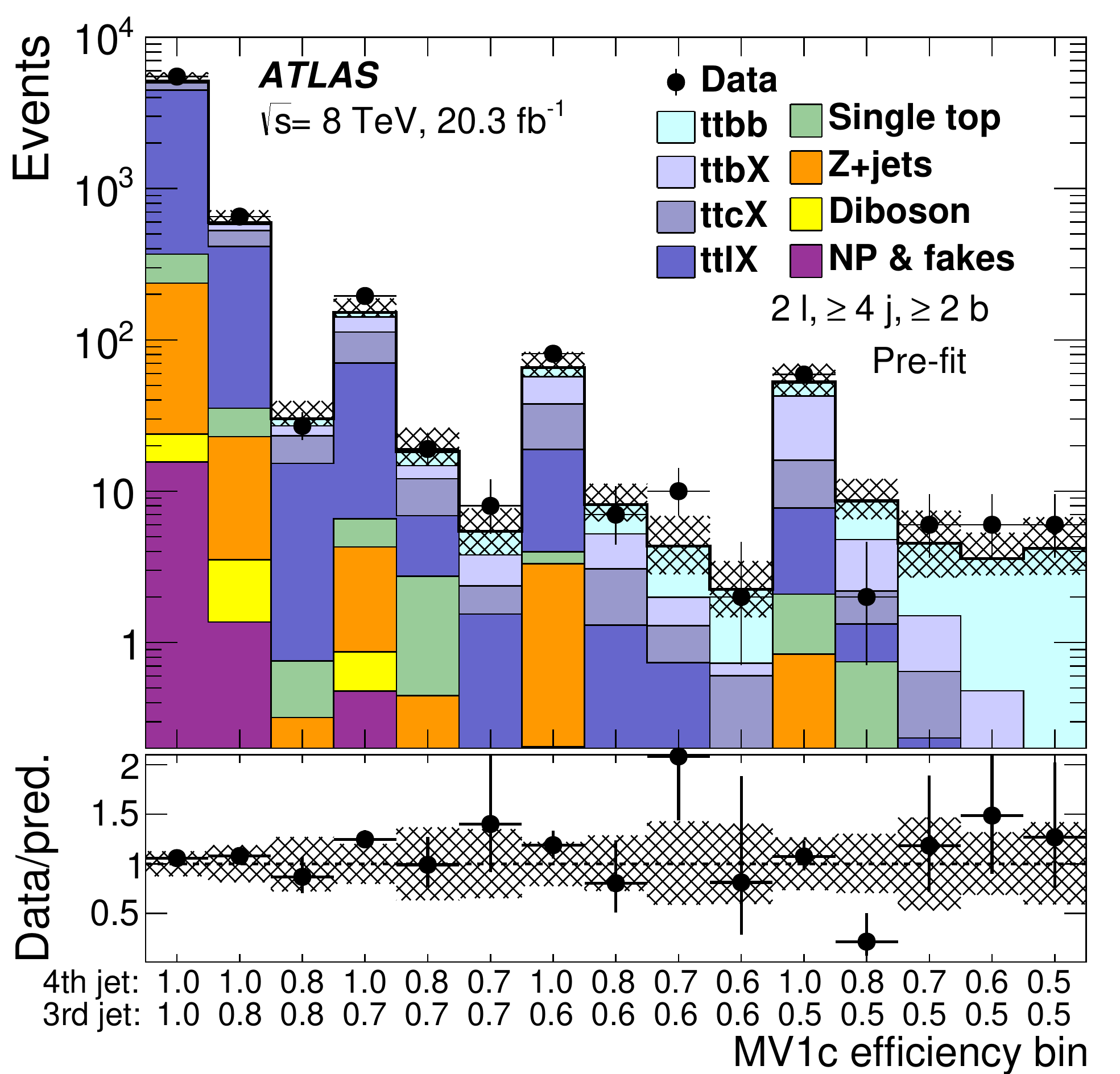}
\includegraphics[width=0.46\linewidth]{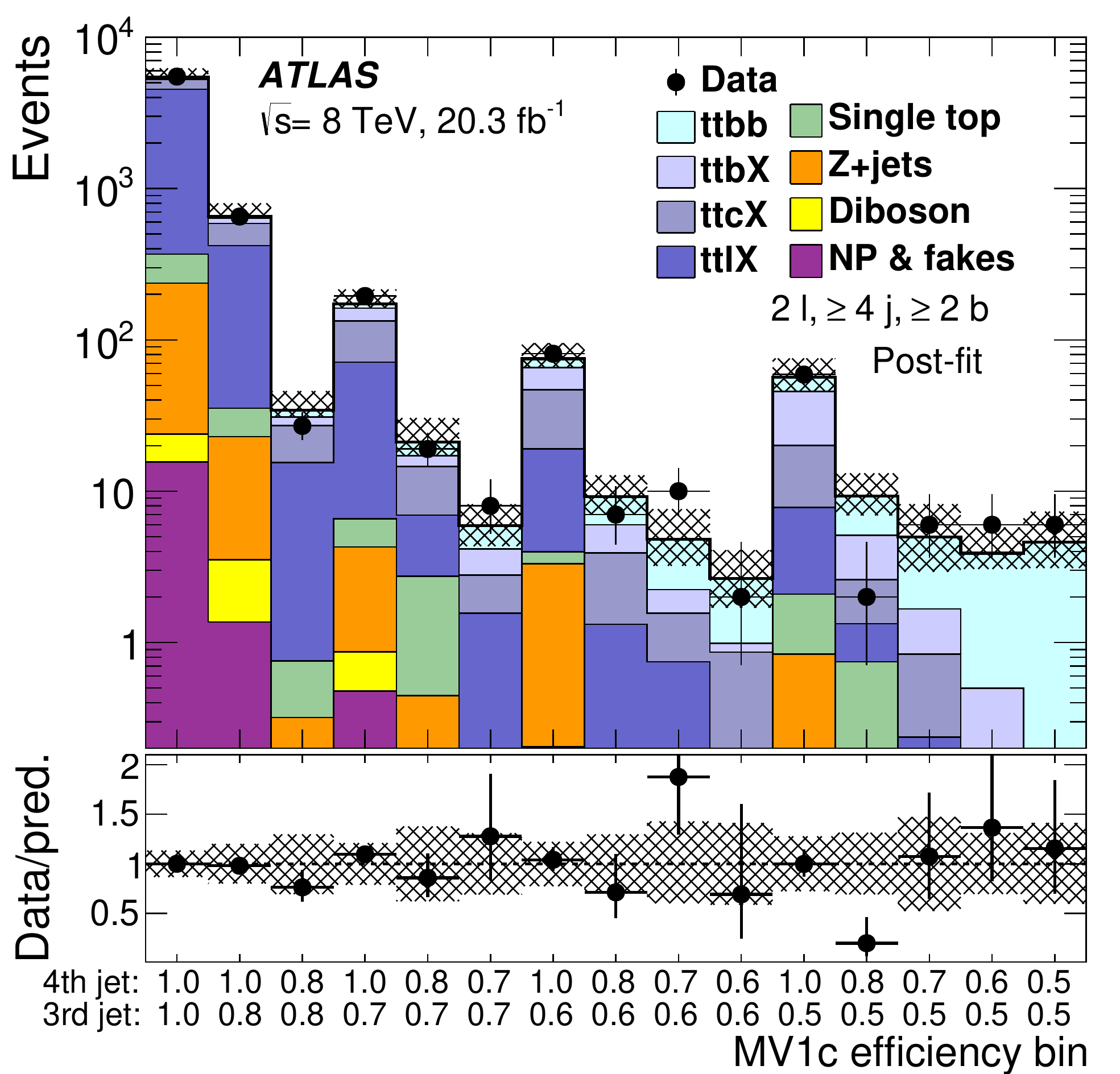}
\caption{The MV1c distribution of jets with the third and fourth highest MV1c weight in the dilepton channel for all signal and background components. 
The bins are labelled with the upper edge of the efficiency point of the third highest and fourth highest MV1c scores in the event.
The data (left) are compared to the nominal predictions (Pre-fit), and (right) to the output of the fit (Post-fit).
The points include the statistical uncertainty on the data.
The hashed area shows the total uncertainties. The bottom sub-plot shows the ratio of the data to the prediction.
The non-prompt and fake lepton backgrounds are referred to as `NP \& fakes'.}
\label{fig:mv1c_bb}
\end{figure}

Figure~\ref{fig:mv1c_bb} shows the MV1c distribution used to fit the number of \ttbbt\ events; the left figure shows the predictions from simulation compared to the observed distribution in data; the right plot shows data compared to the result of the fit.
The fitted cross-sections for each of the components are shown in Table~\ref{tab:ttbb_fit_merged} along with the predictions from {\sc PowhegBox+Pythia\,6}; the uncertainties shown are the statistical uncertainty of each component as obtained from the fit.
The fitted cross-sections are compatible with the predictions within fit uncertainties. The central value for \ttbb\ is 1.1 times the predictions from {\sc PowhegBox+Pythia\,6}, consistent with the $\mu$ values found in the two $ttb$ analyses.
In particular the values for the \ttbXt\, \ttcXt\ and \ttlXt\ may be used to cross-check the assumptions made about the background contributions to the cut-based analysis.

\begin{table}[htb!]
\begin{center}
\begin{small}
\begin{tabular}{l| c | c | c | c | c}
\hline\hline
Process & Observed  & Statistical & Systematic  & Total & Predicted  \\ 
 &  cross-section [fb] &    uncertainty (\%) &  uncertainty (\%) &  uncertainty (\%) & cross-section [fb]\\ \hline
 \ttbbt\ & 13.5		& $\pm$25 	& $\pm$27       & $\pm$36	& 12.3	\\
 \ttbXt\ & 61 		& $\pm$38 	& $\pm$69	& $\pm$79 	& 63		\\
 \ttcXt\ & 270		& $\pm$25 	& $\pm$81 	& $\pm$85	& 180	\\
 \ttlXt\ & 5870	 	& $\pm$4	& $\pm$14	& $\pm$15 	& 5800	\\ \hline
 $R_{ttbb}$ & 1.30\% 	& $\pm$25 	& $\pm$22	& $\pm$33       & 1.27\%	\\
 \hline \hline
\end{tabular}
\caption{Observed and predicted cross-sections for each of the components measured in the $ttbb$ fit analysis and on the $R_{ttbb}$ ratio. 
The statistical, total systematic, and total uncertainties on each component are also shown.
The predicted values are from {\sc Powheg+Pythia 6} \ttbar. 
}\label{tab:ttbb_fit_merged}
\end{small}
\end{center}
\end{table}

For most sources of systematic uncertainty, the templates for signal and background distributions are obtained from the event sample where a $\pm 1\sigma$ shift of the uncertainty was applied. 
The new templates and the old templates are fitted to the nominal MC sample, and the relative difference between the yields is taken as the uncertainty on the number of events. For systematic uncertainties that also affect the fiducial efficiencies, the efficiency is varied coherently and the effect on the final cross-section is obtained.
The effect due to limited number of MC events in the templates is obtained from the mean of 5000 pseudo-datasets obtained from simulation, where the variance of each bin depends on the total MC statistical uncertainty of that bin.
The second to last column of Table~\ref{tab:sys_all} shows the effect on the final $ttbb$ cross-section measurement in this analysis whereas the rightmost column shows the uncertainties on the $R_{ttbb}$ measurement.

The total cross-section uncertainty of each process and on the $R_{ttbb}$ ratio are shown in Table~\ref{tab:ttbb_fit_merged} along with the statistical and total systematic uncertainties.
The uncertainties on the \ttbXt\ and \ttcXt\ processes are large and do not allow the cross-sections of these processes to be constrained significantly.
The signal strength $\mu_{ttbb}$ has a correlation of 0.4 with $\mu_{ttbX}$, $-0.1$ with $\mu_{ttlX}$, and nearly 0 with $\mu_{ttcX}$.

%% file: results.tex
\section{Results}
\label{sec:Results}

The fiducial cross-sections obtained for each analysis in the previous section are shown in Table~\ref{tab:results}. 

The measurements of the $ttb$ cross-section in the lepton-plus-jets and $ttb$ $e \mu$ analyses are both higher than the predicted cross-section from the {\sc Powheg+Pythia} 6 sample, with a best fit value for the signal strength $\mu_{ttb}$ of 1.32 and 1.30, respectively. The total measurement uncertainty in the lepton-plus-jets channel is fractionally smaller than in the $ttb$ $e \mu$ analysis, $\sim$25\% compared to $\sim$32\%, owing to the higher acceptance
times branching ratio of this decay channel. The uncertainty in this channel is dominated by uncertainties on the tagging efficiency due to $c$-jets from events in which the $W$ boson decays to a $c$- and a light quark.

The two measurements of the $ttbb$ cross-section show similar precision despite the different approaches,  with the cut-based and fit-based analyses having a total uncertainty of $\sim$35\% and $\sim$36\%, respectively. 
The cut-based analysis is largely insensitive to the modelling of the non-$ttbb$ background from \ttbar\ events as the selection criteria are very tight.
In contrast, the fit-based analysis uses looser selection criteria in an attempt to obtain a data-driven constraint on these processes.
While the precision of the fit-based analysis does not allow for a measurement of these backgrounds, it does confirm the validity of the simulation, and allows for an explicit measurement of the $R_{ttbb}$ ratio. 
The two $ttbb$ measurements select different events and hence are not fully correlated. A small excess of data with respect to the nominal prediction is seen in the events that are common to both measurements, while a small deficit is seen for events with jets that satisfy the MV1c 80\% criterion but fail the MV1 70\% criterion that is used in the cut-based analysis. These two features explain the difference between the observed cross-section in the two analyses.

\begin{figure}[h]
\centering
\includegraphics[height=8cm]{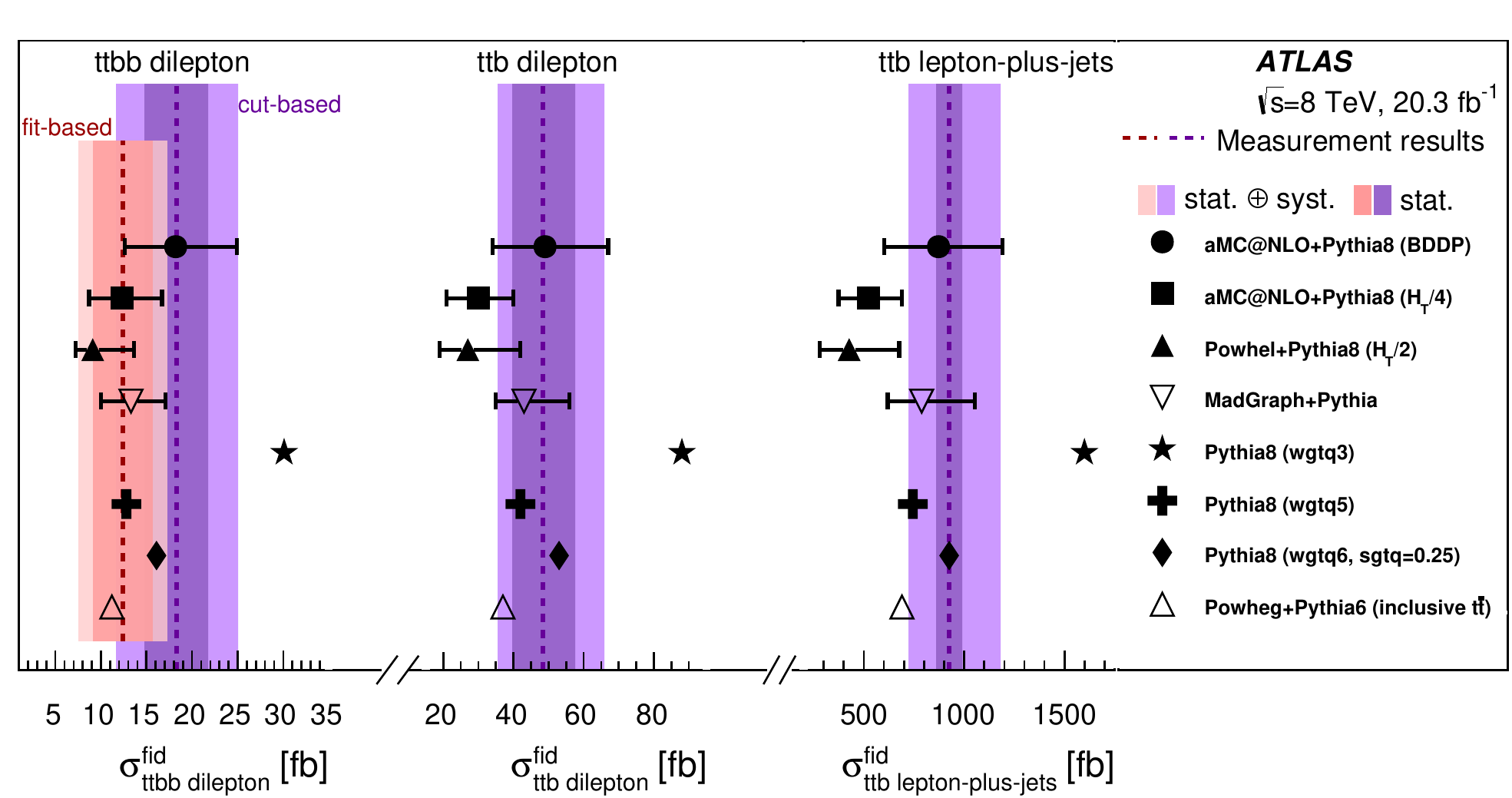}
\caption{Comparison of the measured cross-sections in the three fiducial phase-space regions with theoretical predictions obtained from a variety of different generators. The measurements are shown with the contributions from \ttv\ and \tth\ removed to allow direct comparison to the predictions containing only the pure QCD matrix elements. 
The coloured bands indicate the statistical and total uncertainties of the measurements. 
The errors on the theoretical prediction are obtained by simultaneously varying the renormalisation and factorisation scales by a factor of two. These variations have not been calculated for the LO {\sc Pythia} 8 samples or for the {\sc Powheg+Pythia} 6 sample.}
\label{fig:XsecTheory_ttbb}
\end{figure}

\begin{figure}[h]
\centering
\includegraphics[height=7cm]{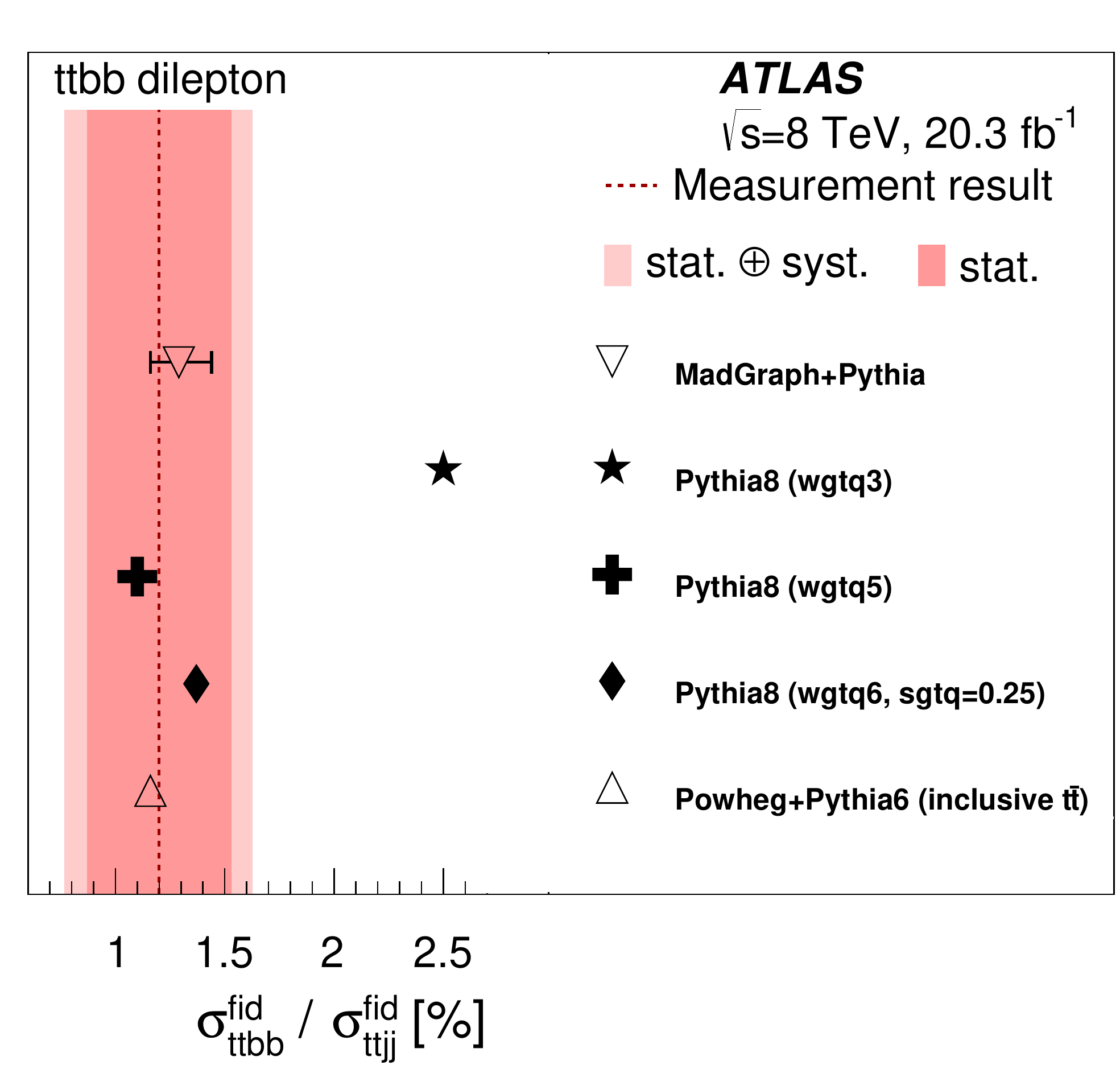}
\caption{Comparison of the measured ratio of the $ttbb$ and $ttjj$ cross-sections in the fiducial phase-space region of the $ttbb$ fit-based analysis with theoretical predictions obtained from a variety of different generators. The measurements are shown with the contributions from \ttv\ and \tth\ removed to allow direct comparison to the pure QCD generators. The coloured bands indicate the statistical and total uncertainties of the measurement. The error on the {\sc MadGraph+Pythia} prediction is obtained by simultaneously varying the renormalisation and factorisation scales by a factor of two. These variations have not been calculated for the LO {\sc Pythia} 8 samples or for the {\sc Powheg+Pythia} 6 sample.}
\label{fig:XsecTheory_ratio}
\end{figure}

An alternative set of results is obtained by subtracting the predicted \ttv\ and \tth\ contribution from the signal; no additional uncertainty due to the cross-section of these processes is considered. This allows a direct comparison of the measurements to QCD-only predictions, although with assumptions about the \ttv\ and \tth\ cross-sections.
These results are summarised in Table~\ref{tab:theory_pred} and Figure~\ref{fig:XsecTheory_ttbb} and compared to theoretical predictions obtained with the generators described in Section~\ref{sec:theory_pred} and shown in Table~\ref{tab:TheoryGenerators}. 
The ratio of the \ttbb\ and $ttjj$ cross-sections as measured in the \ttbb\ fit-based analysis is compared to theoretical predictions in Figure~\ref{fig:XsecTheory_ratio}. The uncertainties on the theoretical predictions are obtained by simultaneously varying the renormalisation and factorisation scales by a factor of two.

\begin{table}[h!]
\begin{center}
\renewcommand{\arraystretch}{1.3}
\begin{tabular}{l| r@{ }l@{ }l@{ }l |  c }
\hline\hline
Analysis  & \multicolumn{4}{c|}{Measured} & Predicted \\
 & \multicolumn{4}{c|}{Cross-section [fb]}  & Cross-section [fb] \\\hline
$\sigma_{ttb\, \mathrm{lepton-plus-jets}}$  	& 950 & $\pm$ 70 &(stat.) $^{+240}_{-190}$ &(syst.) & 720\\
$\sigma_{ttb\, e \mu}$  			&  50 & $\pm$ 10 &(stat.) $^{+15}_{-10}$ &(syst.) & 38 \\
 $\sigma_{ttbb\, \mathrm{cut-based}}$  		& 19.3 & $\pm$ 3.5 &(stat.) $\pm$ 5.7 &(syst.) & 12.3 \\
 $\sigma_{ttbb\, \mathrm{fit-based}}$  		& 13.5 & $\pm$ 3.3 &(stat.) $\pm$ 3.6 &(syst.) & 12.3 \\ \hline
$R_{ttbb}$  			        & 1.30 & $\pm$ 0.33 &(stat.) $\pm$ 0.28 &(syst.)\:\% & 1.27~\rlap{\%} \\
\hline\hline
\end{tabular}
\caption{Measured fiducial cross-section for $ttb$ in the lepton-plus-jets and $e \mu$ channels, and $ttbb$ in the dilepton channel using a cut-based or a fit-based method. Results for the $R_{ttbb}$ ratio measurement from the $ttbb$ fit-based method are also shown. The uncertainties quoted are from the statistical and total systematic uncertainties. The predicted cross-section is from {\sc PowhegBox} with {\sc Pythia 6} for the QCD component, from {\sc Helac} for \tth\, and from {\sc MadGraph} 5 for \ttv.}\label{tab:results}
\end{center}
\end{table}

\begin{table}[htb!]
\begin{center}
\begin{small}
\begin{tabular}{l |c|c|c|c }
\hline\hline
& $ttbb$ & $ttb$ Lepton-plus- & $ttb$ $e \mu$ & $R_{ttbb}$\\[1mm]
& [fb] & jets [fb]  & [fb]  & (\%) \\[1mm]
\hline
& & & \\[0.1mm]
\multirow{2}{*}{Observed} &(cut-based) 18.2 $\pm$3.5 $\pm$5.7 & \multirow{2}{*}{930 $\pm$70 $^{+240}_{-190}$} & \multirow{2}{*}{ 48 $\pm$10 $^{+15}_{-10}$}   & \multirow{2}{*}{1.20 $\pm0.33$ $\pm 0.28$}\\[2mm]
 & (fit-based) 12.4 $\pm$3.3 $\pm$3.6 & &   & \\[3mm]
  \hline
&&&  \\[0.1mm]
{\sc Madgraph5\_aMC@NLO} ($\mu_{\mathrm{BDDP}}$)                     & $18.2^{+6.7}_{-5.6}$ 	& $870^{+320}_{-270}$ 	& $49^{+18}_{-15}$	        & --\\[1mm]
{\sc Madgraph5\_aMC@NLO} ($\mu_{H_{\rm{T}}/4}$) 		        & $12.3^{+4.4}_{-3.6}$	& $520^{+170}_{-150}$ 	& $30^{+10}_{-9}$   	& --\\[1mm]
{\sc Powhel} 							& $9.1^{+4.5}_{-1.9}$ 	& $430^{+250}_{-150}$ 	& $27^{+15}_{-8}$         & --\\ [1mm]
{\sc Madgraph5}+{\sc Pythia} 6 					& $13.3^{+3.8}_{-3.3}$ 	& $790^{+270}_{-170}$  	& $43^{+13}_{-8}$   	& 1.29$^{+0.15}_{-0.13}$ \\[1mm]
{\sc Pythia} 8 (wgtq=3) 					& 30.1 			& 1600 			& 88   			& 2.50 \\[1mm]
{\sc Pythia} 8 (wgtq=5) 					& 12.8 		        & 740 			& 42   			& 1.10 \\[1mm]
{\sc Pythia} 8 (wgtq=6,sgtq=0.25) 				& 16.1 			& 930 			& 53   			& 1.37 \\[1mm]
{\sc Powheg+Pythia} 6 ({\sc hdamp}=$m_{\mathrm{top}}$)               & 11.2 		& 690 			& 37   			& 1.16 \\[1mm]
\hline\hline
\end{tabular}
\caption{Observed and predicted cross-sections for the three fiducial phase-space regions. The measurements are shown with the contributions from \ttv\ and \tth\ removed to allow direct comparison to the predictions containing only the pure QCD matrix elements. Results for the $R_{ttbb}$ ratio measurement from the $ttbb$ fit-based method are also shown. The measurement uncertainties are separated into statistical (first) and systematic (second) uncertainties.
 The uncertainties on the theoretical predictions are obtained by simultaneously varying the renormalisation and factorisation scales by a factor of two up or down. These variations have not been calculated for the LO {\sc Pythia} 8 samples or for the {\sc Powheg+Pythia} 6 sample.}
\label{tab:theory_pred}
\end{small}
\end{center}
\end{table}

The predictions containing NLO matrix elements for the $pp\rightarrow\ttbb$ process, as well as the merged LO+PS prediction from {\sc MadGraph+Pythia}~6 are in agreement with the measured cross-sections within the measurement uncertainties. The cross-sections obtained in the 5FS ({\sc Powhel}) are higher than the 4FS ones ({\sc MadGraph5\_aMC@NLO}) as expected, however the two predictions agree within the respective scale uncertainties. 
The models utilizing softer choices for the renormalisation/factorisation scales show the best agreement with the data.
Different $g\rightarrow b\bar{b}$ splitting models significantly affect the $ttbb$ and $ttb$ cross-sections in the samples where all additional $b$-jets come from the parton shower. The predictions corresponding to wgtq=3 and wgtq=5, which correspond to the extreme models, differ by more than a factor of two. The cross-sections obtained with the wgtq=3 model are significantly higher than the measured ones, thus indicating that this model overestimates the $g\rightarrow b\bar{b}$ rate. The cross-sections obtained with the other models are both in agreement with the data.

\clearpage

%% file: Acknowledgements.tex

We thank CERN for the very successful operation of the LHC, as well as the
support staff from our institutions without whom ATLAS could not be
operated efficiently.

We acknowledge the support of ANPCyT, Argentina; YerPhI, Armenia; ARC, Australia; BMWFW and FWF, Austria; ANAS, Azerbaijan; SSTC, Belarus; CNPq and FAPESP, Brazil; NSERC, NRC and CFI, Canada; CERN; CONICYT, Chile; CAS, MOST and NSFC, China; COLCIENCIAS, Colombia; MSMT CR, MPO CR and VSC CR, Czech Republic; DNRF, DNSRC and Lundbeck Foundation, Denmark; IN2P3-CNRS, CEA-DSM/IRFU, France; GNSF, Georgia; BMBF, HGF, and MPG, Germany; GSRT, Greece; RGC, Hong Kong SAR, China; ISF, I-CORE and Benoziyo Center, Israel; INFN, Italy; MEXT and JSPS, Japan; CNRST, Morocco; FOM and NWO, Netherlands; RCN, Norway; MNiSW and NCN, Poland; FCT, Portugal; MNE/IFA, Romania; MES of Russia and NRC KI, Russian Federation; JINR; MESTD, Serbia; MSSR, Slovakia; ARRS and MIZ\v{S}, Slovenia; DST/NRF, South Africa; MINECO, Spain; SRC and Wallenberg Foundation, Sweden; SERI, SNSF and Cantons of Bern and Geneva, Switzerland; MOST, Taiwan; TAEK, Turkey; STFC, United Kingdom; DOE and NSF, United States of America. In addition, individual groups and members have received support from BCKDF, the Canada Council, CANARIE, CRC, Compute Canada, FQRNT, and the Ontario Innovation Trust, Canada; EPLANET, ERC, FP7, Horizon 2020 and Marie Skłodowska-Curie Actions, European Union; Investissements d'Avenir Labex and Idex, ANR, Region Auvergne and Fondation Partager le Savoir, France; DFG and AvH Foundation, Germany; Herakleitos, Thales and Aristeia programmes co-financed by EU-ESF and the Greek NSRF; BSF, GIF and Minerva, Israel; BRF, Norway; the Royal Society and Leverhulme Trust, United Kingdom.

The crucial computing support from all WLCG partners is acknowledged
gratefully, in particular from CERN and the ATLAS Tier-1 facilities at
TRIUMF (Canada), NDGF (Denmark, Norway, Sweden), CC-IN2P3 (France),
KIT/GridKA (Germany), INFN-CNAF (Italy), NL-T1 (Netherlands), PIC (Spain),
ASGC (Taiwan), RAL (UK) and BNL (USA) and in the Tier-2 facilities
worldwide.

%% file: atlas_authlist.tex
\begin{flushleft}
{\Large The ATLAS Collaboration}

\bigskip

G.~Aad$^{\rm 85}$,
B.~Abbott$^{\rm 113}$,
J.~Abdallah$^{\rm 151}$,
O.~Abdinov$^{\rm 11}$,
R.~Aben$^{\rm 107}$,
M.~Abolins$^{\rm 90}$,
O.S.~AbouZeid$^{\rm 158}$,
H.~Abramowicz$^{\rm 153}$,
H.~Abreu$^{\rm 152}$,
R.~Abreu$^{\rm 116}$
% ...
\bigskip
\\
% ...
$^{11}$ Institute of Physics, Azerbaijan Academy of Sciences, Baku, Azerbaijan\\
% ...
$^{85}$ CPPM, Aix-Marseille Universit{\'e} and CNRS/IN2P3, Marseille, France\\
% ...
$^{90}$ Department of Physics and Astronomy, Michigan State University, East Lansing MI, United States of America\\
% ...
$^{107}$ Nikhef National Institute for Subatomic Physics and University of Amsterdam, Amsterdam, Netherlands\\
% ...
$^{113}$ Homer L. Dodge Department of Physics and Astronomy, University of Oklahoma, Norman OK, United States of America\\
% ...
$^{116}$ Center for High Energy Physics, University of Oregon, Eugene OR, United States of America\\
% ...
$^{151}$ Institute of Physics, Academia Sinica, Taipei, Taiwan\\
$^{152}$ Department of Physics, Technion: Israel Institute of Technology, Haifa, Israel\\
$^{153}$ Raymond and Beverly Sackler School of Physics and Astronomy, Tel Aviv University, Tel Aviv, Israel\\
% ...
$^{158}$ Department of Physics, University of Toronto, Toronto ON, Canada\\
% ...
\end{flushleft}


%% file: ttHf_paper.bbl
\providecommand{\href}[2]{#2}\begingroup\raggedright\begin{thebibliography}{10}

\bibitem{NLO1}
A.~Bredenstein, A.~Denner, S.~Dittmaier,  and S.~Pozzorini, {\em {NLO QCD
  corrections to $pp\rightarrow t\bar{t}b\bar{b}$ + X at the LHC}},
  \href{http://dx.doi.org/10.1103/PhysRevLett.103.012002}{Phys. Rev. Lett.
  {\bfseries 103} (2009) 012002},
\href{http://arxiv.org/abs/0905.0110}{{\ttfamily arXiv:0905.0110 [hep-ph]}}.

\bibitem{NLO2}
A.~Bredenstein, A.~Denner, S.~Dittmaier,  and S.~Pozzorini, {\em {NLO QCD
  Corrections to Top Anti-Top Bottom Anti-Bottom Production at the LHC: 2. full
  hadronic results}}, \href{http://dx.doi.org/10.1007/JHEP03(2010)021}{JHEP
  {\bfseries 03} (2010) 021},
\href{http://arxiv.org/abs/1001.4006}{{\ttfamily arXiv:1001.4006 [hep-ph]}}.

\bibitem{NLO3}
G.~Bevilacqua, M.~Czakon, C.~Papadopoulos, R.~Pittau,  and M.~Worek, {\em
  {Assault on the NLO Wishlist: $pp\rightarrow t\bar{t}b\bar{b}$}},
  \href{http://dx.doi.org/10.1088/1126-6708/2009/09/109}{JHEP {\bfseries 09}
  (2009) 109},
\href{http://arxiv.org/abs/0907.4723}{{\ttfamily arXiv:0907.4723 [hep-ph]}}.

\bibitem{NLOPS1}
A.~Kardos and Z.~Tr\'ocs\'anyi, {\em {Hadroproduction of t anti-t pair with a b
  anti-b pair using PowHel}},
  \href{http://dx.doi.org/10.1088/0954-3899/41/7/075005}{J. Phys. G {\bfseries
  41} (2014) 075005},
\href{http://arxiv.org/abs/1303.6291}{{\ttfamily arXiv:1303.6291 [hep-ph]}}.

\bibitem{NLOPS2}
F.~Cascioli, P.~Maierhoefer, N.~Moretti, S.~Pozzorini,  and F.~Siegert, {\em
  {NLO matching for $t\bar t b \bar b$ production with massive $b$-quarks}},
  \href{http://dx.doi.org/10.1016/j.physletb.2014.05.040}{Phys. Lett. B
  {\bfseries 734} (2014) 210},
\href{http://arxiv.org/abs/1309.5912}{{\ttfamily arXiv:1309.5912 [hep-ph]}}.

\bibitem{NLOPS3}
S.~Hoeche {et~al.}, {\em {Next-to-leading order QCD predictions for top-quark
  pair production with up to two jets merged with a parton shower}},
  \href{http://dx.doi.org/http://dx.doi.org/10.1016/j.physletb.2015.06.060}{Phys.
  Lett. B {\bfseries 748} (2015) 74},
\href{http://arxiv.org/abs/1402.6293}{{\ttfamily arXiv:1402.6293 [hep-ph]}}.

\bibitem{Alwall:2014hca}
J.~Alwall {et~al.}, {\em {The automated computation of tree-level and
  next-to-leading order differential cross sections, and their matching to
  parton shower simulations}},
  \href{http://dx.doi.org/10.1007/JHEP07(2014)079}{JHEP {\bfseries 07} (2014)
  079},
\href{http://arxiv.org/abs/1405.0301}{{\ttfamily arXiv:1405.0301 [hep-ph]}}.

\bibitem{Garzelli:2014aba}
M.~Garzelli, A.~Kardos,  and Z.~Trócsányi, {\em {Hadroproduction of
  $t\bar{t}b\bar{b}$ final states at LHC: predictions at NLO accuracy matched
  with Parton Shower}}, \href{http://dx.doi.org/10.1007/JHEP03(2015)083}{JHEP
  {\bfseries 03} (2015) 083},
\href{http://arxiv.org/abs/1408.0266}{{\ttfamily arXiv:1408.0266 [hep-ph]}}.

\bibitem{ATLASHiggs}
{ATLAS Collaboration}, {\em {Observation of a new particle in the search for
  the Standard Model Higgs boson with the ATLAS detector at the LHC}},
  \href{http://dx.doi.org/{10.1016/j.physletb.2012.08.020}}{Phys. Lett. B
  {\bfseries {716}} (2012) {1}},
  \href{http://arxiv.org/abs/1207.7214}{{\ttfamily arXiv:1207.7214 [hep-ex]}}.

\bibitem{CMSHiggs}
{CMS Collaboration}, {\em {Observation of a new boson at a mass of 125 GeV with
  the CMS experiment at the LHC}},
  \href{http://dx.doi.org/{10.1016/j.physletb.2012.08.021}}{Phys. Lett. B
  {\bfseries {716}} (2012) {30}},
  \href{http://arxiv.org/abs/1207.7235}{{\ttfamily arXiv:1207.7235 [hep-ex]}}.

\bibitem{Khachatryan:2014qaa}
{ATLAS Collaboration}, {\em {Search for the associated production of the Higgs
  boson with a top-quark pair}},
  \href{http://dx.doi.org/10.1007/JHEP09(2014)087}{JHEP {\bfseries 09} (2014)
  087},
\href{http://arxiv.org/abs/1408.1682}{{\ttfamily arXiv:1408.1682 [hep-ex]}}.

\bibitem{ttHMEMCMS}
{CMS Collaboration}, {\em {Search for a standard model Higgs boson produced in
  association with a top-quark pair and decaying to bottom quarks using a
  matrix element method}},
  \href{http://dx.doi.org/10.1140/epjc/s10052-015-3454-1}{Eur. Phys. J. C
  {\bfseries 75} (2015) 251},
\href{http://arxiv.org/abs/1502.02485}{{\ttfamily arXiv:1502.02485 [hep-ex]}}.

\bibitem{atlasttH}
{ATLAS Collaboration}, {\em {Search for the Standard Model Higgs boson produced
  in association with top quarks and decaying into $b\bar{b}$ in pp collisions
  at $\sqrt{s}$ = 8 TeV with the ATLAS detector}},
  \href{http://dx.doi.org/10.1140/epjc/s10052-015-3543-1}{Eur. Phys. J.
  {\bfseries C75} no.~7, (2015) 349},
\href{http://arxiv.org/abs/1503.05066}{{\ttfamily arXiv:1503.05066 [hep-ex]}}.

\bibitem{ATLASttbb7TeV}
{ATLAS Collaboration}, {\em {A study of heavy flavor quarks produced in
  association with top quark pairs at $\sqrt{s} = 7$ TeV using the ATLAS
  detector}}, \href{http://dx.doi.org/10.1103/PhysRevD.89.072012}{Phys. Rev. D
  {\bfseries 89} (2014) 072012},
\href{http://arxiv.org/abs/1304.6386}{{\ttfamily arXiv:1304.6386 [hep-ex]}}.

\bibitem{CMSttbb8TeV}
{CMS Collaboration}, {\em {Measurement of the cross section ratio
  $\sigma_\mathrm{t \bar{t} b \bar{b}} / \sigma_\mathrm{t \bar{t} jj }$ in
  $\mathrm{pp}$ collisions at $\sqrt{s}$ = 8 TeV}},
  \href{http://dx.doi.org/http://dx.doi.org/10.1016/j.physletb.2015.04.060}{Phys.
  Lett. B {\bfseries 746} (2015) 132},
\href{http://arxiv.org/abs/1411.5621}{{\ttfamily arXiv:1411.5621 [hep-ex]}}.

\bibitem{CMSttbb8TeV2}
{CMS Collaboration}, {\em {Measurement of $\mathrm{ t \bar{t} } $ production
  with additional jet activity, including b quark jets, in the dilepton decay
  channel using pp collisions at $\sqrt{s} =$ 8 TeV}},
\href{http://arxiv.org/abs/1510.03072}{{\ttfamily arXiv:1510.03072 [hep-ex]}}.

\bibitem{MV1}
{ATLAS Collaboration}, {\em {Calibration of the performance of b-tagging for c
  and light-flavour jets in the 2012 ATLAS data}}, {ATLAS-CONF-2014-046,
  \href{https://cds.cern.ch/record/1741020}{https://cds.cern.ch/record/1741020}}.

\bibitem{ttjets}
{ATLAS Collaboration}, {\em {Measurement of the $t\bar{t}$ production
  cross-section as a function of jet multiplicity and jet transverse momentum
  in 7 TeV proton-proton collisions with the ATLAS detector}},
  \href{http://dx.doi.org/10.1007/JHEP01(2015)020}{JHEP {\bfseries 01} (2015)
  020},
\href{http://arxiv.org/abs/1407.0891}{{\ttfamily arXiv:1407.0891 [hep-ex]}}.

\bibitem{ref:Cacciari2008}
M.~Cacciari, G.~P. Salam,  and G.~Soyez, {\em The anti-$k_{t}$ jet clustering
  algorithm}, \href{http://dx.doi.org/10.1088/1126-6708/2008/04/063}{JHEP
  {\bfseries 04} (2008) 063},
  \href{http://arxiv.org/abs/0802.1189v2}{{\ttfamily arXiv:0802.1189v2
  [hep-ph]}}.

\bibitem{ref:Cacciari2006}
M.~Cacciari and G.~P. Salam, {\em Dispelling the $N^3$ myth for the $k_t$
  jet-finder}, \href{http://dx.doi.org/10.1016/j.physletb.2006.08.037}{Phys.
  Lett. B {\bfseries 641} (2006) 57},
  \href{http://arxiv.org/abs/0512210v2}{{\ttfamily arXiv:0512210v2 [hep-ph]}}.

\bibitem{ref:fastjet}
M.~Cacciari, G.~P. Salam,  and G.~Soyez, {\em FastJet User Manual},
  \href{http://dx.doi.org/10.1140/epjc/s10052-012-1896-2}{Eur. Phys. J. C
  {\bfseries 72} (2012) 1896}, \href{http://arxiv.org/abs/1111.6097}{{\ttfamily
  arXiv:1111.6097 [hep-ph]}}, \url{{http://fastjet.fr/}}.

\bibitem{Cacciari:2008gn}
M.~Cacciari, G.~P. Salam,  and G.~Soyez, {\em {The Catchment Area of Jets}},
  \href{http://dx.doi.org/10.1088/1126-6708/2008/04/005}{JHEP {\bfseries 04}
  (2008) 005},
\href{http://arxiv.org/abs/0802.1188}{{\ttfamily arXiv:0802.1188 [hep-ph]}}.

\bibitem{atlas-detector}
{ATLAS Collaboration}, {\em {The ATLAS Experiment at the CERN Large Hadron
  Collider}},
\href{http://dx.doi.org/10.1088/1748-0221/3/08/S08003}{JINST {\bfseries 3}
  (2008) S08003}.

\bibitem{atlas-trigger-2010}
{ATLAS Collaboration}, {\em {Performance of the ATLAS Trigger System in 2010}},
  \href{http://dx.doi.org/10.1140/epjc/s10052-011-1849-1}{Eur. Phys. J. C
  {\bfseries 72} (2012) 1849},
\href{http://arxiv.org/abs/1110.1530}{{\ttfamily arXiv:1110.1530 [hep-ex]}}.

\bibitem{powheg}
P.~Nason, {\em {A new method for combining NLO QCD with shower Monte Carlo
  algorithms}}, \href{http://dx.doi.org/10.1088/1126-6708/2004/11/040}{JHEP
  {\bfseries 11} (2004) 040},
  \href{http://arxiv.org/abs/hep-ph/0409146}{{\ttfamily arXiv:hep-ph/0409146}}.

\bibitem{powbox1}
S.~Frixione, P.~Nason,  and C.~Oleari, {\em {Matching NLO QCD computations with
  Parton Shower simulations: the POWHEG method}},
  \href{http://dx.doi.org/10.1088/1126-6708/2007/11/070}{JHEP {\bfseries 11}
  (2007) 070},
\href{http://arxiv.org/abs/0709.2092}{{\ttfamily arXiv:0709.2092 [hep-ph]}}.

\bibitem{powbox2}
S.~Alioli, P.~Nason, C.~Oleari,  and E.~Re, {\em {A general framework for
  implementing NLO calculations in shower Monte Carlo programs: the POWHEG
  BOX}}, \href{http://dx.doi.org/10.1007/JHEP06(2010)043}{JHEP {\bfseries 06}
  (2010) 043}, \href{http://arxiv.org/abs/1002.2581}{{\ttfamily arXiv:1002.2581
  [hep-ph]}}.

\bibitem{ct10}
H.-L. Lai {et~al.}, {\em {New parton distributions for collider physics}},
  \href{http://dx.doi.org/10.1103/PhysRevD.82.074024}{Phys. Rev. D {\bfseries
  82} (2010) 074024}, \href{http://arxiv.org/abs/1007.2241}{{\ttfamily
  arXiv:1007.2241 [hep-ph]}}.

\bibitem{PythiaManual}
T.~Sj{\"o}strand, S.~Mrenna,  and P.~Skands, {\em Pythia 6.4 Physics and
  Manual}, \href{http://dx.doi.org/10.1088/1126-6708/2006/05/026}{JHEP
  {\bfseries 05} (2006) 026},
  \href{http://arxiv.org/abs/hep-ph/0603175}{{\ttfamily arXiv:hep-ph/0603175}}.

\bibitem{cteq6l1}
J.~Pumplin {et~al.}, {\em {New generation of parton distributions with
  uncertainties from global QCD analysis}},
  \href{http://dx.doi.org/10.1088/1126-6708/2002/07/012}{JHEP {\bfseries 07}
  (2002) 012},
\href{http://arxiv.org/abs/hep-ph/0201195}{{\ttfamily arXiv:hep-ph/0201195
  [hep-ph]}}.

\bibitem{perugia}
P.~Z. Skands, {\em {Tuning Monte Carlo Generators: The Perugia Tunes}},
  \href{http://dx.doi.org/10.1103/PhysRevD.82.074018}{Phys. Rev. D {\bfseries
  82} (2010) 074018},
\href{http://arxiv.org/abs/1005.3457}{{\ttfamily arXiv:1005.3457 [hep-ph]}}.

\bibitem{ref:xs1}
{M. Cacciari et al.}, {\em Top-pair production at hadron colliders with
  next-to-next-to-leading logarithmic soft-gluon resummation},
  \href{http://dx.doi.org/10.1016/j.physletb.2012.03.013}{Phys. Lett. B
  {\bfseries 710} (2012) 612}, \href{http://arxiv.org/abs/1111.5869}{{\ttfamily
  arXiv:1111.5869 [hep-ph]}}.

\bibitem{ref:xs2}
{P. B\"arnreuther et al}., {\em Percent Level Precision Physics at the
  Tevatron: First Genuine NNLO QCD Corrections to $q\bar{q}\rightarrow
  t\bar{t}$}, \href{http://dx.doi.org/10.1103/PhysRevLett.109.132001}{Phys.
  Rev. Lett. {\bfseries 109} (2012) 132001},
  \href{http://arxiv.org/abs/1204.5201}{{\ttfamily arXiv:1204.5201 [hep-ph]}}.

\bibitem{ref:xs3}
M.~Czakon and A.~Mitov, {\em NNLO corrections to top-pair production at hadron
  colliders: the all-fermionic scattering channels},
  \href{http://dx.doi.org/10.1007/JHEP12(2012)054}{JHEP {\bfseries 12} (2012)
  054}, \href{http://arxiv.org/abs/1207.0236}{{\ttfamily arXiv:1207.0236
  [hep-ph]}}.

\bibitem{ref:xs4}
M.~Czakon and A.~Mitov, {\em NNLO corrections to top-pair production at hadron
  colliders: the quark-gluon reaction},
  \href{http://dx.doi.org/10.1007/JHEP01(2013)080}{JHEP {\bfseries 01} (2013)
  080}, \href{http://arxiv.org/abs/1210.6832}{{\ttfamily arXiv:1210.6832
  [hep-ph]}}.

\bibitem{ref:xs5}
M.~Czakon, P.~Fiedler,  and A.~Mitov, {\em The total top quark pair production
  cross-section at hadron colliders through $\mathcal{O}(\alpha_S^4)$},
  \href{http://dx.doi.org/10.1103/PhysRevLett.110.252004}{Phys. Rev. Lett.
  {\bfseries 110} (2013) 252004},
  \href{http://arxiv.org/abs/1303.6254}{{\ttfamily arXiv:1303.6254 [hep-ph]}}.

\bibitem{ref:xs6}
M.~Czakon and A.~Mitov, {\em {Top++: A Program for the Calculation of the
  Top-Pair Cross-Section at Hadron Colliders}},
  \href{http://dx.doi.org/10.1016/j.cpc.2014.06.021}{Comput. Phys. Commun.
  {\bfseries 185} (2014) 2930},
\href{http://arxiv.org/abs/1112.5675}{{\ttfamily arXiv:1112.5675 [hep-ph]}}.

\bibitem{madgraph}
J.~Alwall {et~al.}, {\em MadGraph/MadEvent v4: the new web generation},
  \href{http://dx.doi.org/{10.1088/1126-6708/2007/09/028}}{JHEP {\bfseries 09}
  (2007) 028}, \href{http://arxiv.org/abs/0706.2334}{{\ttfamily arXiv:0706.2334
  [hep-ph]}}.

\bibitem{auet2b}
{ATLAS Collaboration}, {\em {ATLAS tunes of PYTHIA 6 and Pythia 8 for MC11}},
  {ATL-PHYS-PUB-2011-009,
  \href{http://cdsweb.cern.ch/record/1363300}{http://cdsweb.cern.ch/record/1363300}}.

\bibitem{ttbarVxs1}
J.~M. Campbell and R.~K. Ellis, {\em {$t \bar{t} W^{\pm}$ production and decay
  at NLO}}, \href{http://dx.doi.org/10.1007/JHEP07(2012)052}{JHEP {\bfseries
  07} (2012) 052},
\href{http://arxiv.org/abs/1204.5678}{{\ttfamily arXiv:1204.5678 [hep-ph]}}.

\bibitem{ttbarVxs2}
M.~V. Garzelli, A.~Kardos, C.~G. Papadopoulos,  and Z.~Trocsanyi, {\em
  {$t\bar{t}W$ and $t\bar{t}Z$ Hadroproduction at NLO accuracy in QCD with
  Parton Shower and Hadronization effects}},
  \href{http://dx.doi.org/10.1007/JHEP11(2012)056}{JHEP {\bfseries 11} (2012)
  056}, \href{http://arxiv.org/abs/1208.2665}{{\ttfamily arXiv:1208.2665
  [hep-ph]}}.

\bibitem{Helac}
G.~Bevilacqua {et~al.}, {\em {\sc HELAC-NLO}},
  \href{http://dx.doi.org/10.1016/j.cpc.2012.10.033}{Comput. Phys. Commun.
  {\bfseries 184} (2013) 986},
  \href{http://arxiv.org/abs/1110.1499v2}{{\ttfamily arXiv:1110.1499v2
  [hep-ph]}}.

\bibitem{pythia8}
T.~Sjostrand, S.~Mrenna,  and P.~Z. Skands, {\em {A Brief Introduction to
  PYTHIA 8.1}}, \href{http://dx.doi.org/10.1016/j.cpc.2008.01.036}{Comput.
  Phys. Commun. {\bfseries 178} (2008) 852--867},
\href{http://arxiv.org/abs/0710.3820}{{\ttfamily arXiv:0710.3820 [hep-ph]}}.

\bibitem{Alioli:2010xd}
S.~Alioli, P.~Nason, C.~Oleari,  and E.~Re, {\em {A general framework for
  implementing NLO calculations in shower Monte Carlo programs: the POWHEG
  BOX}}, \href{http://dx.doi.org/10.1007/JHEP06(2010)043}{JHEP {\bfseries 06}
  (2010) 043},
\href{http://arxiv.org/abs/1002.2581}{{\ttfamily arXiv:1002.2581 [hep-ph]}}.

\bibitem{Powhel-ttH}
M.~Garzelli, A.~Kardos, C.~Papadopoulos,  and Z.~Trocsanyi, {\em {Standard
  Model Higgs boson production in association with a top anti-top pair at NLO
  with parton showering}},
  \href{http://dx.doi.org/10.1209/0295-5075/96/11001}{Europhys. Lett.
  {\bfseries 96} (2011) 11001},
\href{http://arxiv.org/abs/1108.0387}{{\ttfamily arXiv:1108.0387 [hep-ph]}}.

\bibitem{au2}
{ATLAS Collaboration}, {\em {Summary of ATLAS PYTHIA 8 tunes}},
  {ATL-PHYS-PUB-2012-003,
  \href{https://cds.cern.ch/record/1474107}{https://cds.cern.ch/record/1474107}}.

\bibitem{Heinemeyer:2013tqa}
{LHC Higgs Cross Section Working Group, S. Heinemeyer {\it et al}}, {\em
  {Handbook of LHC Higgs Cross Sections: 3. Higgs Properties}},
\href{http://arxiv.org/abs/1307.1347}{{\ttfamily arXiv:1307.1347 [hep-ph]}}.

\bibitem{alpgen}
M.~L. Mangano {et~al.}, {\em {ALPGEN, a generator for hard multiparton
  processes in hadronic collisions}},
  \href{http://dx.doi.org/10.1088/1126-6708/2003/07/001}{JHEP {\bfseries 07}
  (2003) 001}, \href{http://arxiv.org/abs/{0206293}}{{\ttfamily
  {arXiv}:{0206293} [{hep-ph}]}}.

\bibitem{cteq6}
P.~M. Nadolsky {et~al.}, {\em {Implications of CTEQ global analysis for
  collider observables}},
  \href{http://dx.doi.org/10.1103/PhysRevD.78.013004}{Phys. Rev. D {\bfseries
  78} (2008) 013004}, \href{http://arxiv.org/abs/0802.0007}{{\ttfamily
  arXiv:0802.0007 [hep-ph]}}.

\bibitem{mlm}
M.~L. Mangano {et~al.}, {\em {Multijet matrix elements and shower evolution in
  hadronic collisions: $Wb\bar{b}+n$ jets as a case study}},
  \href{http://dx.doi.org/(10.1016/S0550-3213(02)00249-3)}{Nucl. Phys. B
  {\bfseries {632}} (2002) 343}, \href{http://arxiv.org/abs/0108069}{{\ttfamily
  arXiv:0108069 [hep-ph]}}.

\bibitem{vjetsxs}
K.~Melnikov and F.~Petriello, {\em {Electroweak gauge boson production at
  hadron colliders through $\mathcal{O}(\alpha_{s}^{2})$}},
  \href{http://dx.doi.org/10.1103/PhysRevD.74.114017}{Phys. Rev. D {\bfseries
  74} (2006) 114017}, \href{http://arxiv.org/abs/0609070}{{\ttfamily
  arXiv:0609070 [hep-ph]}}.

\bibitem{herwig}
G.~Corcella {et~al.}, {\em {HERWIG 6: An Event generator for hadron emission
  reactions with interfering gluons (including supersymmetric processes)}},
  \href{http://dx.doi.org/10.1088/1126-6708/2001/01/010}{JHEP {\bfseries 01}
  (2001) 010},
\href{http://arxiv.org/abs/hep-ph/0011363}{{\ttfamily arXiv:hep-ph/0011363
  [hep-ph]}}.

\bibitem{Gleisberg:2008ta}
T.~Gleisberg {et~al.}, {\em {Event generation with SHERPA 1.1}},
  \href{http://dx.doi.org/10.1088/1126-6708/2009/02/007}{JHEP {\bfseries 02}
  (2009) 007}, \href{http://arxiv.org/abs/0811.4622}{{\ttfamily arXiv:0811.4622
  [hep-ph]}}.

\bibitem{Hoeche:2009rj}
S.~H{\"o}che, F.~Krauss, S.~Schumann,  and F.~Siegert, {\em {QCD matrix
  elements and truncated showers}},
  \href{http://dx.doi.org/10.1088/1126-6708/2009/05/053}{JHEP {\bfseries 05}
  (2009) 053}, \href{http://arxiv.org/abs/0903.1219}{{\ttfamily arXiv:0903.1219
  [hep-ph]}}.

\bibitem{Gleisberg:2008fv}
T.~Gleisberg and S.~H{\"o}che, {\em {Comix, a new matrix element generator}},
  \href{http://dx.doi.org/10.1088/1126-6708/2008/12/039}{JHEP {\bfseries 12}
  (2008) 039}, \href{http://arxiv.org/abs/0808.3674}{{\ttfamily arXiv:0808.3674
  [hep-ph]}}.

\bibitem{Schumann:2007mg}
S.~Schumann and F.~Krauss, {\em {A Parton shower algorithm based on
  Catani-Seymour dipole factorisation}},
  \href{http://dx.doi.org/10.1088/1126-6708/2008/03/038}{JHEP {\bfseries 03}
  (2008) 038}, \href{http://arxiv.org/abs/0709.1027}{{\ttfamily arXiv:0709.1027
  [hep-ph]}}.

\bibitem{dibosonxs}
J.~Campbell and R.~Ellis, {\em An update on vector boson pair production at
  hadron colliders}, \href{http://dx.doi.org/10.1103/PhysRevD.60.113006}{Phys.
  Rev. D {\bfseries 60} (1999) 113006},
  \href{http://arxiv.org/abs/{9905386}}{{\ttfamily {arXiv}:{9905386}
  [{hep-ph}]}}.

\bibitem{dibosonxs2}
J.~M. Campbell, R.~K. Ellis,  and C.~Williams, {\em {Vector boson pair
  production at the LHC}},
  \href{http://dx.doi.org/10.1007/JHEP07(2011)018}{JHEP {\bfseries 07} (2011)
  018},
\href{http://arxiv.org/abs/1105.0020}{{\ttfamily arXiv:1105.0020 [hep-ph]}}.

\bibitem{MCFM}
J.~M. Campbell, R.~K. Ellis,  and D.~L. Rainwater, {\em {Next-to-leading order
  QCD predictions for $W$ + 2 jet and $Z$ + 2 jet production at the CERN LHC}},
  \href{http://dx.doi.org/10.1103/PhysRevD.68.094021}{Phys. Rev. D {\bfseries
  68} (2003) 094021},
\href{http://arxiv.org/abs/hep-ph/0308195}{{\ttfamily arXiv:hep-ph/0308195
  [hep-ph]}}.

\bibitem{SingletopWtDRScheme}
S.~Frixione, E.~Laenen, P.~Motylinski, B.~R. Webber,  and C.~D. White, {\em
  {Single-top hadroproduction in association with a W boson}},
  \href{http://dx.doi.org/10.1088/1126-6708/2008/07/029}{JHEP {\bfseries 07}
  (2008) 029},
\href{http://arxiv.org/abs/0805.3067}{{\ttfamily arXiv:0805.3067 [hep-ph]}}.

\bibitem{stopxs}
N.~Kidonakis, {\em {Next-to-next-to-leading-order collinear and soft gluon
  corrections for $t$-channel single top quark production}},
  \href{http://dx.doi.org/10.1103/PhysRevD.83.091503}{Phys. Rev. D {\bfseries
  83} (2011) 091503}, \href{http://arxiv.org/abs/{1103.2792}}{{\ttfamily
  {arXiv}:{1103.2792} [{hep-ph}]}}.

\bibitem{stopxs_2}
N.~Kidonakis, {\em {Next-to-next-to-leading logarithm resummation for
  $s$-channel single top quark production}},
  \href{http://dx.doi.org/10.1103/PhysRevD.81.054028}{Phys. Rev. D {\bfseries
  81} (2010) 054028}, \href{http://arxiv.org/abs/{1001.5034}}{{\ttfamily
  {arXiv}:{1001.5034} [{hep-ph}]}}.

\bibitem{stopxs_3}
N.~Kidonakis, {\em {Two-loop soft anomalous dimensions for single top quark
  associated production with a $W^-$ or $H^-$}},
  \href{http://dx.doi.org/10.1103/PhysRevD.82.054018}{Phys. Rev. D {\bfseries
  82} (2010) 054018}, \href{http://arxiv.org/abs/{1005.4451}}{{\ttfamily
  {arXiv}:{1005.4451} [{hep-ph}]}}.

\bibitem{jimmy}
J.~Butterworth, J.~Forshaw,  and M.~Seymour, {\em {Multiparton interactions in
  photoproduction at HERA}}, \href{http://dx.doi.org/10.1007/s002880050286}{Z.
  Phys. C {\bfseries 72} (1996) 637},
  \href{http://arxiv.org/abs/hep-ph/9601371}{{\ttfamily
  {arXiv}:hep-ph/9601371}}.

\bibitem{PhotosPaper}
P.~Golonka and Z.~Was, {\em PHOTOS Monte Carlo: a precision tool for QED
  corrections in $Z$ and $W$ decays},
  \href{http://dx.doi.org/10.1140/epjc/s2005-02396-4}{Eur. Phys. J. C
  {\bfseries 45} (2006) 97},
  \href{http://arxiv.org/abs/hep-ph/0506026}{{\ttfamily arXiv:hep-ph/0506026
  [hep-ph]}}.

\bibitem{TauolaPaper}
S.~Jadach, J.~H. K{\"u}hn,  and Z.~W\c{a}s, {\em TAUOLA - a library of Monte
  Carlo programs to simulate decays of polarized $\tau$ leptons},
  \href{http://dx.doi.org/10.1016/0010-4655(91)90038-M}{Comput. Phys. Commun.
  {\bfseries 64} (1991) 275}.

\bibitem{Martin:2009iq}
A.~Martin, W.~Stirling, R.~Thorne,  and G.~Watt, {\em {Parton distributions for
  the LHC}}, \href{http://dx.doi.org/10.1140/epjc/s10052-009-1072-5}{Eur. Phys.
  J. C {\bfseries 63} (2009) 189--285},
\href{http://arxiv.org/abs/0901.0002}{{\ttfamily arXiv:0901.0002 [hep-ph]}}.

\bibitem{atlas_sim}
{ATLAS Collaboration}, {\em {The ATLAS Simulation Infrastructure}},
  \href{http://dx.doi.org/10.1140/epjc/s10052-010-1429-9}{Eur. Phys. J. C
  {\bfseries 70} (2010) 823},
  \href{http://arxiv.org/abs/{1005.4568}}{{\ttfamily {arXiv}:{1005.4568}
  [{physics.ins-det}]}}.

\bibitem{geant}
S.~Agostinelli {et~al.}, {\em {Geant4: a simulation toolkit}},
  \href{http://dx.doi.org/10.1016/S0168-9002(03)01368-8}{Nucl. Instrum. Meth.
  {\bfseries 506} (2003) 250}.

\bibitem{matrix_method_CONF}
{ATLAS Collaboration}, {\em {Estimation of non-prompt and fake lepton
  backgrounds in final states with top quarks produced in proton-proton
  collisions at $\sqrt{s}$ = 8 TeV with the ATLAS detector}},
  {ATLAS-CONF-2014-058,
  \href{https://cds.cern.ch/record/1951336}{https://cds.cern.ch/record/1951336}}.

\bibitem{Buckley:2010ar}
A.~Buckley {et~al.}, {\em {Rivet user manual}},
  \href{http://dx.doi.org/10.1016/j.cpc.2013.05.021}{Comput. Phys. Commun.
  {\bfseries 184} (2013) 2803--2819},
\href{http://arxiv.org/abs/1003.0694}{{\ttfamily arXiv:1003.0694 [hep-ph]}}.

\bibitem{Maltoni:2012pa}
F.~Maltoni, G.~Ridolfi,  and M.~Ubiali, {\em {b-initiated processes at the LHC:
  a reappraisal}}, \href{http://dx.doi.org/10.1007/JHEP04(2013)095,
  10.1007/JHEP07(2012)022}{JHEP {\bfseries 07} (2012) 022},
  \href{http://arxiv.org/abs/1203.6393}{{\ttfamily arXiv:1203.6393 [hep-ph]}},
[Erratum: JHEP04,095(2013)].

\bibitem{Sjostrand:2014zea}
T.~Sj{\"o}strand {et~al.}, {\em {An Introduction to PYTHIA 8.2}},
  \href{http://dx.doi.org/10.1016/j.cpc.2015.01.024}{Comput. Phys. Commun.
  {\bfseries 191} (2015) 159--177},
\href{http://arxiv.org/abs/1410.3012}{{\ttfamily arXiv:1410.3012 [hep-ph]}}.

\bibitem{Skands:2014pea}
P.~Skands, S.~Carrazza,  and J.~Rojo, {\em {Tuning PYTHIA 8.1: the Monash 2013
  Tune}}, \href{http://dx.doi.org/10.1140/epjc/s10052-014-3024-y}{Eur. Phys. J.
  C {\bfseries 74} (2014) 3024},
\href{http://arxiv.org/abs/1404.5630}{{\ttfamily arXiv:1404.5630 [hep-ph]}}.

\bibitem{ATL-PHYS-PUB-2015-007}
{ATLAS Collaboration}, {\em {A study of the sensitivity to the Pythia8 parton
  shower parameters of $t\bar{t}$ production measurements in $pp$ collisions at
  $\sqrt{s} = 7$ TeV with the ATLAS experiment at the LHC}},
  {ATL-PHYS-PUB-2015-007,
  \href{https://cds.cern.ch/record/2004362}{https://cds.cern.ch/record/2004362}}.

\bibitem{TimeShowerWebpage}
T.~Sj{\"o}strand {et~al.}, {\em {PYTHIA 8 online manual}},
\newblock
  \url{http://home.thep.lu.se/~torbjorn/pythia82html/TimelikeShowers.html}.
  Accessed: 2015-07-14.

\bibitem{ElectronPerformance}
{ATLAS Collaboration}, {\em Electron performance measurements with the ATLAS
  detector using the 2010 LHC proton-proton collision data},
  \href{http://dx.doi.org/10.1140/epjc/s10052-012-1909-1}{Eur. Phys. J. C
  {\bfseries 72} (2012) 1909}, \href{http://arxiv.org/abs/1110.3174}{{\ttfamily
  arXiv:1110.3174 [hep-ex]}}.

\bibitem{ATLASJetEnergyMeasurement}
{ATLAS Collaboration}, {\em {Jet energy measurement with the ATLAS detector in
  proton--proton collisions at $\sqrt{s} = 7\;\mbox{TeV}$}},
  \href{http://dx.doi.org/10.1140/epjc/s10052-013-2304-2}{Eur. Phys. J. C
  {\bfseries 73} (2013) 2304}, \href{http://arxiv.org/abs/1112.6426}{{\ttfamily
  arXiv:1112.6426 [hep-ex]}}.

\bibitem{jvf}
{ATLAS Collaboration}, {\em {Performance of pile-up mitigation techniques for
  jets in $pp$ collisions at $\sqrt{s} = 8$ TeV using the ATLAS detector}},
\href{http://arxiv.org/abs/1510.03823}{{\ttfamily arXiv:1510.03823 [hep-ex]}}.

\bibitem{btagging2}
{ATLAS Collaboration}, {\em {Calibration of $b$-tagging using dileptonic top
  pair events in a combinatorial likelihood approach with the ATLAS
  experiment}}, {ATLAS-CONF-2014-004,
  \href{https://cds.cern.ch/record/1664335}{https://cds.cern.ch/record/1664335}}.

\bibitem{lumi}
{ATLAS Collaboration}, {\em {Luminosity determination in $pp$ collisions at
  $\sqrt{s}=7\tev$ using the ATLAS detector at the LHC}},
  \href{http://dx.doi.org/10.1140/epjc/s10052-011-1630-5}{Eur. Phys. J. C
  {\bfseries 71} (2011) 1630}, \href{http://arxiv.org/abs/1101.2185}{{\ttfamily
  arXiv:1101.2185 [hep-ex]}}.

\bibitem{ATLAS-CONF-2014-032}
{ATLAS Collaboration}, {\em {Electron efficiency measurements with the ATLAS
  detector using the 2012 LHC proton--proton collision data}},
  {ATLAS-CONF-2014-032,
  \href{http://cdsweb.cern.ch/record/1706245}{http://cdsweb.cern.ch/record/1706245}}.

\bibitem{PERF-2014-05}
{ATLAS Collaboration}, {\em {Measurement of the muon reconstruction performance
  of the ATLAS detector using 2011 and 2012 LHC proton--proton collision
  data}}, \href{http://dx.doi.org/10.1140/epjc/s10052-014-3130-x}{Eur. Phys. J.
  C {\bfseries 74} (2014) 3130},
  \href{http://arxiv.org/abs/1407.3935}{{\ttfamily arXiv:1407.3935 [hep-ex]}}.

\bibitem{PERF-2013-05}
{ATLAS Collaboration}, {\em {Electron and photon energy calibration with the
  ATLAS detector using LHC Run 1 data}},
  \href{http://dx.doi.org/10.1140/epjc/s10052-014-3071-4}{Eur. Phys. J. C
  {\bfseries 74} (2014) 3071}, \href{http://arxiv.org/abs/1407.5063}{{\ttfamily
  arXiv:1407.5063 [hep-ex]}}.

\bibitem{jvf2}
{ATLAS Collaboration}, {\em {Pile-up subtraction and suppression for jets in
  ATLAS}}, {ATLAS-CONF-2013-083,
  \href{https://cds.cern.ch/record/1570994}{https://cds.cern.ch/record/1570994}}.

\bibitem{PERF-2012-01}
{ATLAS Collaboration}, {\em {Jet energy measurement and its systematic
  uncertainty in proton--proton collisions at $\sqrt{s} = 7\;\mbox{TeV}$ with
  the ATLAS detector}},
  \href{http://dx.doi.org/10.1140/epjc/s10052-014-3190-y}{Eur. Phys. J. C
  {\bfseries 75} (2015) 17}, \href{http://arxiv.org/abs/1406.0076}{{\ttfamily
  arXiv:1406.0076 [hep-ex]}}.

\bibitem{PERF-2011-04}
{ATLAS Collaboration}, {\em {Jet energy resolution in proton--proton collisions
  at $\sqrt{s} = 7\;\mbox{TeV}$ recorded in 2010 with the ATLAS detector}},
  \href{http://dx.doi.org/10.1140/epjc/s10052-013-2306-0}{Eur. Phys. J. C
  {\bfseries 73} (2013) 2306}, \href{http://arxiv.org/abs/1210.6210}{{\ttfamily
  arXiv:1210.6210 [hep-ex]}}.

\bibitem{gap_fraction}
{ATLAS Collaboration}, {\em {Measurement of $t \bar{t}$ production with a veto
  on additional central jet activity in pp collisions at $\sqrt{s}$ = 7 TeV
  using the ATLAS detector}},
  \href{http://dx.doi.org/10.1140/epjc/s10052-012-2043-9}{Eur. Phys. J. C
  {\bfseries 72} (2012) 2043},
\href{http://arxiv.org/abs/1203.5015}{{\ttfamily arXiv:1203.5015 [hep-ex]}}.

\bibitem{ref:pdf4lhc}
M.~Botje {et~al.}, {\em The PDF4LHC Working Group Interim Recommendations},
  \href{http://arxiv.org/abs/1101.0538}{{\ttfamily arXiv:1101.0538 [hep-ph]}}.

\bibitem{mstw1}
A.~D. Martin {et~al.}, {\em {Parton distributions for the LHC}},
  \href{http://dx.doi.org/10.1140/epjc/s10052-009-1072-5}{Eur. Phys. J. C
  {\bfseries 63} (2009) 189},
  \href{http://arxiv.org/abs/{0901.0002}}{{\ttfamily {arXiv}:{0901.0002}
  [{hep-ph}]}}.

\bibitem{mstw2}
A.~D. Martin {et~al.}, {\em {Uncertainties on $\alpha_S$ in global PDF analyses
  and implications for predicted hadronic cross sections}},
  \href{http://dx.doi.org/10.1140/epjc/s10052-009-1072-5}{Eur. Phys. J. C
  {\bfseries 64} (2009) 653},
  \href{http://arxiv.org/abs/{0905.3531}}{{\ttfamily {arXiv}:{0905.3531}
  [{hep-ph}]}}.

\bibitem{ct102}
J.~Gao {et~al.}, {\em {CT10 next-to-next-to-leading order global analysis of
  QCD}}, \href{http://dx.doi.org/10.1103/PhysRevD.89.033009}{Phys. Rev. D
  {\bfseries 89} (2014) 033009},
\href{http://arxiv.org/abs/1302.6246}{{\ttfamily arXiv:1302.6246 [hep-ph]}}.

\bibitem{nnpdf}
R.~D. Ball {et~al.}, {\em Parton distributions with LHC data},
  \href{http://dx.doi.org/10.1016/j.nuclphysb.2012.10.003}{Nucl. Phys. B
  {\bfseries 867} (2013) 244}, \href{http://arxiv.org/abs/1207.1303}{{\ttfamily
  arXiv:1207.1303 [hep-ph]}}.

\bibitem{mcatnlo_1}
S.~Frixione and B.~R. Webber, {\em {Matching NLO QCD computations and parton
  shower simulations}},
  \href{http://dx.doi.org/10.1088/1126-6708/2002/06/029}{JHEP {\bfseries 06}
  (2002) 029},
\href{http://arxiv.org/abs/0204244}{{\ttfamily arXiv:0204244 [hep-ph]}}.

\bibitem{mcatnlo_2}
S.~Frixione, E.~Laenen, P.~Motylinski,  and B.~R. Webber, {\em {Single-top
  production in MC@NLO}},
  \href{http://dx.doi.org/10.1088/1126-6708/2006/03/092}{JHEP {\bfseries 03}
  (2006) 092}, \href{http://arxiv.org/abs/0512250}{{\ttfamily arXiv:0512250
  [hep-ph]}}.

\bibitem{mcatnlo_3}
S.~Frixione, E.~Laenen, P.~Motylinski, B.~R. Webber,  and C.~White, {\em
  {Single-top hadroproduction in association with a $W$ boson}},
  \href{http://dx.doi.org/10.1088/1126-6708/2008/07/029}{JHEP {\bfseries 07}
  (2008) 029}, \href{http://arxiv.org/abs/0805.3067}{{\ttfamily arXiv:0805.3067
  [hep-ph]}}.

\bibitem{asimov}
G.~Cowan, K.~Cranmer, E.~Gross,  and O.~Vitells, {\em {Asymptotic formulae for
  likelihood-based tests of new physics}},
  \href{http://dx.doi.org/10.1140/epjc/s10052-011-1554-0,
  10.1140/epjc/s10052-013-2501-z}{Eur. Phys. J. C {\bfseries 71} (2011) 1554},
  \href{http://arxiv.org/abs/1007.1727}{{\ttfamily arXiv:1007.1727
  [physics.data-an]}},
[\href{http://link.springer.com/article/10.1140%2Fepjc%2Fs10052-013-2501-z}{Erratum:
  Eur. Phys. J.C73,2501(2013)}].

\end{thebibliography}\endgroup
